# Electron Spin for Classical Information Processing: A Brief Survey of Spin-Based Logic Devices, Gates and Circuits

Supriyo Bandyopadhyay[1,†] and Marc Cahay[*]

†Department of Electrical and Computer Engineering

Virginia Commonwealth University

Richmond, VA 23284, USA

*Department of Electrical and Computer Engineering

University of Cincinnati

Cincinnati, OH 45221, USA

ABSTRACT

In electronics, information has been traditionally stored, processed and communicated using an electron's charge. This paradigm is increasingly turning out to be energy-inefficient, because movement of charge within an information-processing device invariably causes current flow and an associated dissipation. Replacing "charge" with the "spin" of an electron to encode information may eliminate much of this dissipation and lead to more energy-efficient "green electronics". This realization has spurred significant research in spintronic devices and circuits where spin either directly acts as the physical variable for hosting information or augments the role of charge. In this review article, we discuss and elucidate some of these ideas, and highlight their strengths and weaknesses. Many of them can potentially reduce energy dissipation significantly, but unfortunately are error-prone and unreliable. Moreover, there are serious obstacles to their technological implementation that may be difficult to overcome in the near term.

---

[1] Corresponding author. E-mail: sbandy@vcu.edu

This review addresses three constructs: (1) single devices or binary switches that can be constituents of Boolean logic gates for digital information processing, (2) complete gates that are capable of performing specific Boolean logic operations, and (3) combinational circuits or architectures (equivalent to many gates working in unison) that are capable of performing universal computation.

# TABLE OF CONTENTS

# 1. INTRODUCTION

The workhorse of modern digital electronics is the celebrated "transistor" discovered by Bardeen, Brattain and Shockley more than sixty years ago. The transistor is a quintessential *charge based device* where an electron's charge is utilized to encode information in digital form. This is most evident in the case of the metal-insulator-semiconductor-field-effect-transistor (MISFET), which has three terminals: the source, the drain and the gate. When a positive potential is applied on the gate terminal that sits between the source and the drain, it attracts negatively charged electrons into the channel region separating the source and the drain. The incoming electrons make the channel conducting and allow current to flow between the source and drain. This is the ON state of the transistor. If we reverse the polarity of the gate potential, then electrons are repelled from the channel by the negative charge on the gate terminal, which makes the channel non-conducting and drops the device conductance dramatically (usually by ~ 6 orders of magnitude). This turns the transistor OFF. The two conductance states - ON and OFF – are used to encode the binary bits 0 and 1, which, in turn, allows one to store and process digital information using binary bit streams.

The problem with this strategy is that switching between the ON and OFF states always requires moving an amount of charge $Q$ in and out of the channel, which needs an amount of energy equal to $Q\Delta V$, where $\Delta V$ is the change in the gate potential necessary to change the channel charge by the quantity $Q$. This energy is dissipated as heat when the switching action is over.

Heat dissipation is a serious spoiler in modern integrated circuits. Today's transistors dissipate between 1,500 and 2,500 eV (0.24-0.4 fJ) of energy when they switch from the ON to the OFF state, or vice versa [1]. The Pentium IV chip has a transistor density of $10^8$ cm$^{-2}$ and with a 2.8 GHz clock should dissipate between $2.4 \times 10^{-16} \times 10^8 \times 2.8 \times 10^9$ = 67 Watts/cm$^2$ and $4 \times 10^{-16} \times 10^8 \times 2.8 \times 10^9$ = 112 Watts/cm$^2$ if all the transistors switch simultaneously. With a 5% activity level, i.e. if only one in twenty transistors switch at any given time, the power dissipation should be between 3.35 and 5.6 Watts/cm$^2$. In actuality, the Pentium IV chip, which has an area of 13 cm$^2$ (Northwood generation), dissipates about 50 Watts [2], which translates to a power dissipation of 3.8 Watts/cm$^2$. This level of power dissipation is not particularly worrisome since removal of 1000 Watts/cm$^2$ from a chip using conventional technology was demonstrated almost 30 years ago [3].

The chip making industry however is driven by an empirical law known as Moore's law [4] that mandates doubling of the transistor density on a chip every 18 months or so. If Moore's law holds in perpetuity, we should reach a transistor density of ~ $10^{13}$ cm$^{-2}$ by 2025, given that the Pentium IV chip was released in ca. 2000. Furthermore, if the clock speed increases to 10 GHz by then, the power dissipation on a chip will reach 2 MW/cm$^2$ (still with 5% activity level) unless we can reduce the energy dissipation in a transistor during switching. The thermal load associated with a heat dissipation of 2 MW/cm$^2$ approaches that in a rocket nozzle. Therefore, unless revolutionary new heat sinking technologies are pressed into service, thermal management on a chip will inevitably fail and continued miniaturization of transistors in accordance with Moore's law should stop well before 2025. The International Technology Roadmap for Semiconductors has termed this impending catastrophe the "Red Brick Wall" [1].

As long as we use charge to encode information, we will always be menaced by the Red Brick Wall since charge based electronics is intrinsically energy-inefficient and dissipative. This is due to the fact that charge is a *scalar* quantity which has only magnitude and no direction or any other attribute. Therefore, if we intend to encode binary bit information in charge, then we must do so by using two different *magnitudes* or amounts of charge – $Q_1$ and $Q_2$ – to represent the two bits. We cannot use anything else, other than a difference in the magnitude, to demarcate the logic levels. If that is the case, then switching will always dissipate an amount of energy equal to $|\Delta V (Q_1 - Q_2)|$, which is never zero since $Q_1 \neq Q_2$. In fact, we would prefer to make $Q_1$ very different from $Q_2$ so that the logic bits are well-separated and clearly distinguishable from each other in a noisy environment, which will reduce the error associated with mistaking one for the other. Thus, there is a fundamental connection between dissipation and bit error probability. Increasing one reduces the other and vice versa. This fundamental truism is enshrined in the famous Landauer-Shannon result for the ultimate limits on dissipation that we will have occasion to discuss later in this article.

## 2. SPIN BASED TRANSISTORS AND LOGIC SWITCHES

Since the fundamental source of dissipation in all charge based transistors is charge motion within the device mandated by the requirement that $Q_1 \neq Q_2$, it is natural to seek ways where the switching action of the transistor, i.e. switching from a high conductance state to a low conductance state (or vice versa), can be accomplished *without* changing the amount of charge in the device. In other words, we wish to switch between two well distinguished logic levels while maintaining $Q_1 = Q_2$. To our knowledge, the first proposal that found a route to realizing this objective is that of the celebrated Spin Field Effect Transistor (SPINFET) where the spin degree of freedom of an electron is utilized to switch the conductance of a MISFET-like structure from high to low, or vice versa, *without* changing the amount of charge in the channel. We describe this device next.

## 2.1. The Spin Field Effect Transistor (SPINFET)

In a seminal paper published in 1990 [5], Datta and Das proposed a remarkable idea for changing the conductance of a MISFET-like device without changing the channel charge at all. The way they accomplished this was to use the spin degree of freedom of channel electrons, instead of the charge degree of freedom, to switch the channel conductance. Their device, which has since been dubbed a SPINFET (Spin Field Effect Transistor), has the exact same physical structure as a traditional MISFET, except that the source and drain contacts are ideal (half-metallic) ferromagnets that are magnetized in the direction of current flow in the channel. We call this the "parallel configuration" since both contacts are magnetized in the same direction, i.e. the north pole of the source magnet faces the south pole of the drain magnet, or vice versa. This device is represented schematically in Fig. 1.

The SPINFET has been discussed in many papers dealing with spintronics since it truly inspired the entire field. Here, we will describe its operation by invoking only classical physics since there is a misperception among some that it relies on quantum mechanical interference between two orthogonal spin states, and is therefore a quantum interference device. That is not correct. Although, in ref. [5], Datta and Das described their device in a language that might convey this impression, in reality this device is purely classical and its operation can be described classically. That should be welcome news since quantum interference devices are very delicate [6, 7], and while they certainly have a role as laboratory curiosities, they are unlikely to be practical as operating devices in a circuit environment. The Datta-Das device, if ever realized, will actually be quite robust because it is classical.

To keep the ensuing discussion focused on the essential elements, we will assume that the channel of the SPINFET in Fig. 1 is a strictly one-dimensional quantum wire. A one-dimensional channel performs better than the more conventional two-dimensional channel for many reasons. First, the two dimensional device can never be turned off completely, even in the ideal case, because the effect of the gate potential on the spin of an electron in the channel depends on the direction of the electron's velocity [5]. In a two-dimensional channel, the electron's velocity spans two dimensions, so that even when the current is shut off completely for electrons with velocity along one direction, it is not shut off for electrons whose velocities have non-zero components in the orthogonal direction. Hence, a transistor with a two-dimensional channel will invariably have a significant leakage current flowing in the OFF state. In contrast, electrons in a one-dimensional channel have their velocity vectors always pointing along one direction only. Such a transistor can be completely turned off (provided everything else is ideal), resulting in zero leakage current. A second reason for preferring a one-dimensional channel is that the SPINFET works best if there is no random spin relaxation in the channel. In semiconductor channels, random spin relaxation usually occurs via two modes: the D'yakonov-Perel' mode [8] and the Elliott-Yafet mode [9]. The former is completely suppressed in a one-dimensional channel [10-12] and the latter is also significantly reduced by one-dimensional confinement. Therefore, the one-dimensional device is not just a convenient idealization, it actually happens to be the ideal prototype.

We will now make three assumptions that will allow us to describe the SPINFET's operation lucidly.

- First, the ideal half-metallic ferromagnetic source contact is assumed to be a perfect spin injector that injects only its majority spin, at the complete exclusion of its minority spin. Similarly, the half-metallic ferromagnetic drain contact is assumed to be an ideal spin detector (or transmitter) that transmits only its own majority spin and completely blocks its minority spin. In other words, the spin injection and detection efficiencies at the source and drain respectively, defined as

$$\xi_S = \xi_D = \frac{I_{maj} - I_{min}}{I_{maj} + I_{min}}, \qquad (1)$$

  are both 100%. Here $I_{maj}$ is the injected or detected current due to electrons with the majority spin and $I_{min}$ is that due to electrons with the minority spin.

- Second, we will assume that the ferromagnetic contacts do not cause any magnetic field in the channel. A magnetic field can cause "spin mixing" in the channel that has deleterious effects which will adversely affect the transistor's operation [13, 14].

- Finally, we will assume that there is no random spin relaxation in the channel, i.e. neither the D'yakonov-Perel' nor the Elliott-Yafet mechanism (nor any other spin relaxing mechanism for that matter) is operative. There can be plenty of scattering in the channel due to phonons, other electrons or holes, and non-magnetic impurities, but they should not relax spin.

As shown in Fig. 1, the source and drain contacts are magnetized in the +z direction so that the majority spins in both contacts are polarized in the +z-direction. Hence all spins injected from the source into the channel under a source-to-drain bias are initially polarized in the +z-direction. This is also the direction of current flow or electron velocity in the channel.

Without any gate voltage, the injected spins arrive at the drain with their polarizations intact and transmit completely through the drain contact (since the majority spin in the source is also majority spin in the gate), resulting in maximum channel current. When the gate voltage is turned on, it causes an electric field to appear in the +y-direction. This electric field induces a Rashba spin-orbit interaction [15] in the channel which gives rise to an effective magnetic field in the direction mutually perpendicular to the electron velocity and the electric field, i.e. in the +x-direction. The flux density of this field is [16]

$$\vec{B}_{Rashba} = \left[\frac{2m^*a_{46}}{g\mu_B\hbar}\mathrm{E}_y v_z\right]\hat{x}, \qquad (2)$$

where $m^*$ is the effective mass of electrons in the channel, $a_{46}$ is a material constant indicative of the strength of the Rashba spin-orbit interaction in the channel material, $\mathrm{E}_y$ is the electric field in the y-direction induced by the applied gate voltage, $v_z$ is the z-directed velocity of electrons in the channel and $\hat{x}$ is the unit vector in the x-direction.

Because we assumed 100% spin injection efficiency, the source injects electrons into the channel with their spins polarized exclusively in the +z-direction. The x-directed magnetic field $\vec{B}_{Rashba}$, caused by the gate potential, will make these spins precess about $\vec{B}_{Rashba}$ in the y-z plane (Larmor precession) as they travel towards the drain. The angular frequency of this spin precession is given by the Larmor formula:

$$\Omega = \frac{d\phi}{dt} = \frac{g\mu_B B_{Rashba}}{\hbar} = \frac{2a_{46}m^*}{\hbar^2}\mathrm{E}_y v_z \qquad (3)$$

and therefore the corresponding spatial rate of precession is

$$\frac{d\phi}{dz} = \frac{d\phi}{dt}\bigg/\frac{dz}{dt} = \frac{\Omega}{v_z} = \frac{2a_{46}m^*}{\hbar^2}\mathrm{E}_y. \qquad (4)$$

Note from the last equation that the spatial rate of precession is *independent* of the electron velocity and therefore is the *same* for every electron, regardless of its velocity or kinetic energy. Consequently, every electron precesses by exactly the same angle in traversing the channel. A scattering event due to a phonon or impurity in the channel can change an electron's velocity and scatter it backwards, but as long as it reverses direction once again (owing to another scattering event or because of the electrostatic potential gradient between the source and drain) and finally arrives at the drain, its spin would have precessed by exactly the same angle as every other electron that traversed the channel, irrespective of its scattering history! Thus, the angle by which every spin precesses as it travels from the source to the drain is constant and given by

$$\Phi = \frac{2a_{46}m^*}{\hbar^2}\mathrm{E}_y L, \tag{5}$$

where $L$ is the channel length (distance between source and drain). Note, once again, that as long as the channel is one-dimensional, transport in the SPINFET channel does *not* have to be ballistic to keep $\Phi$ constant. There can be frequent scattering in the channel, but as long as these scattering events conserve spin, $\Phi$ will be the same for every electron.

Using the so-called Bloch sphere concept [16], and noting that every electron entered the channel with +z-polarized spin, the 2×1 component spinor that represents the spin of any arbitrary electron arriving at the drain contact can be written as

$$[\psi]_{drain} = e^{i\gamma}\begin{bmatrix}\cos\left(\frac{\Phi}{2}\right) \\ \sin\left(\frac{\Phi}{2}\right)e^{i\varphi}\end{bmatrix}, \tag{6}$$

where $\nu$ and $\varphi$ are two arbitrary phase angles and $\Phi$ is the spin precession angle given in Equation (5).

The probability amplitude for this arriving spin to transmit through the drain (whose majority spins are +z-polarized and hence described by the spinor $\begin{bmatrix}1\\0\end{bmatrix}$) is given by

$$t = e^{-i\gamma}\left[\cos\left(\frac{\Phi}{2}\right)\quad \sin\left(\frac{\Phi}{2}\right)e^{-i\varphi}\right]\begin{bmatrix}1\\0\end{bmatrix} = e^{-i\gamma}\cos\left(\frac{\Phi}{2}\right). \qquad (7)$$

Note that this transmission amplitude $t$ is the same for every electron since every one of them entered the channel with the same spin orientation (+z) from the source and rotated by the same angle by the time it arrived at the drain. Therefore $\Phi$ and consequently $t$ is the same for every electron.

The current through a one-dimensional conductor subjected to a bias of $V_{bias}$ is given by the Tsu-Esaki formula [17] $I = \frac{2e}{h}\int_0^\infty dE\,|t(E)|^2\left[f(E-E_F) - f(E+eV_{bias}-E_F)\right]$ [2], where $E_F$ is the Fermi energy in the injecting contact. Therefore, the source-to-drain current in a SPINFET with one-dimensional channel subjected to a source-to-drain bias of $V_{SD}$ is given by

$$\begin{aligned}
I_{SD} &= \frac{2e}{h}\int_0^\infty dE\,|t(E)|^2\left[f(E-E_F) - f(E+eV_{SD}-E_F)\right] \\
&= \underbrace{\left\{\frac{2e}{h}\int_0^\infty dE\,\cos^2\left(\frac{\Phi}{2}\right)\left[f(E-E_F) - f(E+eV_{SD}-E_F)\right]\right\}}_{I_0} \\
&= \cos^2\left(\frac{\Phi}{2}\right)\underbrace{\left\{\frac{2e}{h}\int_0^\infty dE\left[f(E-E_F) - f(E+eV_{SD}-E_F)\right]\right\}}_{I_0} \quad \text{(since }\Phi\text{ is energy-independent)} \\
&= I_0\cos^2\left(\frac{\Phi}{2}\right).
\end{aligned} \qquad (8)$$

---

[2] This result is strictly valid when there is no inelastic scattering in the channel. Hence we are assuming the absence of inelastic scattering processes. Elastic scattering processes can be present.

where $I_0 = \dfrac{2e}{h} \int_0^\infty dE \left[ f(E - E_F) - f(E + eV_{SD} - E_F) \right]$.

The above relation clearly shows that the source-to-drain current is maximum when the gate voltage (or, equivalently, the gate induced electric field $E_y$) is such that $\Phi$ is an even multiple of π, and it falls to zero when the gate voltage is such that $\Phi$ is an odd multiple of π. Since $\Phi$ depends on $E_y$ (see Equation (5)), therefore, by changing the gate voltage (and hence $E_y$), one can modulate the channel current and realize transistor action. Basically, we can change the current from maximum to zero and vice versa, i.e. turn the transistor ON and OFF with the gate voltage, by rotating the spin orientation of channel electrons *without having to change their number or concentration*. This is the basic principle of the SPINFET, whose ideal transfer characteristic (channel current versus gate voltage) is schematically depicted in Fig. 2.

### 2.1.1. The bane of spin injection efficiency

Fig. 2 and Equation (8) seem to indicate that the ratio of the ON-to-OFF conductance, which is also the ratio of the ON-current to OFF-current at a fixed source-to-drain bias, will be infinity since the OFF-current (corresponding to $\Phi = (2n+1)\pi$) is zero[3]. In reality, this can be true only if the source spin injection efficiency and the drain spin detection efficiency are both 100%, i.e. the source injects only +z-polarized spins and the drain also transmits only +z-polarized spins, at the complete exclusion of −z-polarized spins. This is what we had tacitly assumed in deriving Equation (8). However, if either efficiency is less than 100%, then −z-polarized spins will also be injected by the source and/or transmitted by the drain, which will degrade the on-to-off conductance ratio. It has been shown [16, 18] that when the non-ideality of spin injection and detection is taken into account, the maximum conductance ratio is given by

$$\eta = \frac{G_{ON}}{G_{OFF}} = \frac{I_{ON}}{I_{OFF}} = \frac{1+\xi_S\xi_D}{1-\xi_S\xi_D}, \tag{9}$$

where $\xi_S$ and $\xi_D$ are the source injection and drain detection efficiencies, respectively. This ratio becomes infinity only when $\xi_S = \xi_D = 1$ and decreases rapidly with falling $\xi_S$ and $\xi_D$. Obviously, in order to achieve a conductance on-off ratio of at least $10^5$, typical of today's transistors, these injection and detection efficiencies will have to be as high as 99.9995%, which is clearly impractical (at least at room temperature) given that the maximum spin injection efficiency demonstrated at a ferromagnetic/semiconductor junction to date is only ~ 70% [19], which would make the conductance on-off ratio $\eta$ a paltry 2.9. That is clearly inadequate for any mainstream application. In the near term, it is unlikely that spin injection and detection efficiencies of 99.9995% can be achieved at room temperature, which means that: (1) the conductance ratio will remain small, and (2) the leakage current during the OFF state will be large. The former leads to a large bit error probability (since the ON and OFF states will not be all that distinguishable in a noisy environment if the conductance ratio is small) and the latter leads to unacceptable standby power dissipation in a circuit (the transistor will continuously dissipate energy since it is flowing current even when it is OFF). Ultimately, the poor spin injection and detection efficiencies are the show-stopper; they will make the SPINFET energy-inefficient and error-prone (i.e. unreliable), even if this device could be built.

---

[3] The reader should appreciate that the OFF-current is zero only because $\Phi$ is energy-independent.

In Fig. 3, we plot the conductance on-off ratio as a function of the spin injection or detection efficiency, assuming that the two efficiencies are equal. The ratio drops off precipitously with decreasing efficiency and drops to ~ 10 when the injection or detection efficiency reaches 90%, which is a rather high efficiency yet to be achieved at room temperature. Therefore, the conductance on-off ratio of the SPINFET is likely to remain low in the near term, which translates to high bit error rate and large standby power dissipation – both undesirable traits.

### 2.1.2. Other types of SPINFETs

A number of modifications of the original Datta-Das idea have appeared in the literature over the last decade. All of them share the property that they switch conductance ON and OFF by modulating spin instead of channel charge concentration. One replaces the Rashba interaction in the channel with the Dresselhaus spin-orbit interaction with everything else the same [20], another uses a delicate balance between the Rashba and Dresselhaus interactions [21, 22], and a third relies on inducing spin relaxation in the channel with a gate voltage [23] to switch between ON and OFF states. None of these ideas engender any improvement in the conductance on-off ratio since all of them still require ideal spin injection and detection at ferromagnet/channel interfaces. Therefore, they all have too low a conductance ON/OFF ratio to be of much use. None of them is any more energy-efficient than the original Datta-Das device, either.

### 2.1.3. Is the SPINFET any more energy efficient than the traditional MISFET?

Let us put aside the issue of conductance on-off ratio for the time being and investigate if it had not been a problem, would the SPINFET been competitive with the traditional MISFET. It would have been competitive if it consumed less energy when it switched ON or OFF. Since the SPINFET switches without requiring a change in the electron concentration in the channel, it may at first appear to be more energy-efficient than the MISFET because $\Delta V(Q_1 - Q_2) = 0$. However, in reality, it is not. First of all, it should be recognized that there is still charge flowing through the channel of the SPINFET in the ON state since there is a source-to-drain current. This causes dissipation just as it does in a traditional MISFET. Therefore, there is no advantage as far as *static* power dissipation (during the ON state) is concerned. During the OFF state, the SPINFET actually has much more static power dissipation if we consider the fact that it has a much larger leakage current than a MISFET because of the poor conductance ON/OFF ratio. Finally, if we consider the *dynamic* power that is dissipated during switching, then the winner is determined by which device – the SPINFET or the MISFET – requires a smaller turn-on or turn-off voltage. The turn-on voltage is the potential applied at the gate to turn the device ON (if the device is normally OFF) while the turn-off voltage is the gate potential required to turn the device OFF (if it is normally ON). These potentials cause charge movement somewhere in the circuit and the associated energy dissipation is almost always $\sim C_G V_G^2$, where $C_G$ is the gate capacitance and $V_G$ is the turn-on (or turn-off) voltage. Assuming that $C_G$ is the same for the MISFET and the SPINFET, since it is determined by structural parameters, what finally determines the winner among them is $V_G$.

Refs. [18] and [24] have carried out detailed comparisons between the turn-on or turn-off voltages of a SPINFET and a traditional MISFET. It should be obvious from Equation (5) that the gate electric field required to turn a SPINFET off is $\mathrm{E}_y^{off} = \frac{\pi \hbar^2}{2 a_{46} m^* L}$ (corresponding to $\Phi = \pi$) so that the turn-off voltage will depend on the channel length $L$. Longer-channel devices require a smaller turn-off voltage and therefore are more energy-efficient. Refs. [18] and [24] concluded that the turn-off voltage of a SPINFET with reasonable channel length (smaller than 100 nm), is actually much larger than that of a comparable normally-on MISFET, primarily because the strength of Rashba spin-orbit interaction (denoted by the quantity $a_{46}$) is usually too small in the conduction band of technologically important semiconductors to make $E_y^{off}$ small enough to yield a small turn-off voltage in a short channel device of length $\leq 100$ nm. Only a SPINFET with a channel length larger than a few μm could be more energy efficient than a MISFET. Such large devices are no longer practical or affordable. Therefore, the SPINFET does not make a better digital switch and hence is not particularly desirable for digital information processing.

### 2.1.4. Analog applications of the SPINFET

If the SPINFET would not yield any significant advantage in digital electronics, does it have a role to play in analog electronics? Unfortunately, the answer again is no. For analog applications, the two most important metrics are power gain and bandwidth. Both are determined by the transconductance of the transistor. Ref. [16] showed that the maximum transconductance of SPINFET with a one-dimensional channel is

$$|g_m| = \frac{2e^2}{h} V_{SD} \xi_S \xi_D \frac{m^* a_{46}}{\hbar^2} \frac{\kappa_i}{\kappa_s} \frac{L}{d} \sin \Phi, \tag{10}$$

where $\kappa_i$ and $\kappa_s$ are the dielectric constants of the gate insulator and semiconductor channel, respectively, and $d$ is the width of the gate insulator layer (see Fig. 1).

In the common-source or common-gate configurations, a single stage transistor amplifier has a voltage gain given by [25]

$$a_v = \frac{g_m}{g_0}, \tag{11}$$

where $g_0$ is the channel conductance. This quantity is equal to $I_{SD}/V_{SD}$ when the transistor is operated at small source-to-drain biases so that it operates in the linear response regime. Ref. [16] showed that in the linear response regime, $I_{SD} = \frac{e^2}{h} V_{SD} \left(1 + \xi_S \xi_D \cos \Phi \right)$. Therefore, the expression for the voltage gain becomes

$$|a_v| = \xi_S \xi_D \frac{2m^* a_{46}}{\hbar^2} \frac{\kappa_i}{\kappa_s} \frac{L}{d} V_{SD} \frac{\sin \Phi}{1 + \xi_S \xi_D \cos \Phi}, \qquad (12)$$

which becomes

$$|a_v| = \frac{2m^* a_{46}}{\hbar^2} \frac{\kappa_i}{\kappa_s} \frac{L}{d} V_{SD} \tan\left(\frac{\Phi}{2}\right) \text{ if } \xi_S = \xi_D = 1. \qquad (13)$$

Note that the voltage gain depends on both the source-to-drain bias $V_{SD}$ and the gate bias (through $\Phi$) and becomes quite large (approaching infinity) as the transistor approaches cut-off condition ($\Phi = \pi$). Therefore it is not at all true that the device has no gain. It *does have gain* contrary to what was casually claimed in ref. [26]. The voltage gain however vanishes when the transistor is fully on ($\Phi = 0$) because there the transconductance $g_m$ is zero while the channel conductance $g_0$ is not. Therefore, it is not possible to maintain adequate voltage gain while supplying enough current through the device to drive several succeeding stages in wired analog circuits. In other words, it is not possible to have simultaneously a large voltage gain and a large fan out, which makes this device unsuitable for mainstream applications in analog electronic circuits.

### 2.1.5. Experimental status of the SPINFET

To our knowledge, the SPINFET has never been experimentally demonstrated despite a serious and concerted worldwide effort spanning nearly two decades. Gate control of the Rashba interaction in a SPINFET-like structure was however demonstrated many years ago [27], but it failed to produce any noticeable modulation of the source-to-drain current. The experiment in ref. [27] however showed unambiguously that the quantity $\frac{\partial \Phi}{\partial V_G}$ is very small, even in an InAs channel that should have strong Rashba interaction, indicating that very large gate voltages will be required to precess spins by $180^0$ which will turn the transistor off. This tells us that the SPINFET is not a low power device and therefore not energy-efficient, in agreement with the claim of ref. [24].

Ref. [23] made the bold claim that their SPINFET device will be extremely energy-efficient (much more efficient than a MISFET) since a very small gate voltage (~ 100 mV) can supposedly turn it on. This device is different from the Datta-Das construct in that it does not rely on spin precession and its source and drain magnetizations are anti-parallel, instead of parallel[4]. The source is assumed to inject spins into the channel with nearly 100% efficiency. When the gate voltage is zero, the injected spins will not relax in the channel since the spin orbit interaction in the channel will be weak or non-existent. The drain therefore will block all the spins from transmitting (assuming near 100% spin detection efficiency) – since it is anti-parallel with the source - and the channel current should be ideally zero. When a small voltage (~ 100 mV) is applied on the gate, it supposedly causes strong enough spin-orbit interaction in the channel to ensure rapid spin relaxation. The injected spins begin to flip in the channel and the flipped spins transmit through the drain, turning the device on. Ref. [23] claimed that the turn on voltage (the transistor will be turned fully on when ~ 50% of injected spins transmit through the drain) will be ~ 100 mV.

---

[4] The Datta-Das device works equally well with parallel and anti-parallel magnetizations of the source and drain contacts. The parallel configuration results in a normally-on device and the anti-parallel configuration results in a normally-off device.

Experiments however strongly contradict such claims. Ref. [28] has studied the dependence of spin diffusion length (and hence spin relaxation rate) on gate voltage in a semiconductor structure and found that a gate voltage of 3 V decreases the spin diffusion length by a mere 2.5% which belies the claim in ref. [23] that a small gate voltage can significantly increase the spin relaxation rate. Therefore, it is very unlikely that a device of the type proposed in ref. [23] can be turned on or off with 100 mV applied to the gate. In fact, even assuming 100% efficient spin injection and detection efficiencies, the conductance ON/OFF ratio in this device will be the ratio of the spin relaxation rate with the gate voltage on to the spin relaxation rate with the gate voltage off. To make this ratio $10^5$ as claimed in ref. [23], the gate voltage needs to increase the spin relaxation rate by five orders of magnitude instead of a mere 2.5%. That might require a gate voltage of several kilovolts, which makes this device impractical, let alone energy-efficient. This device too, like any other SPINFET, has not been demonstrated.

### 2.1.6. Obstacles to experimental realization of SPINFETs

There are serious obstacles to demonstrating the Datta-Das SPINFET and its various cousins, primary among which is the inability to inject and detect spins with high enough efficiencies at the source/channel and drain/channel interfaces. This makes the conductance modulation of the transistor very weak and probably undetectable in a noisy environment. The second obstacle is the weak spin-orbit interaction in the conduction band of semiconductors which makes it difficult to precess the spin by $180^0$ with a reasonable gate voltage. Spin-orbit interaction can be stronger in the valence band of some semiconductors, but spin precession of holes is a more complicated business because of the presence of two different types of holes (heavy and light) and possible mixing between them. Therefore, it is not clear whether a p-channel SPINFET is any easier to demonstrate than an n-channel SPINFET. The third obstacle is the inevitable magnetic field in the channel caused by the source and drain contacts. Since these are two ferromagnets facing each other, they will invariably generate a magnetic field in the channel This field, like the Rashba field, also causes Larmor spin precession and the spatial rate of precession due to it is not velocity-independent unlike that due to the Rashba field of Equation (2). As a result, electrons with different velocities in the channel undergo different additional spin precessions and ensemble averaging over these electrons will dilute the conductance modulation. Finally, there is also the possibility of Ramsauer resonances occurring in the channel of the SPINFET which may cause current oscillation [13, 14]. Under some circumstances, these oscillations may be mistaken for current modulation due to the Rashba effect [13, 14] and therefore complicate matters. The channel magnetic field also causes a leakage current [29]. As a result, the experimental demonstration of the Datta Das SPINFET (or any other related device) has remained elusive despite nearly 20 years of effort.

The other types of SPINFET that we have discussed are even harder to demonstrate. The device in ref. [20] avoids a channel magnetic field, but employs the Dresselhaus interaction which is typically weaker than the Rashba interaction in technologically important semiconductors. It also requires a more complicated structure that is more vulnerable to fabrication defects. Therefore, it is harder to implement. The devices in refs. [21, 22], on the other hand, require a very delicate balance between the Rashba and the Dresselhaus interactions, which is difficult to achieve given the numerous imperfections in fabrication. Therefore, these devices have remained theoretical curiosities and eluded experimental realization.

### 2.2. Other Types of Spin-Based Transistors

**2.2.1. The transit time spin field effect transistor (TTSFET)**

The Datta-Das SPINFET and related devices are not the only spin-based transistors that have been explored in the literature. A different genre was proposed by Appelbaum and Monsma, which they termed "transit time spin field effect transistor" (TTSFET) [30]. This device employs silicon – the most technologically developed semiconductor - which unfortunately also has very weak spin-orbit interaction. Therefore, this device could not – and does not - rely on gate controlled spin-orbit interaction to precess spins and modulate current as in the Datta-Das SPINFET. Instead it uses a fixed magnetic field in the channel and a bias voltage to modulate electron velocity. The velocity modulation modulates spin precession. This, together with spin-selective injection and extraction of carriers in the channel, realizes transistor action very much like in the Datta-Das SPINFET.

The TTSFET is a four terminal device and consists of six material layers with current flowing perpendicular to the heterointerfaces. The structure of this device is shown in Fig. 4. The principle of operation of this transistor can be explained in five steps: First, a tunnel junction, composed of the first three layers on the left, injects unpolarized spins from a paramagnetic metal emitter (PM) into the ferromagnetic base (Ferro 1) under the emitter bias $V_e$ applied between terminals 1 and 2. Second, the ferromagnetic base preferentially scatters hot electron minority spins and allows the hot electron majority spins to go through relatively unscattered [31, 32] – a phenomenon known as "hot electron ballistic spin filtering". Third, the electrons that enter the semiconductor layer by thermionically emitting over the Schottky barrier at the Ferro1/Semiconductor interface are spin-polarized since only the unscattered electrons (which are majority spins in Ferro1) have enough kinetic energy to transcend the Schottky barrier and enter the semiconductor layer. There is a static magnetic field in the semiconductor layer pointing in the direction of current flow. As the entering spins drift through this layer under the applied bias $V_b$ applied between terminals 2 and 3, they precess about this field with an angular frequency given by the Larmor formula:

$$\Omega = \frac{d\phi}{dt} = \frac{g\mu_B B}{\hbar}. \tag{14}$$

The angle by which a given spin precesses in traveling from the ferromagnetic source to the ferromagnetic drain is

$$\Phi = \frac{g\mu_B B}{\hbar} \tau_t = \frac{g\mu_B B}{\hbar} \frac{L}{v}, \tag{15}$$

where $L$ is the width of the semiconducting layer and $v$ is the electron's velocity in this layer.

Upon reaching the second ferromagnetic layer Ferro2, spins which are parallel to this ferromagnet's magnetization are transmitted while the anti-parallel spins are blocked. The transmission amplitude will be once again given by Equation (7), except this time, it is not the same for every electron since $\Phi$ depends on electron velocity $v$. This step, namely spin detection, is the fourth step in device operation. In the final (fifth) step, the transmitted electrons are collected by the collecting layer which results in a current between terminals 3 and 4.

The angle $\Phi$ in Equation (15) can be varied by varying the average electron velocity $\langle v \rangle = v_d$, which, in turn, is varied by changing the bias voltage $V_b$ across the semiconducting layer. Note that because of the Schottky barriers at the Ferro1/Semiconductor and Ferro2/Collector interfaces, the bias $V_b$, by itself, does not cause a current to flow. Instead, it modulates the current caused by $V_e$, i.e. the current flowing between terminals 3 and 4, by controlling the spin precession in the semiconducting layer by varying the drift velocity $v_d$. Since $V_b$ controls the current flowing between terminals 3 and 4, transistor action has been realized.

This device shares one feature with the Datta-Das SPINFET and all its clones, namely that current modulation is achieved via spin-precession and not via charge modulation. In all other respects, it is very different from the Datta-Das SPINFET since it (1) does not rely on modulating spin-orbit interaction with a voltage (which is an advantage since it takes a lot of voltage to change spin-orbit interaction strength even slightly), (2) does not rely on spin-injection at a semiconductor/ferromagnet interface since it uses ballistic spin filtering instead[5], and (3) it is a four-terminal device instead of a three-terminal device. The only disadvantage is that the spin-precession angle $\Phi$ is now no longer velocity- or energy-independent, unlike in the case of the Datta-Das SPINFET. As a result, ensemble averaging over the electron energy (or velocity) will reduce the current modulation and adversely affect the transconductance of the transistor as well as causing some leakage current in the OFF state. The saving grace is that because of hot electron transport across the semiconducting layer, the spread in the electron velocity is likely to be relatively small and hence the deleterious effect of ensemble averaging over the electron velocity may not be drastic.

---

[5] It, however, does rely on spin detection at a ferromagnet/semiconductor interface.

### 2.2.2. Is the TTSFET an energy-efficient device?

Since the TTSFET does not require charge modulation to achieve conductance modulation, and furthermore since the conductance modulation does not require modulating spin-orbit interaction which is weak in technologically important semiconductors, it may appear that the TTSFET will be more energy-efficient than both the MISFET and the Datta-Das SPINFET. However, this may not be true. It is a hot-electron device and therefore necessarily a high power device. Hot electron transport is needed for both the spin-filtering effect and to ensure that the energy spread in the transiting electrons is small so that energy averaging over the spin precession angle does not reduce the conductance modulation (ON/OFF ratio) too much. The energy of the hot electrons that transit the device is dissipated in the collecting contact. This transistor may or may not turn out to be more energy-efficient than the Datta-Das SPINFET, but it is unlikely to be more energy-efficient than the MISFET.

In terms of conductance ON/OFF ratio, that quantity is once again determined by the efficiencies of ballistic spin filtering and spin detection at ferromagnet/paramagnet interfaces. Since these efficiencies are typically low, the ON/OFF ratio is likely to be much lower than that of a MISFET and of the same order as that of the Datta-Das SPINFET.

**2.2.3. Experimental status of the TTSFET**

There has been admirable progress towards the demonstration of the TTSFET. Huang, et al. and Appelbaum, et al. have shown spin injection and detection in this transistor as well as transistor operation [33, 34]. In particular, ref. [33] showed a 37% spin injection efficiency and clear modulation of the transistor current by varying the bias voltage across the semiconductor layer which causes a variation in the spin precession angle $\Phi$. The modulation however is small – the collector current changes by a factor of 7 or so, indicating that the conductance ON/OFF ratio of this device is currently of the order of 7. This small ratio is most likely due to inefficient ballistic spin filtering effect and spin detection, as well as perhaps some deleterious effect of ensemble averaging over electron velocity. Thus, it seems that all SPINFET type devices that require spin injection and detection at ferromagnet/paramagnet interfaces, including the TTSFET, suffer from the curse of having a very small conductance ON/OFF ratio. The only solution to this problem is to enhance spin injection and detection efficiencies, but that seems to be a tall order.

**2.3. Comparison Between the SPINFET, TTSFET and MISFET for Device Density, Speed and Cost**

The device density and cost for the SPINFET and the MISFET are about the same since both are identical structures, except that the source and drain are ferromagnetic for the SPINFET. The TTSFET will have a slightly lower device density and a slightly higher cost because it is a 4-terminal device as opposed to a 3-terminal device.

The speeds of these devices are determined by the transit time of carriers through the active region (channel in the cases of MISFET and SPINFET, and the semiconducting layer in the case of the TTSFET). Since the thickness of the semiconducting layer in TTSFET is determined by film growth techniques while the channel lengths in SPINFET and MISFET are determined by lithography, and furthermore since the velocity of carriers in TTSFET is higher because they are hot carriers, the TTSFET is likely to have a slightly higher switching speed because carriers will travel through it faster.

Table I presents a comparison between the SPINFET, the TTSFET and the traditional MISFET for digital electronic applications.

### 2.4. Spin Bipolar Junction Transistor

The devices that we have discussed so far rely on unipolar transport, where only one type of carrier – either electrons or holes, but not both – carries current. There are, however, spin *bipolar* junction transistor (SBJT) proposals as well [35, 36] which were preceded by the proposal for a spin unipolar junction transistor (SUJT) [37] that was supposed to mimic a bipolar junction transistor.

These devices do *not* bypass charge modulation by using spin properties of carriers; hence, they are not any more energy efficient than traditional (charge-based) bipolar junction transistors. However, in the case of SBJT, there may be some additional *functionality* available because this device could, in principle, act as a four-terminal device that gives it more flexibility. We discuss this device next.

The SBJT is identical to a traditional BJT except that the base of the transistor is ferromagnetic and therefore carriers residing in it are spin-polarized (i.e. there are more of majority spins than minority spins). The conduction band profile of a heterostructured SBJT, where the emitter, base and collector layers are composed of three different materials, is shown in Fig. 5, assuming that the transistor is n-p-n and is biased in the active region (emitter-base junction is forward biased while base-collector junction is reverse biased). The expressions for the emitter, base and collector currents – $I_E$, $I_B$ and $I_C$ - were derived in ref. [35] and reproduced below.

$$I_C = qA \frac{D_{nb}}{L_{nb}} \frac{1}{\sinh(W/L_{nb})} n_{be} \left[ e^{V_{EB}/kT} - 1 \right] - qA \frac{D_{nb}}{L_{nb}} \coth(W/L_{nb}) n_{bc} \left[ e^{V_{CB}/kT} - 1 \right] - qA \frac{D_{pc}}{L_{pc}} \coth(W_c/L_{pc}) p_{oc} \left[ e^{V_{CB}/kT} - 1 \right],$$

$$I_E = qA \frac{D_{nb}}{L_{nb}} \coth(W/L_{nb}) n_{be} \left[ e^{V_{EB}/kT} - 1 \right] - qA \frac{D_{nb}}{L_{nb}} \frac{1}{\sinh(W/L_{nb})} n_{bc} \left[ e^{V_{CB}/kT} - 1 \right] + qA \frac{D_{pe}}{L_{pe}} \coth(W_e/L_{pe}) p_{oe} \left[ e^{V_{EB}/kT} - 1 \right]$$

$$I_B = -I_E - I_C$$

(16)

where $A$ is the cross-sectional area of the transistor, $W_c$ is the collector width, $W_e$ is the emitter width, $D_{nb}$ ($L_{nb}$) is the minority carrier diffusion coefficient (length) of electrons in the base, $D_{pc}$ ($L_{pc}$) is the minority carrier diffusion coefficient (length) of holes in the collector, $D_{pe}$ ($L_{pe}$) is the minority carrier diffusion coefficient (length) of holes in the emitter, $n_{be} = (n_i^2/N_{AB})(1+\alpha_e \alpha_{0b})/\sqrt{1-\alpha_{0b}^2}$, $n_{bc} = (n_i^2/N_{AB})(1+\alpha_c \alpha_{0b})/\sqrt{1-\alpha_{0b}^2}$, $p_{oc} = (n_i^2/N_{DC})$, $p_{oe} = (n_i^2/N_{DE})$, $N_{AB}$ is the acceptor concentration in the base, $N_{DC}$ is the donor concentration in the collector, $N_{DE}$ is the donor concentration in the emitter, $\alpha_e$ and $\alpha_c$ are the non-equilibrium spin polarizations in the emitter and collector, and $\alpha_{0b} = \tanh(\Delta/kT)$ is the equilibrium spin polarization in the base, while $2\Delta$ is the spin-splitting energy in the base.

A routine small signal analysis carried out in ref. [38] has shown that the SBJT has about the same voltage and current gains as a conventional (non-spin-based) BJT. The short-circuit current gain $\beta = \partial I_C / \partial I_B$ however is not constant and depends on the spin-splitting energy $\Delta$ in the base. This last quantity can be modulated with an external magnetic field, which can therefore act as a fourth terminal. Thus, this device has a non-linear property and will be suitable for a frequency mixer. If the base current is an ac sinusoid with an angular frequency $\omega_1$ and the modulating magnetic field is another sinusoid with an angular frequency $\omega_2$, then the collector current will contain frequency components $\omega_1 \pm \omega_2$. This non-linear functionality makes the SBJT a more powerful device than a conventional BJT.

The SUJT of ref. [37] unfortunately turns out to be a device which simply cannot work as a transistor. Transistors are benchmarked by two properties: the output conductance $g_0$ which determines the fan out and the voltage gain given by $a_v = g_m / g_0$ (see Equation 11); and the so-called feedback conductance $g_\mu$ which determines the isolation between the input and output terminals. Isolation is an extremely important property which we discuss later in Section 4. Isolation ensures that the output signal of any device is controlled by the input signal, but not the other way around. Both $g_0$ and $g_\mu$ should be vanishingly small to yield a large voltage gain and good isolation between input and output terminals. Ref. [38] showed that both these quantities are actually extremely large in the SUJT; in fact, $g_\mu$ is 34 *orders of magnitude* larger in the SUJT than in a conventional BJT, meaning that the SUJT practically has no isolation, not to mention that it may not have any gain either because of the very large $g_0$. Such a "transistor" is obviously not suitable for electronics.

There is a plethora of other spin based transistors which have been proposed and some have been experimentally demonstrated. This review cannot possibly discuss all of them. However, there are two transistors that deserve special mention because of the significant experimental progress made in fabricating and demonstrating them. These two devices are the all-metal spin transistor proposed by Johnson [39, 40] and the spin valve transistor demonstrated by a large number of researchers including Monsma, et al. [31], Mizushima, et al. [41], and LeMinh et al. [42]. Their performances are probably not superior to that of the SBJT, but they have attracted considerable attention from experimentalists. We do not discuss them here, but instead refer the reader to an excellent review of these transistors by Jansen, et al. [43].

## 3.  SPIN BASED LOGIC GATES

The devices discussed in Section 2 are single devices that act like binary switches with two stable conductance states – ON and OFF. These two states encode the binary bits 0 and 1. A switch, by itself, does not implement a Boolean logic operation. In order to do that, one has to build a logic "gate" by interconnecting multiple switches in specific ways [25], while ensuring that logic signal flows unidirectionally from the driving switch to the driven switch. That will require the switch to possess isolation between its input and output terminals.

There are two fundamentally different types of logic gates: universal and non-universal. A universal gate is sufficient by itself to implement any arbitrary Boolean logic operation, which a non-universal gate cannot. Examples of universal gates are NAND and NOR, while NOT, AND and OR gates are non-universal.

A number of researchers have proposed and/or demonstrated spintronic logic gates that are capable of executing Boolean logic operations [44-49]. The last group [49] demonstrated a gate using only a *single* switch. Furthermore, the gate was "reconfigurable", i.e. it could be changed from a NAND gate to a NOR gate to an AND gate to an OR gate by adding an additional electrical input. The switch was implemented by a magnetic tunnel junction, which we briefly describe below.

## 3.1. A Magnetic Tunnel Junction (MTJ) Switch

A magnetic tunnel junction (MTJ) is a spintronic device that can act as a binary logic switch. It can have two conductance states ON and OFF which will encode the binary bits 0 and 1, just like the various spin transistors discussed in the previous section. Unlike a transistor, which has at least three electrical terminals, this device has only two electrical terminals. A magnetic field acts like the third terminal to switch the current flowing between the two electrical terminals (and hence the device conductance) from high to low, or vice versa. This realizes the switching action.

The MTJ is a tri-layered structure where the two outer layers are ferromagnetic and the central layer (called the "spacer layer") is paramagnetic. One outer layer acts as a spin injector and the other is the spin detector. Assuming that there is very little spin relaxation in the spacer layer (which is typically very thin and the injected electrons tunnel through it), the spins injected by the injecting layer will transmit through the device if the injector's and detector's magnetizations are parallel, i.e. the majority spins in one are also majority spins in the other[6]. In this case, the device conductance will be high.

---

[6] This assumes that the two ferromagnets have the same sign of spin polarization, like in the case of cobalt and nickel. In that case, the majority spins in one will be majority spins in the other. However, if the ferromagnets have opposite signs of spin polarization, as in the case of iron and cobalt, then majority spins in one will be minority spins in the other. In that case, the MTJ will have high conductance when the magnetizations of the two ferromagnetic layers are anti-parallel and low conductance when they are parallel.

One of the ferromagnets is "hard" and has a high coercivity, while the other is "soft" and has a lower coercivity. When a magnetic field whose strength exceeds the coercive field of the soft magnet, but not that of the hard magnet, is applied to the device, it selectively flips the magnetization of the soft layer and places the two ferromagnets in an anti-parallel configuration. Now the spins injected by the injector cannot transmit through the detector and the device conductance falls. Thus, the magnetic field switches the device conductance from ON to OFF by switching the magnetization direction of the soft layer.

Unfortunately, just like the SPINFET, this switch's conductance ON/OFF ratio is determined by the efficiencies of spin injection and detection at ferromagnetic/spacer interfaces. Since these efficiencies have never been very high, the conductance ON/OFF ratio stays low and the maximum value demonstrated as of this writing is about 7:1 [50], which, once again, is insufficient for logic circuits since it will result in an unacceptably high bit error rate if conductance states are used to encode binary bits 0 and 1.

### 3.2. A Reconfigurable Magnetic Tunnel Junction Logic Gate

Ney, et al [49] demonstrated a reconfigurable logic gate utilizing MTJs. The gate structure, with an MTJ at its core, is shown in Fig. 6(a). The lower ferromagnetic layer is softer than the upper ferromagnetic layer, i.e. it has a smaller coercivity. There are three current lines $I_1^{in}$, $I_2^{in}$ and $I_3^{in}$ beneath the lower layer. Current in any one line is not enough to generate a magnetic field of sufficient strength to flip the magnetization of either ferromagnetic layer. Current in lines 1 and 2 together (when they are flowing in the same direction) can flip the magnetization of the lower (softer) layer, but not that of the upper (harder) layer. Current in all three lines together (when they are flowing in the same direction) can flip the magnetization of either layer.

This system can realize either an AND or an OR gate with just lines 1 and 2. Adding the third line 3 allows one to realize the NAND and NOR as well. Thus, this is an example of a *reconfigurable* logic gate, where the behavior of the gate (i.e. whether it performs as an AND, NAND, OR or NOR) can be programmed with currents.

We explain how this gate can act as an AND gate. The operation actually takes place in two steps: SET and LOGIC, which are preceded by an INITIALIZATION step. First, in the INITIALIZATION step, currents in all three lines flow to the left and align the magnetizations of both layers pointing inward as shown in Fig. 6(b). Next, in the SET step, the currents in lines 1 and 2 flow to the right and flip the magnetization of the soft layer, placing the two layers in the anti-parallel configuration, as shown in Fig. 6(c). Finally, in the LOGIC step, the currents in lines 1 and 2 are interpreted as the two inputs to the gate and the loop current (flowing between the top and bottom ferromagnetic layers as shown in Fig. 6(a)) is interpreted as output. When the current in line 1 or 2 flows to the left it encodes input bit 1, and when it flows to the right, it encodes input bit 0. When both inputs are 1, the softer layer's magnetization flips inwards, placing both ferromagnetic layers in the parallel configuration. The MTJ turns ON and a loop current now flows through it, so that the output is interpreted as logic 1. When only one input (or neither) is logic 1, the magnetization of the softer layer remains anti-parallel to that of the harder layer and therefore no loop current flows since the MTJ is OFF. Accordingly, the output is logic 0. Therefore, the output is logic 1 only when both inputs are logic 1; otherwise, it is at logic 0. This realizes the AND gate whose truth table is in Table II. Every time the gate acts as an AND gate, it must be reset using the SET step, before the next LOGIC step can be executed. This is an inconvenience, since, instead of just one step, we always need two steps for logic operation. Ultimately, it reduces the clock speed by a factor of 2.

In order to convert the AND gate to an OR gate, we do not need to repeat the INITIALIZATION step, but need to repeat the SET step and use a different SET strategy. In the SET step, we now make currents in lines 1 and 2 flow to the left and that places both ferromagnetic layers in the parallel configuration. In this case, unless both inputs are 0, i.e. currents in both lines 1 and 2 are flowing to the right, the softer layer will remain parallel to the harder layer. Consequently, a loop current will flow because the MTJ resistance is low, and the output will be in logic state 1. Thus, the output is 0 only when both inputs are 0, and it is 1 otherwise. This realizes an OR gate.

In order to convert the AND gate to a NAND and the OR to a NOR, we have to reverse the INITIALIZATION step and align both layers' magnetizations pointing outward by making current in all three lines flow to the right. For the NAND gate, the SET step will flow currents in lines 1 and 2 to the right, which will leave the layers in the parallel state. During the succeeding LOGIC step, only if the currents in both lines 1 and 2 flow to the left, then the soft layer will flip, placing the two layers in the anti-parallel state and cutting off the loop current. Thus, only when both inputs are 1, the output will be 0; otherwise it will be 1. This replicates a NAND gate whose truth table is in Table II.

For the NOR operation, the SET step will flow currents in lines 1 and 2 to the left, which will leave the ferromagnetic layers anti-parallel. Subsequently, during the LOGIC state, only when both inputs are logic 0, so that currents in lines 1 and 2 are flowing to the right, the two layers will become parallel and a loop current will flow. Thus, the output will be logic 1 only when both inputs are logic 0, and it is 0 otherwise. This is a NOR gate.

### 3.2.1. Speed, energy efficiency and cost of MTJ logic gates

In the MTJ logic gate, input and output logic bits are encoded in currents. Therefore, this is *not* a low power paradigm since current flow in needed for logic operation. In fact, the power dissipation can be quite large since relatively large currents are needed to generate strong enough magnetic fields to switch the magnetization of (even soft) ferromagnetic layers. This same problem afflicts the logic gates proposed or demonstrated in refs. [44-48] which use the giant magnetoresistance effect, instead of the tunneling magnetoresistance effect, to implement the basic logic switch. Giant magnetoresistance effect also needs the generation of a magnetic field to change the resistance of a device and thus is energy-hungry.

The speed of MTJ logic gates will be determined by how fast one can flip the magnetization of the soft layer. This will take at least 1 nanosecond, so that the clock speed is limited to 1 GHz.

MTJ gates may however cost a little less than traditional transistor gates since they are relatively simple to manufacture.

## 4. SPIN BASED LOGIC ARCHITECTURES AND CIRCUITS FOR COMPUTATION

So far, we have talked of individual switches and gates which are essential elements of digital information processing circuits, but which, by themselves, cannot compute or process arbitrary information. For that purpose, we need logic architectures or circuits. Several universal logic gates, each comprising a number of logic switches, can be interconnected in specific ways to act as an arithmetic logic unit (ALU) capable of processing digital information in any desired fashion. In this section, we describe a few such spin-based architectures, which are capable of universal computation. Because of the way they implement the basic switch, they can be *extremely energy-efficient*. The switch is implemented in a way very different from the individual switches or logic gates that we discussed in Sections 2 and 3, which are not likely to be any more energy efficient, cost effective, or faster than traditional charge based devices.

### 4.1         Single Spintronics: An Energy-Efficient Paradigm

We mentioned earlier that charge based electronics squander energy since logic levels must be demarcated by a difference in the magnitude of charge stored in a device. Altering this magnitude to switch between logic states invariably entails current flow and therefore excessive dissipation. This dissipation can be avoided if we somehow eliminate current flow when we switch. That can be accomplished if we encode logic bits 0 and 1 in *bistable* spin polarization of a single electron (or the spin polarization of a collection of interacting electrons) confined in a region of space. We can then switch by simply flipping spins, without causing charge motion and current flow. In that case, there will not be any dynamic energy dissipation during switching associated with current flow; nor will there be any static dissipation in the ON or OFF state since no current flows in those states either. This is potentially a very energy-efficient scheme.

The reason why it can work is that the spin of an electron, unlike its charge, is a pseudo-vector that has a fixed magnitude of $\hbar/2$ but a variable direction. If we place a lone electron in a magnetic field, then the direction of the spin (or spin polarization) can point either parallel or anti-parallel to the field. No other direction would be stable. This can be shown easily from the Pauli equation that governs the electron's spin property:

$$\left\{[H_0] - \frac{g}{2}\mu_B \vec{B} \bullet [\vec{\sigma}]\right\}[\psi] = E[\psi], \tag{17}$$

where $[H_0]$ is the spin-independent Hamiltonian, $g$ is the Landé g-factor, $\vec{B}$ is the flux density of the magnetic field, $[\vec{\sigma}]$ is the Pauli spin matrix and $[\psi]$ is the 2×1 component spinor describing the spin. Assume now that the magnetic field is in the z-direction which makes $\vec{B} \bullet [\vec{\sigma}] = B_z [\vec{\sigma}_z]$. In that case, diagonalization of the total Hamiltonian in Equation (17) immediately yields that the eigenspinors are

$$[\psi] = \begin{bmatrix} 1 \\ 0 \end{bmatrix} \text{ or } \begin{bmatrix} 0 \\ 1 \end{bmatrix}, \tag{18}$$

which are +z and –z polarized states, meaning that they are oriented parallel and anti-parallel to the magnetic field. Therefore, the electron's spin can point only in *two* directions: either parallel to the field (which we call the "down" state, or $|\downarrow\rangle$), or anti-parallel (which we call the "up" state, or $|\uparrow\rangle$). These two "polarizations" could represent the binary bits 0 and 1[7]. Switching between them would simply require flipping the spin *without moving the electron in space and causing a current to flow*. Moreover, there is no change in the magnitude of the charge which would have caused an amount of dissipation equal to $|\Delta V (Q_1 - Q_2)|$, as we have seen before. Thus, encoding logic bits in spin can drastically cut down on dynamic energy dissipation during switching.

Flipping the spin, however, is not completely dissipationless. The two spins polarizations do not have the same energy since the two eigenenergies of Equation (17) are separated by an amount $g\mu_B |B|$ which is the Zeeman splitting energy [51]. At the very least, this amount of energy must be dissipated in flipping the bits from 0 to 1, or vice versa. But this energy can be made very small by using a weak magnetic field that has a small flux density *B*. It cannot be made arbitrarily small in the presence of noise and we certainly cannot overcome the Landauer-Shannon limit that we alluded to earlier, but ultimately what matters is whether this approach could be more energy-efficient than the traditional charge based transistor paradigm, or the SPINFET/TTSFET paradigm, or the MTJ paradigm in an actual circuit environment. That is the all-important question which will determine whether "single spintronics" – where bit information in encoded in single spins - has a future and any serious role in digital information processing.

---

[7] One might wonder why it is necessary to make the spin polarization "bistable" in order to encode binary bits. After all, conventional electronics encodes binary bits in two distinct levels of voltage, current or charge, which are not intrinsically bistable. The reason why spin has to be intrinsically bistable is because there is no such thing as a "spin amplifier" that can restore the separation between logic levels if the levels get corrupted by noise and begin to merge with each other. In the case of voltage and current, amplifiers with non-linear transfer characteristics restore the separation between the voltage and current levels, but that is not possible with spin since the equivalent of the current or voltage amplifier does not exist. Therefore, spin polarization must be made intrinsically bistable so that no amount of noise and distortion can ever make the two levels merge. This fact is often not understood or appreciated.

It is intuitive to think that if we encode binary information in anti-parallel spin states in a magnetic field, we must ensure that $g\mu_B|B| \gg kT$ where *kT* is the thermal energy, since the two spin levels might be broadened in energy by ~ *kT* and therefore the above condition needs to be fulfilled in order to keep the two spin states distinguishable at the temperature *T*. That line of thinking would be incorrect. Spin-phonon coupling is very weak, much weaker than charge phonon coupling, and it may be made even weaker in quantum confined systems like quantum dots because of the phonon-bottleneck effect [52]. Consequently, spin levels are not broadened by ~ *kT* unlike energy states coupled to the charge degree of freedom. As a result, $g\mu_B|B|$ can be much less than the thermal energy *kT* and yet the spin levels can remain distinguishable at the temperature *T*. A case in point is Electron Spin Resonance (ESR) which involves transitions between two Zeeman split levels separated in energy by $g\mu_B|B|$ which is typically a few tens of µeV (microwave photon energy) in most solids. In spite of that, when ESR experiments are carried out at room temperature (when $kT \gg g\mu_B|B|$), the signals remain well-resolved, meaning that the spins-split levels remain distinguishable even when $g\mu_B|B| \ll kT$. This can happen only if the spin levels are not broadened by ~*kT*. Therefore, *in principle*, a single spin device could dissipate considerably less energy during switching than a single charge device. Whether this advantage is ultimately realized depends on the device design.

Spin has two other advantages over charge. First, because of weaker spin-phonon coupling, a spin system can be maintained in a non-equilibrium state longer and more easily than a charge system since it is the coupling to the thermal bath that restores equilibrium. In a non-equilibrium system, the occupation probability of the two spin-split states encoding the binary bits 0 and 1 is *not* governed by Boltzmann statistics or Fermi-Dirac statistics, so that the relative occupation probability of the two states is *not* $e^{\frac{-g\mu_B B}{kT}}$. Therefore, the bit error probability *p* associated with occupation of the wrong state is not $e^{\frac{-g\mu_B B}{kT}}$, but could be considerably less. Also, since $p \neq e^{\frac{-g\mu_B B}{kT}}$, the energy dissipated in switching between logic levels, which is $g\mu_B B$, is not the Landauer-Shannon limit of *kTln(1/p)*, but could be much less. The possibility of overcoming the Landauer-Shannon limit by using non-equilibrium systems has been discussed earlier by Zhirnov, et al. [53] and Cavin, et al. [54]. Energy dissipation in some non-equilibrium systems has been discussed by Nikonov, et al. [55]. We do not dwell on those here since non-equilibrium dynamics is not employed in any of the paradigms discussed in this article. The reader should however understand that maintaining a system out of equilibrium needs a continuous supply of energy and the additional energy cost associated with it may wipe out any energy saving accruing from non-equilibrium dynamics.

A second advantage of spin over charge is that it does not couple easily to stray electric fields (except through spin-orbit interaction). Hence spin is more robust against electrical noise. These two features make "spin" intrinsically superior to "charge".

The shortcomings of charge as a state variable to encode information were anticipated (although not expounded) in ref. [53, 54] which advocated investigating alternate state variables, different from charge, to encode logic states (see also [56]). Spin appears to be a good choice in this regard. Later work [57] that questioned this wisdom did not take into account any of the advantages (weaker spin-phonon coupling, easier enforcement of non-equilibrium dynamics, stronger immunity against noise, etc.) that spin has. Therefore, the claim of ref. [57] that spin and charge are "equal" is ultimately not correct; spin can be superior, if properly used.

### 4.2. The Single Spin Logic (SSL) Family

A well-known paradigm where bistable spin polarizations of an electron (placed in a magnetic field) are used to represent binary bits 0 and 1 is *Single Spin Logic* (SSL) [58, 59] which is a single-spin based approach to combinational and sequential logic.

In SSL, single electrons are confined in semiconductor quantum dots that are delineated on a wafer which is placed in a dc magnetic field. This field may be generated by a permanent magnet or an electromagnet, with the former requiring no energy budget. The magnetic field defines the spin quantization axis and makes the spin polarization of every electron bistable, i.e. only polarizations parallel and anti-parallel to the field are stable or metastable in each dot. These two polarizations represent logic bits 1 and 0, respectively. Each spin can interact only with its nearest neighbor via exchange[8]. This is ensured by maintaining a small separation (~10 nm) between nearest neighbor dots so that the wavefunctions of their resident electrons overlap in space. Wavefunctions of electrons in second nearest neighbor dots do not overlap and therefore there is no second nearest neighbor interaction. Input data are provided to the quantum dot array by aligning the spins in certain chosen dots (designated as input ports) in desired directions (parallel or anti-parallel to the global magnetic field) using external agents, such as local magnetic fields[9] (**writing**). The arrival of the inputs takes the interacting array into a many-body excited state. The system is then allowed to relax to the thermodynamic ground state by coupling to the thermal bath[10]. When the ground state is reached by emitting phonons, magnons, etc., the spin orientations in certain other chosen dots (designated as output ports) will represent the result of a specific computation in response to the input bits. This result (spin orientations in output ports) can be read using a variety of schemes, all of which have been experimentally demonstrated [60-62] (**reading**). The reader should appreciate that this is an "all-hardware" computer with no involvement of any "software". Hence it is extremely fast. The disadvantage is that a particular computer can do only one computation since the computer is entirely hard wired. This is an extreme example of application-specific-integrated-circuit (ASIC).

---

[8] Second nearest neighbor interaction is much weaker than nearest neighbor interaction since exchange interaction tends to decay exponentially with distance.

Note that unlike the SPINFET, which is a single device or switch, and unlike the MTJ gate which is a single gate, the SSL is a complete circuit or architecture that goes well beyond a single device or gate. A single device can perform only one operation, namely switch on and off, while a single gate can carry out a specific Boolean operation only. Both are incapable of performing universal computation. In contrast, SSL is capable of universal computation since (as we will show later) we can implement universal Boolean logic gates (the NAND gate) and connect them in any desired fashion using spin wires to realize arbitrary combinational or sequential circuits. Hence, SSL is capable of universal computation.

---

[9] Spin can also be aligned using electrical currents via the spin transfer torque effect. A current can move domain walls in ferromagnets and therefore is able to generate local magnetic fields over the size of a domain.

[10] The coupling of a single isolated electron to the thermal bath is weak, but the collective coupling of many interacting electrons (a many-body system) to the thermal bath is much stronger. Hence the many body system relaxes to ground state much faster than an isolated spin.

Note also that SSL is an equilibrium system where the spins are *not* maintained out of equilibrium; in fact, computation is performed by relaxation of excited spins to the ground state by coupling with the thermal bath (phonons)[11]. Therefore, this paradigm does not exploit any possible advantage of the non-equilibrium dynamics discussed in ref. [53-55]. At the same time, no energy is expended to maintain the system out of equilibrium. Therefore, it is not clear if SSL would have benefitted from non-equilibrium dynamics if it could be brought to bear on the operation of the system.

Equilibrium statistics mandates that the absolute minimum energy dissipated in a single irreversible logic operation should be the Landauer-Shannon limit

$$E_{diss}^{min} = kT \ln(1/p),  \qquad (19)$$

where $p$ is the bit error probability in the gate. Indeed it turns out that the energy dissipated in any irreversible logic operation in the SSL NAND gate is given by the above expression. This is actually a remarkable result since it shows that no paradigm can better the SSL in single bit flip dissipation, since SSL operates at the Landauer-Shannon limit.

---

[11] Although the coupling of a single spin with phonons is weak, the coupling of a many spin system with phonons is much stronger to that the many-body system relaxes to equilibrium relatively fast.

### 4.2.1. The SSL NAND gate and spin wire

So far, we have describe the working of application specific integrated circuits (ASIC) which can be very fast, but not always very useful since it can do only one type of computation. In order to do a different type of computation, we have to build a different array of dots, which is not very convenient. In order to overcome this problem and carry out general purpose computing (GPC) with the same array of dots, we must construct general purpose Boolean logic circuits by employing universal logic gates (e.g. the NAND gate) realized with few interacting single spins. We will then interconnect them with "spin wires" that ferry spin signals between them unidirectionally. The two ingredients – NAND gates and unidirectional spin wires – are all that are required to implement a universal computing machine.

A NAND gate is implemented with a linear array of three quantum dots with nearest neighbor exchange interaction. All three dots are placed in a global static magnetic field which defines the spin quantization axis. Spin polarized parallel to the field encodes the binary bit 1 and that polarized anti-parallel to the field encodes the binary bit 0.

The two peripheral dots are treated as input ports (whose spins are aligned to conform to input bits 0 or 1 with external agents that can generate local magnetic fields) and the central dot is the output port. This system is shown in Fig. 7(a). It was rigorously shown in ref. [63] that the ground state spin configuration in this system is *anti-ferromagnetic*, i.e. spins in nearest neighbor quantum dots will be anti-parallel as long as the exchange interaction strength between nearest neighbors is greater than one-half the Zeeman splitting energy in any dot due to the global magnetic field, and the local magnetic fields applied to the input dots is much stronger than the global magnetic field. In that case, whenever the two inputs are 1, the output must be 0 to preserve the anti-ferromagnetic ordering, and similarly whenever the two inputs are 0, the output must be 1. When one input is 1 and the other is 0, a tie seemingly occurs. This tie however is broken by the global magnetic field, which will generate a slight preference for the spin in the output dot to be aligned parallel to the field when the two inputs are dissimilar. Since the spin orientation parallel to the global magnetic field encodes logic bit 1, the output will be 1. Thus, the input-output relation of this system obeys the truth table of the NAND gate shown in Table II.

To treat this system using rigorous quantum mechanics and show that it indeed acts as described (i.e. performs the NAND logic operation), one must consider the Hamiltonian of the three-dot system. This can be described by a Hubbard Hamiltonian which will have 29 independent basis states. However, if we assume single electron occupancy in each dot[12], then the Hubbard Hamiltonian can be reduced to a much simpler Heisenberg Hamiltonian [59, 63] which has only 8 independent (orthonormal) basis states. This Hamiltonian is given by

$$H_{Heisenberg} = \sum_{<ij>} J_{ij}^{\parallel} \sigma_{zi}\sigma_{zj} + \sum_{<ij>} J_{ij}^{\perp} \left(\sigma_{xi}\sigma_{xj} + \sigma_{yi}\sigma_{yj}\right) + \sum_{\text{input dots}} \sigma_{zi} h_{zi}^{inputs} + \sum_{i} \sigma_{zi} h_{zi}^{global} \quad (20)$$

where the $\sigma$-s are Pauli spin matrices. We adopt the convention that the direction of the local and the global magnetic fields is the $z$-direction. The last two terms in the Hamiltonian account for the Zeeman energies associated with these fields. The first two terms account for exchange interaction between nearest neighbors (the angular brackets denote summation over nearest neighbors). We will assume the isotropic case when $J_{ij}^{\perp} = J_{ij}^{\parallel} = J$, where $J$ is the exchange energy, which is non-zero if the wavefunctions in dots $i$ and $j$ overlap in space.

---

[12] This requires that the dots are so small and have such small capacitance that the energy cost to add a second electron to any dot is prohibitive.

The spins in the quantum dots are polarized in either the +z or –z direction by the global magnetic field (conforming to bits 1 or 0), and we designate the corresponding states as "upspin" $(\uparrow)$ and "downspin" $(\downarrow)$ states, respectively. Recall that the upspin state (aligned anti-parallel to the global magnetic field) encodes bit 0 and the downspin state (parallel to the global field) encodes bit 1.

There are 8 independent 3-spin basis states representing the spin configurations in the 3-dot array, which are $|\downarrow\downarrow\downarrow\rangle, |\downarrow\downarrow\uparrow\rangle, |\downarrow\uparrow\downarrow\rangle, |\downarrow\uparrow\uparrow\rangle, |\uparrow\downarrow\downarrow\rangle, |\uparrow\downarrow\uparrow\rangle, |\uparrow\uparrow\downarrow\rangle, |\uparrow\uparrow\uparrow\rangle$. In each state, the first entry is the spin polarization in the left dot, the second in the central dot and the third in the right dot. These eight basis functions form a complete orthonormal set. The matrix elements $<\phi_m | H_{Heisenberg} | \phi_n >$ are given in the matrix below, where the $\phi$-s are the 3-electron basis states enumerated above.

$$\begin{pmatrix} 2J - h_L - h_R - 3Z & 0 & 0 & 0 & 0 & 0 & 0 & 0 \\ 0 & -h_L + h_R - Z & 2J & 0 & 0 & 0 & 0 & 0 \\ 0 & 2J & -2J - h_L - h_R - Z & 0 & 2J & 0 & 0 & 0 \\ 0 & 0 & 0 & -h_L + h_R + Z & 0 & 2J & 0 & 0 \\ 0 & 0 & 2J & 0 & h_L - h_R - Z & 0 & 0 & 0 \\ 0 & 0 & 0 & 2J & 0 & -2J + h_L + h_R + Z & 2J & 0 \\ 0 & 0 & 0 & 0 & 0 & 2J & h_L - h_R + Z & 0 \\ 0 & 0 & 0 & 0 & 0 & 0 & 0 & 2J + h_L + h_R + 3Z \end{pmatrix}$$

In the above matrix, Z is one-half of the Zeeman splitting energy associated with the global magnetic field (i.e. $2Z = g\mu_B B_{global}$), while $2h_L$ and $2h_R$ are Zeeman splitting energies in the left and right input dots caused by the local magnetic fields that write input data ($2h_L = g\mu_B B_{local}^{left}$; $2h_R = g\mu_B B_{local}^{right}$). If the local magnetic field is in the same direction as the global field and writes bit 1, then the corresponding h is positive; otherwise, it is negative. The quantity J is always positive.

Ref. [63] diagonalized the above Hamiltonian for the 4 possible inputs to the NAND gate corresponding to both inputs being logic 1 ($h_L = h_R = h$), left input logic 1 and right input logic 0 ($h_L = -h_R = h$), left input logic 0 and right input logic 1 ($h_L = -h_R = -h$) and both inputs being logic 0 ($h_L = h_R = -h$). It was found that the ground state wavefunctions in the four cases approach the states $|\downarrow\uparrow\downarrow\rangle, |\downarrow\downarrow\uparrow\rangle, |\uparrow\downarrow\downarrow\rangle, |\uparrow\downarrow\uparrow\rangle$, respectively, provided $h_L, h_R > J$ and $J > Z/2$. Thus, the ground state spin polarization in the output dot is always the NAND function of the spin polarizations in the input dots, provided the Zeeman splitting caused by the local magnetic fields that "write" input bits in the input dots is much larger than the strength of exchange coupling between nearest neighbors, and the latter, in turn, is larger than one-fourth of the Zeeman splitting caused by the global magnetic field. Therefore, the NAND gate is indeed realized by three spins with nearest neighbor exchange coupling if we satisfy the conditions $h_L, h_R > J$ and $J > Z/2$. Since the NAND gate is universal, any arbitrary combinational or sequential circuit can be implemented by interconnecting NAND gates with a "spin-wire" shown in Fig. 7(b).

A spin-wire is a linear array of quantum dots each containing a single electron that interacts with its nearest neighbors via exchange. We will describe in the next subsection how a spin polarization state can be unidirectionally propagated from left to right along the wire using a 3-phase clock. The clock signal is a sequence of positive voltage pulses that are applied to the gates interposed between each pair of dots. The arrival of a positive voltage pulse temporarily lowers the potential barrier between two adjacent dots and exchange couple their resident spins. By sequentially exchange coupling three adjacent dots at a time using a 3-phase clock, the spin state can be propagated *unidirectionally* down the wire in a bucket-brigade fashion. In conventional circuits, the wires are bidirectional but the logic devices (transistors) are unidirectional since they possess isolation between their input and output terminals. Here, the logic devices (single spins in quantum dots) are completely bidirectional so that the wires have to be unidirectional to implement logic circuits. We discuss this in more detail in the next subsection.

A spin wire can obviously also perform the role of fan-out where a signal is split into multiple paths in order to drive multiple stages with the output of one stage. Fig. 7(b) shows how this is accomplished.

Finally, one last requirement that wires must satisfy is the function of "cross-over" where two wires cross each other in space without interfering with one another. This is the most challenging requirement and normally will require multiple layers of dots where a dot in one layer is sufficiently distant from the nearest dot in the closest layer to preclude any significant exchange coupling between them.

**4.2.2. Unidirectional signal transfer along a spin wire**

Not only should a spin wire be able to ferry spin state from one dot to a remote dot, but it must do so *unidirectionally,* so that signal always flows from the input stage to the output stage and not the other way around. In conventional circuits, the wires are not unidirectional since the devices (transistors) themselves are inherently unidirectional. For example, in a field effect transistor, the input signal (in the form of a voltage) is applied to the gate terminal which affects the current flowing through the channel and this current is the output signal. However, the channel current does not affect the gate voltage. Thus, a master-slave relationship exists between the input and output, where the former dictates the latter, but the latter has no influence over the former. Device physicists and engineers call this property "isolation" between the input and output.

Unfortunately, there is no isolation in a quantum dot. Exchange interaction that connects two neighboring dots and plays the role of wires in this "wireless" architecture also happens to be entirely bidirectional. As a result, among any two neighboring dots, the first dot affects the second just as much as the second affects the first. Neither is the master or the slave. Since there is no unidirectionality in the device, we are left with no choice but to make the wires unidirectional.

There is indeed no easy way to impose unidirectionality in *space*, but there is a way to impose it in *time* via sequential clocking [64]. This is actually a standard technique used to steer signal unidirectionally through devices that lack isolation between their input and output terminals. A well-known example of this is the charge-coupled-device shift register, frequently used in digital imagers, where a charge packet carrying bit information is steered unidirectionally through a linear array of charge coupled devices (which are bidirectional themselves) using 3-phase clocks [65]. A similar scheme is adopted for SSL. By sequentially raising and lowering potential barriers between two pairs of dots at a time using a 3-phase clock, one can sequentially turn exchange interactions on and off between any two adjacent pairs and steer spin signal unidirectionally along a spin wire. This paradigm was explained in ref. [66].

In order to carry out the sequential clocking in practice, a gate pad is inserted between each pair of dots. When the electrostatic potential applied to a pad is zero, the potential barrier between the two dots is so high that the exchange coupling between them is negligible. When a positive potential is applied to a gate pad, the barrier is lowered and the two electrons on either side are exchange coupled. When this act is carried out by raising the potentials of two succeeding gate pads simultaneously, three consecutive electrons are exchange coupled (nearest neighbor only) and the third electron begins to assume the polarization of the first[13] as the array approaches the anti-ferromagnetic ground state. It is assumed that the many-body ground state is reached in a time shorter than the clock period, which is a reasonable assumption since a coupled spin system can relax to the ground state much faster than a single isolated spin. This relaxation transfers the signal from the first electron to the third, and ultimately to every odd-numbered electron, thus transmitting signal unidirectionally.

### 4.2.3 Energy dissipation in SSL

There are two sources of dissipation in SSL: dynamic dissipation in the gate while it switches between bits, and dissipation in the clock. We examine both below:

**Gate dissipation**

Ref. [63] showed that the energy dissipated in switching the NAND gate is of the order of $g\mu_B |B_{global}|$ which also happens to be the energy difference between the two anti-parallel spin states in any isolated dot. Furthermore, it was shown that if the coupled spin system is in thermal equilibrium and governed by Boltzmann statistics, then the energy $g\mu_B |B_{global}|$ is also equal to $kT\ln(1/p)$ where $p$ is the probability of gate error caused by spins straying from the many-body ground state (which represents the correct gate result) into many-body excited states by absorbing phonons or magnons[14]. Remarkably, this energy - $kT\ln(1/p)$ - is the minimum energy that any irreversible gate must dissipate in a single logic operation as long as the gate is in thermodynamic equilibrium with the environment, according to the Landauer-Shannon theorem [67, 68].

---

[13] This is not completely correct, but is nearly correct. When the spin state of one of the edge dots is fixed by an external agent, the ground state of the 3-spin array will be an entangled state which is not the desired state $|\downarrow\uparrow\downarrow\rangle$ or $|\uparrow\downarrow\uparrow\rangle$, but the probability of the third spin being parallel to the first is much higher than its being anti-parallel.

The energy dissipated in a gate operation, as well as the strength of the global magnetic field, is therefore determined by how much gate error probability can be tolerated at a given temperature. If the error probability cannot be allowed to exceed $10^{-9}$, then the energy dissipated in a gate operation will be $kT\ln(10^9) = 21kT$ at any temperature. Since this energy is also equal to $g\mu_B |B_{global}|$, one must choose the global magnetic field strength such that $|B_{global}| = \frac{kT}{g\mu_B}\ln(1/p)$. With $p = 10^{-9}$, $|B_{global}| = \frac{21kT}{g\mu_B}$.

**Clock**

The clock in SSL causes additional dissipation. If we use non-adiabatic clocking to maintain sufficient speed of operation and adequate noise margin, then the energy dissipated in the clock will be $\sim CV^2$ where $C$ is the capacitance of the clock pad and $V$ is the amplitude of the clock pulse. This energy should be considerably larger than $kT$ to protect against thermal noise [69]. Let us assume that the clock amplitude $V$ is 10 times larger than the thermal voltage fluctuation on the clock pad which is $\sqrt{kT/C}$, resulting in a signal-to-noise ratio of 20 dB. Therefore, the clock dissipation is 100 $kT$ per cycle.

It should be clear now that there are two dissipaters in an SSL circuit – the clock and the device. The former could dissipate about 100 $kT$ per clock cycle and the latter dissipates $kTln(1/p)$ per bit flip, which will be 21 $kT$ if we operate with a bit error probability of $10^{-9}$. Therefore, the total dissipation per clock cycle per bit is potentially ~ 121 $kT$, which is considerably less than the ~ 50,000 $kT$ that present CMOS transistors dissipate [1]. This extreme energy efficiency makes the SSL paradigm worthy of further exploration.

---

[14] This result, although obvious for an isolated spin, is not obvious for a 3-spin system forming a NAND gate. Reference [11] proved this result rigorously.

### 4.2.4. The speed of SSL

The speed of SSL (i.e. the maximum allowable clock frequency) is determined by four factors: (1) the speed with which an input bit can be written in an input port by the writing agent, (2) the speed with which an output bit can be read in an output port by the reader, (3) the gate switching speed, and (4) whether or not the architecture is pipelined. If the architecture is pipelined, then the clock speed is limited by the lowest of the other three speeds; otherwise, the clock speed will be much slower. In a non-pipelined architecture, the clock speed will be $M$ times slower than in a fully pipelined architecture, where $M$ is the number of bit flips required to execute the most complex calculation. Needless to say, the number $M$ can be extremely large, which makes all non-pipelined architectures impractical.

**Pipelining in SSL**

Fortunately, SSL is a pipelined architecture. The clock in SSL not only propagates signals unidirectionally, but it is also responsible for making the architecture pipelined. To understand this concept, consider the spin-wire in Fig. 8. The input bit is applied to the leftmost (first) dot by aligning its spin in the down-direction (Fig. 8a). This is done during cycle 1. In the next cycle (cycle 2), the potentials in the first two gate pads are raised to cause nearest neighbor exchange coupling between the first three dots which then orders their spin in the anti-ferromagnetic configuration as shown in Fig. 8b. In the third cycle, the potential in the first gate pad is lowered, while that in the second gate pad is held, and that in the third gate pad is raised to cause nearest neighbor coupling between the second, third and fourth dots (Fig. 8c). This ensures anti-ferromagnetic ordering within this latter trio which successfully orients the fourth dot's spin anti-parallel to the input spin. In the fourth cycle, the potential at the second gate pad is lowered, that in the third gate pad is held and that in the fourth gate pad is raised, which successfully transfers the input bit applied at the first dot to the fifth dot (Fig 8d), thereby ensuring unidirectional signal propagation along the wire.

The point to note here is that as soon as the potential in the first gate pad is lowered in the third cycle (Fig. 8c), the first dot is decoupled from the chain and the input applied to this dot *can then be changed without affecting successful replication of the original input bit in the fifth dot as described above*. In other words, the input can be changed during the fourth cycle regardless of how long the chain is. During the fifth clock cycle, when the first and second gate pad's potentials are raised again to exchange-couple the first three dots, the original input bit has already propagated down the chain (to the sixth dot) and is decoupled from the input side since the third gate potential has been lowered in the fifth cycle, which decouples the input side from the output side. Thus, the traveling bit will not be affected by the new input. In other words, a new input bit can be fed to the spin wire *before* the earlier input makes it to the very end of the wire. In other words, the input bits can be *pipelined*. The reader should be able to determine that in this case, the input bit rate will be only one-third the clock frequency and not $1/M$ times the clock frequency where $M$ is the number of dots in the chain ($M \gg 3$).

The pipelining however comes with a cost since gate pads must now be inserted between *every* pair of dots in order to apply a local potential between any chosen pair to exchange-couple them. We call this scheme of clocking "granular clocking" since every pair has its own clock pad. This increases the fabrication complexity and cost, and limits the bit density on a chip. However, the alternate is a non-pipelined architecture which will be extremely slow and hence unacceptable. When we discuss magnetic quantum cellular automata architecture later, we will show that it is possible to apply a global clock to ensure unidirectional propagation of signal in SSL. This would have doubtlessly relieved much of the fabrication burden and increased the bit density, but it would have also made the architecture non-pipelined and extremely slow.

**The clock speed in SSL**

Once it has been established that SSL is a pipelined architecture, we have to next determine the writing speed, the reading speed and the gate switching speed to ascertain which is the slowest among them. The slowest speed will also be the clock speed.

**Writing speed**

The speed with which an input bit can be written in an input port (by aligning the spin of the lone electron in a designated quantum dot using a local magnetic field) depends on the flux density of the local field $B_{input}$. The stronger this field, the faster is the writing speed. The energy in this field, however, need *not* be dissipated[15]. Ref. [63] showed that this field must be strong enough that the Zeeman splitting it causes is at least 20 times larger than the exchange coupling strength between dots. The latter can be about 1 meV in semiconductor dots [70]. Therefore, in InSb quantum dot systems,

$$g\mu_B B_{input} \geq 20 \text{ meV};$$
$$B_{input} \geq 6.94 \text{ Tesla}$$
(21)

---

[15] The energy dissipated in switching an input bit is still $g\mu_B B$ where $B$ is the flux density of the global magnetic field that has no relation with $B_{input}$.

where we have assume the g-factor of bulk InSb which is -51. The g-factor in quantum dots can be less than in bulk, which will increase $B_{input}$. There are some materials like $InSb_{1-x}N_x$ which reportedly have a g-factor as large as 900 in the bulk [71]. Assuming that the same g-factor can be retained in quantum dots, the value of $B_{input}$ needs to be only ~0.4 Tesla, if one employs $InSb_{1-x}N_x$ quantum dots as hosts for the spins. Generating field strengths of this magnitude locally is still not easy.

The time required to complete the "writing" of input bits in isolated input dots is of the order of $\sim h/(2g\mu_B B_{input})$ [72]. Using the g-factors and associated local magnetic fields that we have discussed, we find that the writing time is ~ 0.1 ps, which is indeed very fast.

### Reading

There are at least three different schemes to ascertain the spin polarization of single electrons in quantum dots [60-62], among which the scheme of ref. [62] is best suited to SSL. In ref. [62], the reading time was of the order of a millisecond. This time is determined by the speed with which electrons can tunnel in and out of the dot and therefore one should be able to increase this speed dramatically with better engineered structures. We see no fundamental barriers to reducing the reading time to about 1 nanosecond.

### Gate switching speed

The gate switching speed is determined by how long it takes for a gate to complete a logic operation. That, in turn, depends on how fast the coupled spin system can relax to the ground state when coupled with the external thermal bath. This time is much shorter than the spin relaxation time of a single isolated spin for essentially the same reasons that the ensemble averaged spin dephasing time of many interacting spins is orders of magnitude shorter than the dephasing time of a single isolated spin [73]. We are not aware of any measurement of spin relaxation times in *coupled* (as opposed to isolated) quantum dots. However, it is obvious that the coupled system will relax much faster than an isolated spin since a many-body wavefunction, spread out over multiple dots, will interact much more strongly with long-wavelength phonons (whose wavelengths span many dots) than the wavefunction of a single electron in an isolated quantum dot. At any temperature, longer wavelength phonons are more numerous than shorter wavelength ones since phonons obey Bose Einstein statistics at equilibrium. If we assume that the coupled spin system decays to the ground state in ~ 1 nanosecond[16], then the gate operation time will be 1 nanosecond.

It is clear now that among all the three switching speeds, the gate switching speed and the reading speed are the slowest and therefore will determine the clock speed. Assuming reading times and gate switching times of ~ 1 nanosecond, the maximum clock frequency will be

$$f_{clock}^{max} \approx 1 \text{ GHz}. \qquad (22)$$

### 4.2.5. The gate error probability in SSL

There are two types of gate error in SSL. The intrinsic error is caused by the coupled spin system in a gate occupying thermally excited states instead of the ground state and the associated probability is *p*. Extrinsic error is caused by a spin in a dot flipping spontaneously during a clock period and its probability is given by (assuming a Markovian process)

---

[16] The ensemble averaged transverse spin relaxation time $T_2^*$ measured in an ensemble of CdS dots in the author's laboratory was found to be ~ 1 nanosecond at 2 K temperature. We have no measurement of ensemble averaged longitudinal spin relaxation time $T_1^*$. We are tacitly assuming that it is of the same order.

$$p_{extrinsic} = 1 - e^{-\frac{T_c}{T_1}}, \qquad (23)$$

where $T_c$ is the clock period and $T_1$ is the spin flip time of an *isolated* spin. Spin flip times of an isolated spin as long as 1 second has been demonstrated in GaAs quantum dots at very low temperatures of 120 mK [74] and in organic nanostructures at much higher temperatures of 100 K [75]. Assuming $T_c$ = 1 nsec and $T_1$ = 1 sec at the operating temperature, $p_{extrinsic} = 10^{-9}$, which is very encouraging.

### 4.2.6. The temperature of operation of SSL

Ref. [63] showed that if we want a fixed intrinsic error probability $p$, then the temperature of operation is determined by the condition[17]

$$2J = g\mu_B |B_{global}| = kT \ln(1/p), \qquad (24)$$

where $J$ is the energy of exchange coupling between neighboring dots. Assuming $J$ = 1 meV, which is achievable with today's quantum dot technology [70], the maximum operating temperature turns out to be

$$T_{max} \approx 1 \text{ K}, \qquad (25)$$

if we operate with an intrinsic error probability of $10^{-9}$. This is very low temperature and requires $He^3$ cooling, which is a distinct inconvenience. Room temperature operation with such low error probability would have required exchange coupling strengths in excess of 300 meV, which is not presently achievable with semiconductor quantum dot technology.

---

[17] The condition for SSL to work is that $J > g\mu_B B/4$. Equation (2.4) satisfies that condition.

Had we operated at room temperature with the presently achievable $J = 1$ meV, then the bit error probability would have been $p = e^{-\frac{2J}{kT}} = 92.6\%$, which is clearly unacceptable. At 4.2 K temperature (which requires He$^4$ cooling instead of the more elaborate He$^3$ cooling), the bit error probability would have been $4 \times 10^{-3}$ which may be acceptable if significant error correction resources are available.

A recent development has raised some hopes regarding higher temperature operation. It has been shown that graphene nanoflakes can implement SSL type logic gates with much higher exchange interaction strength ($2J = 180$ meV) which allows room-temperature operation with a bit error probability $p = e^{-\frac{2J}{kT}} = 0.1\%$ [76]. This is a very exciting and promising route for SSL.

Equation (24) also yields the value of the global dc magnetic field required for operating at 1 K with an error probability of $10^{-9}$. In an InSb quantum dot with $g = -51$, $|B_{global}|$ will be 0.7 Tesla, which is easily achieved. If the quantum dot material has a g-factor of 900 [71], then the required strength of $|B_{global}|$ is only 0.04 Tesla. These field strengths can be easily achieved with permanent magnets.

The expected performance figures for SSL are summarized in Table III. This table is based on the preceding discussion.

### 4.2.7. Current experimental status of SSL

Like the SPINFET, SSL has also never been demonstrated, but the pathways to its realization are clear. This architecture requires the delineation of an array of quantum dots, each containing a single electron, in specific topological patterns on a wafer. Neighboring dots must be spaced within ~ 10 nm to allow significant exchange coupling between nearest neighbor spins, and gate pads must be inserted between every pair of dots to allow clocking. Such systems are typically fabricated with fine-line lithography. Self-assembly, which is usually preferable over lithography for delineating quantum dots with high density, is unfortunately not easily adaptable to SSL since self assembly is not capable of generating arbitrary geometries.

Numerous groups have demonstrated arrays of quantum dots with single electron occupancy [77] and manipulation of single electron spins in isolated quantum dots has also been demonstrated by a number of groups recently [78-86]. These results inspire hope that SSL, which only requires single electron dots with nearest neighbor exchange coupling, is within the reach of current technology. The only major challenge is the alignment of gate pads between every pair of dots with a high degree of reliability. Recent demonstration of field effect transistors with 6 nm gate length [87] shows that lithography is advancing to the level where such challenges can be met.

### 4.2.8. Organic molecules for SSL?

A more exotic vision for SSL was inspired by the demonstration of exceptionally long spin relaxation time (~ 1 second) in the organic molecule tris(8-hydroxyquinolinolato aluminum) [*Alq$_3$*] at 100 K [75]. Organics typically have very weak spin-orbit interaction since they are normally composed of light elements and the strength of spin-orbit interaction scales with the fourth power of the atomic number of elements. Therefore, the spin relaxation times in organic semiconductors can be very long. With such a long relaxation time, the bit error probability in SSL (see Equation (23)) will be $10^{-9}$ with a 1 GHz clock. *Alq$_3$* is a very suitable organic semiconductor where the lowest unoccupied molecular orbital (LUMO) plays the role of the conduction band and the highest occupied molecular orbital (HOMO) plays the role of the valence band. Isolated nanostructures of such molecules have been demonstrated [52, 75, 88] by trapping one or two molecules in nanocavities and nanopores, and it is likewise possible to delineate nanopores lithographically, impregnate them with few molecules, and use them as hosts for spin bits. While this is possible in principle, it has never been demonstrated to our knowledge. If it does become experimentally accessible at a future date, then organic semiconductors may play a major role in SSL.

## 4.3. Extending SSL to Room Temperature Operation: Replacing a Single Spin with a Collection of Spins

The most serious drawback of SSL is that it requires strictly cryogenic operation (~ 1 K) for low bit error probability, which makes it inconvenient and probably impractical. This shortcoming can be overcome if we replace a single spin with a collection of many (~$10^4$) spins in a ferromagnetic grain (which we call a "nanomagnet") of few nm diameter. Such a ferromagnetic grain will typically contain only a *single* ferromagnetic domain in which all the electron spins are aligned in the same direction. The grain however cannot be too small, since then it may become super-paramagnetic instead of ferromagnetic at room temperature, i.e. the blocking temperature may fall below the room temperature[18]. As along as the grain remains

---

[18] The "blocking" temperature is the temperature above which a ferromagnet transitions into a super-paramagnet. In a super-paramagnet, all electron spins do not point in the same direction, i.e. there is no domain formation. Consequently, a super-paramagnet will neither possess a well-defined direction of magnetization, nor will its magnetization be bistable if we make its shape anisotropic. Thus, a super-paramagnet cannot be used for magnetic

ferromagnetic and contains a single domain, all the $10^4$ spins in it can effectively form one giant classical spin [89].

If we make the geometrical shape of the ferromagnetic grain anisotropic (such as ellipsoid or rectangular parallelopiped), then the long axis of the grain will become the easy axis of magnetization [90]. Only magnetization along the easy axis is stable. Since there are only two possible orientations along the easy axis – parallel and anti-parallel - the magnetization of the nanomagnet becomes *bistable*. These two magnetization orientations, which we once again can designate as "up" and "down", can encode the logic bits 0 and 1.

Unlike in the case of SSL, where the two states encoding the binary bits were not degenerate in energy but separated by an amount $g\mu_B |B_{global}|$, the up and down-magnetization states in a ferromagnetic grain are degenerate in energy if the long axis is an axis of symmetry. As a result, it may appear that we might be able to switch between the logic states without dissipating any energy at all. In principle, this is possible using adiabatic switching schemes, but such schemes are known to be error-prone and not self-correcting. Non-adiabatic switching schemes, which are more reliable, must contend with the fact that there is an energy barrier separating the two magnetization states which is equal to $2Ku_2V$, where $Ku_2$ is the anisotropy energy per unit volume and $V$ is the volume of the nanomagnet. This energy barrier must be transcended when a bit switches from 0 to 1, or vice versa. Accordingly, an amount of energy, that is at least equal to $2Ku_2V$, will be dissipated per bit flip if we employ non-adiabatic schemes.

The minimum allowed magnitude of $2Ku_2V$ (and therefore the energy dissipation) will be determined by the maximum bit error probability that we can tolerate. The extrinsic bit error

---

quantum cellular automata discussed here. The blocking temperature decreases rapidly with decreasing size of the magnetic grain, but even grains with a few nm diameter can remain ferromagnetic at room temperature.

probability is still given by Equation (23), except now the time $T_1$ is the so-called magnetic *retention time*, which is the duration over which the grain retains its magnetization (in the "up" or "down" orientation) before thermal fluctuations destabilize it. The retention time is determined by $Ku_2V$ and is given by $T_1 \approx \tau_0 e^{\frac{Ku_2V}{kT}}$ [91] where $1/\tau_0$ is the so-called attempt frequency, which is in the range $10^9$ -$10^{12}$ Hz [92]. Therefore, we obtain

$$E_{diss}^{min} = 2Ku_2V\big|_{min} = 2kT \ln\left[-\frac{T_c}{\tau_0 \ln(1-p_{max})}\right]. \tag{26}$$

We also find that the extrinsic bit error probability is given by

$$p_{extrinsic} = 1 - e^{\frac{-T_c}{T_1}} = 1 - e^{\frac{-T_c}{\tau_0 e^{\frac{Ku_2V}{kT}}}}. \tag{27}$$

In spherical Fe-Pt alloy clusters, $Ku_2$ can be made as large as $10^6$ Joules/m$^3$ even in grains of diameter as small as 4 nm [93]. Therefore in clusters of diameter 5 nm, the value of $Ku_2V$ is 0.4 eV. Consequently, the retention time $T_1$ at room temperature is in the range $10^{-5}$ – $10^{-2}$ seconds and the extrinsic bit error probability $p_{extrinsic}$, according to Equation (27), should be in the range $10^{-7}$ - $10^{-4}$ for a 1 GHz clock at *room temperature*. This is a very encouraging result since it shows that replacing the single spin with ~ $10^4$ spins in a magnetic grain *allows room temperature operation with still an acceptable bit error probability*.

The cost that we pay for room temperature operation with acceptable bit error probability is the much increased energy dissipation during bit flip compared to SSL. That energy will now be $2Ku_2V = 2\times 0.4 = 800$ meV. Compare that with the energy dissipated in SSL at 1 K temperature, which is ~ 21 $kT$ = 1.82 meV. Thus, the magnetic grain is ~450 times more dissipative than the single spin, and the bit error probability is also a few orders of magnitude larger. However, all this may still be a small price to pay for room temperature operation. Therefore, replacing a single spin with a collection of spins in a magnetic grain to encode binary bit data is a worthwhile idea.

If the electron concentration in the magnetic grain is $8\times10^{22}$ cm$^{-3}$ (typical of metals), then the number of spins (or electrons) in a grain of volume $(5nm)^3$ is $\sim 10^4$. What is amazing is that $\sim 10^4$ spins do not dissipate $10^4$ times as much as a single spin when they collectively flip their polarizations, but dissipate only 450 times more! That happens because *interaction* between the spins, which makes them work collectively and cooperatively as one giant spin, reduces the degrees of freedom from $10^4$ to something much smaller than $10^4$. That, in turn, reduces energy dissipation. The physics behind this striking and counter-intuitive result was expounded in a recent paper [94].

### 4.3.1. Magnetic quantum cellular automata (MQCA) logic gates

Cowburn and Welland [90] reported the first experimental demonstration of a logic gate employing nanomagnets or magnetic grains whose magnetizations encoded binary bits. Each grain has anisotropic shape and hence an easy axis of magnetization. Consequently, each has two possible magnetization directions which encode the logic bits 0 and 1. These grains interact via nearest neighbor dipole-dipole interaction and that interaction is exploited to realize the functionality of a Boolean logic gate. In ref. [90], the authors called their gate "magnetic quantum cellular automata" (MQCA) although it has really no connection with conventional cellular automata architectures that are intrinsically non-Boolean [95]. The device they demonstrated was, in fact, an individual logic gate and not a complete combinational or sequential circuit.

The MQCA logic gate of ref. [90] consists of a linear array of (say, elliptical) magnetic grains. This system is shown in Fig. 9. Magnetization along the major axis of the ellipse is stable since the major axis is the easy axis of magnetization. A strong magnetic pulse, indicated by the black arrow in Fig. 9(a), magnetizes the grains in the direction shown (magnetization pointing to the right). The array is then subjected to an oscillating (square wave) magnetic field that has a large negative dc component. Its pulse shape is shown in Fig. 9(a). During the positive cycle, the field in this wave points to the right, and during the negative cycle, it points to the left. Because of the large negative dc component, the positive amplitude of the wave is weaker than the coercive field of the nanomagnets, but the negative amplitude is stronger.

The two logic inputs to this "gate" are encoded in the direction of the magnetic pulse and the direction of the oscillating square wave magnetic field. When the pulse field is directed to the right (as shown in Fig. 9a), the corresponding input is interpreted as binary bit 1, and when it is directed to the left, the input is binary bit 0. If the square wave field is in the positive cycle, it points to the right and the corresponding input is interpreted as binary bit 1. When it is in the negative cycle, it points to the left and the input is interpreted as binary bit 0.

The logic output of the gate is encoded in the magnetization direction of the grains themselves. If it is pointing to the right, it is interpreted as binary bit 1, and if it is pointing to the left, it is interpreted as binary bit 0.

A right-pointing pulse field (corresponding to input1 = 1) sets the output logic bit to 1 since the grains are magnetized to the right by the pulse. The pulse is then removed and the square wave field is turned on. During the positive cycle of the square wave (i.e. when input2 = 1), the grains' magnetizations continue to point to the right, since the square wave field is directed to the right along the grain's original magnetization, so that the output logic bit remains at logic 1. This situation is shown in Fig. 9a.

Now, if the square wave goes into the negative cycle, so that input1=1 and input2=0, the direction of the applied field on the nanomagnets switches to the left and since this field is much stronger than the coercive field of the grains, the nanomagnets flip their magnetization which then begins to point to the left. At this point, the output state becomes logic bit 0. This situation is shown in Fig. 9b.

The two situations corresponding to the pulse field pointing to the left (input1=0) are shown in Figs. 9c and 9d. Note that in these two cases, the grains are initially magnetized in the left direction by the pulse. The square wave field cannot flip the magnetization during its positive cycle (i.e. when input2=1) since the positive amplitude of the wave is too small to exceed the coercive field of the grain. Therefore, the grains remain magnetized to the left (i.e. output remains in state 0). During the negative cycle of the square wave field, the grains obviously remain magnetized to the left so that the output bit is once again 0. It is easy to see from the above discussion that the relations between the output bit and input bits obeys the truth table of the AND gate in Table II. Therefore, the system in Fig. 9 indeed realizes the Boolean AND operation and serves as an MQCA AND gate.

An alternate MQCA realization of an AND gate was demonstrated in ref. [96] and its principle of operation is elucidated in Fig. 10. Here, instead of using a magnetic field pulse as the first input to the gate (input1), the magnetization of an elongated magnetic grain, placed at the left of the linear array, is used as input1. This grain is a *hard* magnet, whose magnetization cannot be flipped by an external field. The magnetization of this grain is of course bistable since the grain's shape is anisotropic with an easy axis of magnetization along the long axis. When the magnetization of this grain points to the right, input1 is interpreted as logic bit 1, and when the magnetization points to the left, input1 is logic bit 0.

Once again, the direction of the field in the square wave pulse acts as input2 and the magnetizations of the elliptical grains serve as the output. When both inputs are 1, i.e. the elongated grain is magnetized to the right and the square wave is in the positive cycle, the net magnetic field appearing over the elliptical grains is pointing to the right and hence the grains are magnetized to the right. Consequently, the output is 1. This is shown in Fig. 10a. When input1=1 but input2=0, the square wave field is in the negative cycle and points to the left. Its amplitude is large enough to negate the right-oriented field due to the elongated grain and flip the magnetization of the elliptical grains to the left so that the output becomes 0. This is shown in Fig. 10b. When input1=0 and input2=1, the field due to the elongated gain is pointing to the left, while the square wave field is in the positive cycle and points to the right. However, the positive amplitude is too small to overcome the field due to the elongated grain. Therefore, the total field on the elliptical grain is either still pointing to the left, or if it is pointing to the right, then it is simply not strong enough to flip the magnetization of the elliptical grains to the right. Therefore, the grains remain magnetized to the left and the output is logic bit 0. This situation is shown in Fig. 10c. Finally, when both inputs are at logic 0, i.e. the magnetization in the elongated grain and the field in the square wave pulse both point to the left, the total field on the elliptical grains is definitely pointing to the left and the hence the magnetization in these grains is oriented to the left, corresponding to the output bit being 0. This situation is shown in Fig. 10d. Once again, the input-out relation obeys the truth table of the AND gate in Table II.

There are some obvious problems with such realizations. The inputs and the output are all *dissimilar entities*, i.e. one input is a magnetic pulse (or the magnetization of a grain), another is a square wave, and the output is the magnetization of a grain. This makes cascading of successive stages a near impossibility since it is not possible to make the output of one stage serve as an input to another stage and so on. It is also not clear how logic signal can be made to travel unidirectionally from the driving stage to the driven stage, if such logic gates were to be interconnected to form a logic circuit. These difficulties arise because the two inputs and the output are dissimilar quantities. This problem does not arise in conventional electronic logic circuits since inputs and outputs are all encoded in the *same* physical quantity such as voltage or current. The heterogeneous approach of encoding input and output bits in different physical quantities is not conducive to circuit implementation. Therefore, such ideas can result in isolated gates, but are not likely to go far beyond that (without substantial modification) and yield a useful circuit that is capable of universal computation.

**4.3.2. Magnetic quantum cellular automata (MQCA) circuits**

In order to extend the MQCA logic *gate* idea to logic *circuits* where many gates work in tandem, one has to accomplish at least two feats: (1) replace the heterogeneous bit encoding scheme with a homogeneous one, i.e. encode all input and output bits in the same physical quantity (e.g. magnetization of the grains), so that the output of one stage can directly serve as the input to another, and (2) devise a way to steer logic signal unidirectionally from an input stage to an output stage. This was accomplished in ref. [97] which proposed an MQCA architecture that mimics SSL (with the magnetization of a grain – or collection of spins – replacing the single spin, and dipole-dipole interaction between adjacent grains replacing the exchange interaction between adjacent spins). Bit encoding is homogeneous, i.e. both input and output bits are encoded in the magnetization of nanomagnets and unidirectional steering of logic bits is accomplished with a global clock which is a pulsed magnetic field.

Ref. [97] considered an array of anisotropic nanomagnets with nearest neighbor dipole-dipole interaction that makes the ground state of the array anti-ferromagnetic, i.e. the magnetizations of immediate neighbors are anti-parallel. As usual, the nanomagnets have shape anisotropy which results in an easy axis of magnetization along the major axis and therefore two stable magnetization directions (parallel and anti-parallel to the easy axis). These two magnetization orientations – shown in the top panel of Fig. 11 - encode logic bits 0 and 1. Note that while a dc magnetic bias field is required in SSL to define the spin quantization axis and make the spin polarization of every electron bistable, no such magnetic field is required in MQCA; shape anisotropy defines the magnetization quantization axis and it also makes the magnetization orientation of every grain bistable.

The absence of the magnetic field is a boon, but it also makes the implementation of a NAND gate (in the manner of SSL) impossible. To understand this, consider the three nanomagnets with nearest neighbor dipole-dipole interaction shown in the bottom panel of Fig. 11. Once again, the two peripheral nanomagnets host the input bits, and the central one hosts the output bit. When both inputs are logic 1 (corresponding to the magnetization pointing up), the output will be in logic 0 in order to maintain the anti-ferromagnetic ordering. Similarly, when the two inputs are in logic 0, the output will be logic 1 to maintain the anti-ferromagnetic ordering.

However, when one input is logic 1 and the other logic 0, there will be a tie that cannot be resolved in the manner of SSL since there is no dc bias magnetic field here (although there can be) to break the tie. Therefore, the NAND gate cannot be implemented easily with nanomagnets in the absence of a global magnetic field.

Instead of the NAND gate, MQCA utilizes a majority voting gate to implement arbitrary Boolean logic circuits. In such a gate, the output bit replicates the majority of an odd number of input bits. Ref. [97] presented the design of a majority logic gate utilizing anisotropic nanomagnets. Dipole-dipole interaction between the grains enforces anti-ferromagnetic ordering and ensures that the output bit replicates the majority among the input bits, when the array reaches ground state. This gate is compatible with multi-gate circuits since, unlike the MQCA logic gates of refs. [90, 96], the bit encoding scheme here is homogeneous; both inputs bits and output bit are encoded in the same physical quantity, namely the magnetization orientations of the grains.

**4.3.3. Unidirectional signal propagation in MQCA circuits; Granular and global clocks**

In order to steer logic bits undirectionally from the input to the output of a circuit, one will need a clock. A popular clocking scheme that can satisfy this need is Bennet clocking [98] which we describe next. Consider an "MQCA wire" consisting of an array of four nanomagnets. The objective is to steer the magnetization orientation of the first grain (input bit) unidirectionally from left to right along this wire. In the ground state, the ordering of the magnetizations is anti-ferromagnetic as shown in the first row of Fig. 12. This ordering faithfully transfers the input bit to every odd numbered grain so that the third grain (which is near the right end of the MQCA wire) replicates the input bit. If the input bit at the left end is changed (flipped), then the array temporarily goes into the state shown in the second row of Fig.12. At this point the second magnet's magnetization is in an indeterminate state since the dipole interaction with its left neighbor is telling it to switch, but the dipole interaction with its right neighbor is telling it to stay put. Since both dipole interactions are *equally* strong, the second magnet's magnetization orientation remains in an indeterminate state and signal propagation stops[19]. To resume propagation to the *right* (i.e. unidirectionally), we apply a local magnetic pulse *selectively* to the third magnet to set its magnetization along the hard axis (perpendicular to the easy axis) as shown in the third row of Fig. 12. This magnetic pulse acts as a *local* clock. During this time, the second magnet finds its left and right neighbor's influences are *unequal* since the clock has broken the symmetry. Since the second magnet will prefer to conform to anti-ferromagnetic ordering, and its magnetization must align along the easy axis, it will now flip down to satisfy both needs. This is shown in the fourth row of Fig. 12. Signal has now propagated through the second magnet. In the next cycle, the local clock pulse applied to the third magnet subsides and the fourth magnet is clocked *locally* to place its magnetization along the hard axis. This is shown in the fifth row. During this time, the third magnet's magnetization flips up to minimize energy. By now, the signal has propagated through the first two magnets completely. Thus, by sequentially clocking the nanomagnets, we can make the signal propagate unidirectionally along the nanomagnet-wire. This is the essence of Bennet clocking.

This is not the clocking scheme proposed in ref. [97], but we propose it here since, as we will show later, it makes the architecture pipelined. A very similar clocking scheme has been discussed in ref. [99] which motivated the scheme here. We will later show that the original clocking scheme in ref. [97] would have allowed signal to propagate unidirectionally, but would *not* have permitted pipelining of data.

---

[19] This is actually a metastable state. Thermal fluctuations can drive the system out of the metastable state into a global ground state, but the same thermal fluctuations can also take it out of the global ground state and drive it back to the metastable state.

We term the clocking scheme that we proposed here "granular" since it requires making a connection to every individual nanomagnet in order to apply the clock pulse *locally* to that magnet alone. A local current flowing around a magnetic grain can reorient the magnetization of that grain via the spin transfer torque effect [100-104] and thus serve as the local clock. The granular scheme however carries a significant fabrication overhead since it mandates a separate electrical connection to a current-carrying path around *every* nanomagnet. Perhaps to avoid this overhead, ref. [97] considered a *global* clock for unidirectional signal propagation. To understand how a global clock works, refer to Fig. 13. A global clock applies a magnetic field simultaneously over every nanomagnet which resets the magnetization of all magnets along the hard axis as shown in the first row of Fig. 13. Next, an input bit is applied at the left end which magnetizes the first nanomagnet along the easy axis ("up" or "down" depending on whether the input bit is 0 or 1). This is shown in the second row, where we assumed that the input bit was "up". The second nanomagnet is now in an asymmetric environment since its left neighbor is magnetized in the vertical (easy axis) direction and the right neighbor is magnetized in the horizontal (hard axis) direction. The left neighbor's influence is obviously stronger since it is magnetized along the easy axis. Consequently, the second magnet's magnetization flips down to assume anti-ferromagnetic ordering as shown in the third row. Then the third magnet finds itself in an asymmetric environment and its magnetization flips up to conform to anti-ferromagnetic ordering as shown in the fourth row, and so on. This will proceed in a domino-like fashion till all the magnets have flipped to assume magnetizations along their easy axes and in accordance with global anti-ferromagnetic ordering. In the end, the input bit is repeated in every odd-numbered magnet. Note that exactly the same scheme could have been adopted in SSL where a global clock field, that is perpendicular to the global dc field that defines the spin quantization axes, would have achieved the same result and guaranteed unidirectional spin propagation down a spin wire. This would have been very attractive since it would have eliminated the need to make a clock connection to every pair of dots. However, as mentioned earlier, it was shunned because such a global clocking scheme does not allow pipelining of input data, as we show below.

**4.3.4. Global clocking in MQCA leads to a non-pipelined architecture**

Global clocking has one serious shortcoming, namely that it makes the architecture non-pipelined and hence very slow. To understand this, compare the granular clocking scheme of Fig. 12 and the global clocking scheme of Fig. 13. In Fig. 12, where granular clocking is used, once the third nanomagnet's magnetization has flipped (as shown in the last row), we can change the input bit in the first nanomagnet, i.e. flip its magnetization, if needed. This will place the second magnet temporarily in an indeterminate state and it will not emerge from this state until the third magnet's neighborhood receives the clock signal again. Until that happens, the second magnet will prevent the new input bit from propagating to the right and catching up with the old bit. By the time the second magnet emerges from the indeterminate state and propagates the new input bit to the right, the old input bit has already traveled down the line. The reader will understand that the new bit never catches up with the old bit. Hence, the new input can be fed to the MQCA wire before the old input has made it to the end of the wire. In other words, data can be pipelined. *This is possible only because every magnet is clocked locally and independently*, i.e. we have employed "granular" clocking. Moreover, the minimum clock period is the time it takes for a single nanomagnet to flip magnetization, which is typically on the order of ~ 1 nsec[20]. Therefore, the clock frequency can be as high as ~ 1 GHz.

Now consider the global clocking scheme of Fig. 13. Here, magnets are not clocked locally and independently. They are all clocked *simultaneously* by one global magnetic field. In Fig. 13, where a *global* clock has been used, we cannot apply the second clock pulse until the input bit has propagated down the *entire chain* in a domino-fashion. Otherwise, the clock will disrupt the already ordered magnetizations and unidirectional propagation in the chain (and hence generation of the correct output bit in the very last magnet) cannot be guaranteed. Since domino-like switching is serial, the minimum clock period is now $M$ times the time it takes for a single nanomagnet to flip, where $M$ is the number of magnetic grains in the chain. Therefore, the clock period is ~ $M$ nsec if we still assume, as before, that each nanomagnet takes 1 nanosecond to switch. Since we are not applying a new clock pulse till the final output has been produced, we also cannot change the input bit before the final output has emerged because each application of input must be preceded by an application of the global clock to reset every nanomagnet. In other words, this architecture is now *not* pipelined and the input bit rate cannot exceed $1/M$ GHz.

The trade off between the two clocking schemes is a much faster data processing speed at the expense of a much more expensive and complex fabrication requirement. Granular clocking provides an improvement in the clock speed by a factor of $M$, but requires many more internal connections on the wafer. Reference [97] claimed that nanomagnets can be fabricated with a density of $10^{10}$ cm$^{-2}$, so that in a chip of 1 cm$^2$, the longest line of grains will have $\sqrt{2} \times 10^5$ grains, i.e. $M = 1.41 \times 10^5$. Therefore, in the *worst-case* scenario, when an MQCA wire is as long as the diameter of the chip, the period of the global clock cannot be shorter than 0.141 msec, so that the clock frequency is limited to only ~ 7 kHz on a 1 cm$^2$ chip with a bit density of $10^{10}$ cm$^{-2}$. On the other hand, a granular clock would have allowed a clock frequency of ~ 1 GHz, but obviously would have required $10^{10}$ internal connections.

---

[20] The time to produce a logic operation has been estimated as 0.7 nsec and the time to flip the magnetization of a nanomagnet has been estimated as 0.15 nsec [see F. M. Spedalieri, A. P. Jacob, D. Nikonov and V. P. Roychowdhury, arXiv: 0906.5172 [cond-mat.mes-hall], (2009).].

**4.3.5. The bit error probability in MQCA with global clock and granular clock**

In Section 4.3, we estimated that the bit error probability in nanomagnetic architectures like MQCA will be in the range of $10^{-7} - 10^{-4}$ if the clock frequency is 1 GHz. That is true for a granular clock which can sustain a frequency of 1 GHz. For a global clock, the clock period will be much larger. If it is 0.14 msec, then the bit error probability according to Equation (27) will be $p_{extrinsic} = 1 - e^{\frac{-T}{T_1}} = 1 - e^{\frac{-0.14 \text{ msec}}{10 \text{ msec}}} = 1.4\%$ to $1 - e^{\frac{-T}{T_1}} = 1 - e^{\frac{-0.14 \text{ msec}}{0.01 \text{ msec}}} \approx 100\%$ in the *worst case*. Thus, the use of a global clock in MQCA architecture may not be practical. A granular clock, on the other hand, reduces the clock period to 1 nanosecond and the bit error probability to $10^{-7} - 10^{-4}$, which is more practical.

**4.3.6. The "misalignment" problem associated with the use of a vector (magnetic field) as the clock in MQCA**

As we have seen, MQCA architecture needs a *vector* clock, namely a magnetic field. This causes the following problem.

Consider the globally clocked architecture first. In the initialization (or resetting) phase, each magnetic grain is magnetized along the hard axis by a magnetic field pulse. Let us assume that this field is not exactly aligned with the hard axis, but subtends an angle $\theta$ with it, which we call the misalignment angle (see Fig. 14(a)). Then the component of the field along the hard axis is $H_{pulse} \cos\theta$ and the component of the field along the easy axis is $H_{pulse} \sin\theta + H_{neighbor}$ where $H_{pulse}$ is the amplitude of the clock pulse and $H_{neighbor}$ is the magnetic field caused by the neighbors on a nanomagnet. Let $r$ be the ratio of the magnetic fields required to magnetize the nanomagnet along the easy axis and along the hard axis respectively. Therefore, we need

$$\frac{H_{pulse} \sin\theta + H_{neighbor}}{H_{pulse} \cos\theta} < r, \qquad (28)$$

which mandates that

$$\tan\theta < r \ll 1. \qquad (29)$$

The above inequality shows that MQCA cannot tolerate much misalignment between the magnetic field pulse of the clock and the hard axis, since $\theta$ must be a very small angle.

It may appear that we can limit $\theta$ to a very small value using sophisticated lithography, but this is actually not correct. Even if advanced lithography techniques can limit the misalignment angle between different grains to less than $1^0$ across an entire 10 cm×10 cm chip containing perhaps $10^{12}$ grains, which is already a tall order, it cannot guarantee that the misalignment angle between the hard axis and the clock field is still less than $1^0$ in every grain, since internal magnetic structure of the grain need not always make the easy axis aligned exactly along the long axis of the grain. In other words, an uncertainty of $1^0$ in fabrication does not necessarily translate to an uncertainty of $1^0$ in $\theta$; the latter uncertainty may be still quite a bit larger than $1^0$. Therefore, misalignment is a very real menace.

In the granular clocking scheme, each nanomagnetic grain has its own private clock so that one could, in principle, align the clock field around each grain *individually* to make $\theta$ very small in every grain (as shown in Fig. 14(a)). This will require a testing and recalibration step, which is an additional burden. The problem however is much worse in the case of a global clock because only one magnetic field pulse is used globally and it cannot possibly be aligned with the hard axis of every single magnetic grain among the $\sim 10^{12}$ grains in a 10 cm×10 cm chip. This problem is elucidated in Fig. 14(b). Grains that are misaligned by relatively large angles do not switch with the clock pulse and ultimately lead to unacceptable bit error rates.

### 4.3.7. Experimental status of MQCA

Unlike the SPINFET and SSL, isolated MQCA logic gates have been experimentally demonstrated [90, 96, 105], but there has been no demonstration of actual logic circuits where several gates work in unison to elicit any kind of computation. The experiment in [105] is qualitatively different from those in [90, 96] since it employed the homogeneous bit encoding scheme (as opposed to the heterogeneous schemes of [90, 96]) and therefore, in principle, can be extended to circuits. This experiment used a global clock and reported that only ~ 25% of the devices worked because of fabrication variability and the run-to-run reproducibility turned out to be only 50% because of *clock field misalignment* [105]. In this experiment, there were only few grains and yet the error introduced by clock field misalignment was large and the reproducibility poor. The problem would have been much worse if there were many more grains. Thus, the "misalignment problem" is indeed a serious handicap for MQCA.

Unlike MQCA, SSL uses a *scalar* clock where the clock variable is an electrostatic potential that is applied to lower potential barriers between neighboring quantum dots and turn on the exchange coupling between them. There is no issue of misalignment here since the clock is scalar and not vector, but there is of course still an issue of clock synchronization in time, which is a challenging problem in today's ultra-large-scale-integrated circuits. In MQCA, the clock is a vector quantity which creates a twofold problem; there is a problem of clock synchronization in time and also in space (direction). That makes MQCA implementation extremely challenging. In spite of that, MQCA does offer an interesting solution to energy-efficient computing at room temperature since the energy dissipation per bit flip (that we calculated to be about 800 meV at room temperature) can be potentially 1,500 to 2,000 times smaller than in today's transistors. The estimated performance figures for MQCA are summarized in Table IV.

**4.3.8. Reading and writing of bits in MQCA and associated power dissipation**

Reading of bits (magnetization of grains) can be accomplished using either giant magnetoresistance devices or magnetic tunnel junctions that are ultra-sensitive magnetic field sensors and can read the magnetization of grains. This strategy is commonly used in reading data stored in magnetic hard disks of computers and is a mature technology [106-108]. This technology is not exceptionally power hungry, but it does consume some energy.

Writing bits requires the use of currents to generate local magnetic fields to magnetize selected grains in desired directions. Spin transfer torques, generated by currents, can also serve the purpose of writing data. Since typically large currents are required for this purpose, writing will tend to consume much more energy than reading. The clock may also require significant energy since it requires the generation of an on-chip magnetic field. Energy estimates for these operations are not available, but the energy dissipated during the writing and clocking operations can exceed the energy dissipated during bit flip (~ 800 meV) considerably. Writing is not the major problem since it is done infrequently, but clocking is much more frequent since a clock is needed for every computational step. Thus the major energy dissipater in MQCA may well turn out to be the clock.

## 4.4. Domain Wall Logic

A scheme closely related to MQCA is "domain wall logic" which is described in this section. A domain wall is the interface between two neighboring domains in a ferromagnet, each of which contains (ideally) perfectly aligned spins. In a bulk ferromagnet, the spin alignments in two domains may subtend any arbitrary angle between them, but in a nanowire, the strong shape anisotropy ensures that the magnetization (or spin alignment direction) in any domain is either parallel or anti-parallel to the nanowire axis. In that case, the domain wall is the boundary between regions of *oppositely* aligned magnetization.

Ref. [109] proposed and demonstrated a logic family based on domain wall motion in nanowires and called it "domain wall logic". Here, the two possible directions of magnetization in any domain encode the binary logic bits 0 and 1. Domain walls can propagate through a nanowire under the influence of an external magnetic field which acts as a clock signal. Such propagation can be harnessed to implement Boolean logic operations.

A set of different Boolean logic gates were demonstrated in ref. [109]. For example, a cusp in a nanowire will reverse the direction of propagation of a domain wall under a driving magnetic field [109]. Now, if we define magnetization parallel to the initial direction of propagation as logic 1, and magnetization anti-parallel to the initial direction as logic 0, then the cusp essentially acts as an inverter or NOT gate, whose output is the logic complement of the input. The cusp is shown in Fig. 15(a). Similarly, a simple fork in a nanowire provides the fan out function, and is shown in Fig. 15(b).

### 4.4.1. Writing and reading of binary data in domain wall logic

Writing of data (or providing input bits to a domain wall logic gate) can be accomplished in two different ways: (1) the direct way is to align the direction of domain propagation in any logic element by applying a local magnetic field generated by a current carrying conductor placed in the vicinity of the logic element. This will set the input bit to either 0 or 1, as desired. Ref. [109] showed how this is accomplished in a shift register comprising eight NOT gates. The NOT gate designated as the input gate has an enlarged cusp which lowers the magnetization reversal field in this gate selectively. Therefore, when a magnetic field that exceeds the reversal field of this gate is generated in the vicinity of the shift register, only this gate responds and the input bit is set to 0 or 1, which is then propagated through the other seven NOT gates in the fashion of a shift register. (2) A second method described in ref. [109] is to modulate the amplitude of the *global* rotating magnetic field which acts as the clock. The modulation amplitude will be such that it will adjust the direction of propagation of the domain wall selectively in an input gate which has an enlarged cusp, while leaving all other gates in the circuit unaffected. This will write input data to the first gate, which will subsequently drive other gates. However, this same rotating magnetic field also acts as the global clock that drives computation. Therefore, any perturbation or modulation of its amplitude may have undesirable effects on all other gates. Typically, the undesirable effects can be avoided if we do not feed a new input bit to the input gate *before* the final output bit has been produced following domain wall propagation through all other gates in the entire circuit. However, in this case, input data *cannot be pipelined* and therefore this architecture will be very slow. The slowness does not arise from the fact that domain wall velocity may not itself be sufficiently high – that would be an additional problem – instead, the slowness comes about from the fact that data cannot be pipelined, regardless of how fast domain walls can propagate in a nanowire.

The output of a domain wall logic gate is read by optical measurements, typically magneto-optical-Kerr-effect. That makes this system difficult to integrate on a chip since reading and writing are accomplished by dissimilar means. As a result, current implementation strategy may not be scalable to logic chips comprising some $10^9$ logic gates.

### 4.4.2. The misalignment problem in domain wall logic

Domain wall logic also uses a vector clock since the clock driver is a magnetic field. However, in this case, magnetization never has to be oriented along the hard axis, which is difficult to do. Therefore, even if the clock field is somewhat misaligned with the intended propagation direction, it may still induce domain wall motion in the desired direction, which makes the misalignment problem significantly less menacing. Therefore, domain wall logic may work with a much smaller bit error rate than magnetic quantum cellular automata.

### 4.4.3. Experimental status of magnetic domain wall logic

The magnetic domain wall logic family has been experimentally demonstrated [109-111] for single or few gates. However, no data on device density and speed, or even estimates of these quantities, have been presented, which precludes a technological assessment at this point. Our own view is that its major drawback will be its relatively low speed. It will probably also be less energy-efficient than MQCA simply because in addition to the energy dissipated during magnetic reversal, and energy dissipated in the clock, there will be energy dissipated during domain wall propagation in a nanowire, which can be viewed as viscous flow.

## 5. SPIN ACCUMULATION LOGIC (SAL)

The spin accumulation logic concept [112] is a relatively new idea that is yet to be experimentally demonstrated. Its proponents recognized what we stated in Section 2, .i.e. the primary reason for the failure of spin transistors is inefficient spin injection/detection efficiencies at ferromagnet/semiconductor interfaces. That results in low conductance ON/OFF ratios for the transistors, which, in turn, leads to low noise margin and high bit error rate in a noisy environment. To circumvent this problem, ref. [112] proposed an idea where efficient spin injection or detection across ferromagnet/semiconductor interfaces is not needed; instead, spin accumulation in a semiconductor region flanked on either side by two ferromagnets is exploited to encode and process logic bits.

The cross-section of the proposed logic gate is shown schematically in Fig. 16. The input logic bits are encoded in the magnetization directions of the ferromagnetic contacts 1 and 5, which are set by current carrying wires inducing the spin-transfer-torque effect. Current flows between contacts 1 and 2, as well as 4 and 5 under the action of two batteries denoted by $V_{bat}$. If magnetizations in neighboring ferromagnets (1 and 2, or 4 and 5) are anti-parallel, spins injected by one into the underlying semiconductor layer are not immediately extracted by the other, leading to spin accumulation in the semiconductor layer. Regardless of how efficiently or inefficiently the ferromagnets inject and extract spins, the spin accumulation in the intervening semiconductor region is much larger when the ferromagnets are anti-paraellel, compared to the case when they are parallel.

To understand how this gate can act as a NAND gate, consider the situation when ferromagnets 2 and 4 are both magnetized in the inward direction as shown, which we designate as encoding logic 1 states. The magnetizations of contacts 1 and 5 are supposed to encode the input bits. When the inputs to the logic gate are both bits 0, then contacts 1 and 5 are magnetized in the outward direction, so that neighboring contacts (1,2 and 4,5) are mutually anti-parallel. In this case, the semiconductor regions pinched between contacts 1 and 2, and between contacts 4 and 5, experience significant spin accumulation. This spin accumulation will affect the electrochemical potential underneath contact 3 and cause a transient current to flow through the capacitor connected between contact 3 and ground. This transient current is the output of the logic gate and flows only when the input bits switch. It will depend on the electrochemical potential underneath contact 3, which is determined by the degree of spin accumulation between contacts 1,2 and 4,5, which, in turn, is determined by the magnetization orientations of contacts 1 and 5 that are the input bits. Thus, the transient output is a logic function of the two input bits, and ref. [112] showed that the logic function happens to be that of the NAND gate.

## 5.1. The Logic Separation Between Bits 0 and 1 in SAL

In SPINFETs and MTJ-gates, the logic levels are encoded in conductance states. For example, the high conductance state of a SPINFET or MTJ could encode logic bit 0 and the low conductance state could encode logic bit 1. The separation between the logic levels is then determined by the conductance ON/OFF ratio, which is currently quite low (~10 or less) because of the inefficiency of spin-injection and detection across a ferromagnet/paramagnet interface. This separation determines the bit error rate in a noisy environment.

The SAL paradigm bypasses the need for efficient spin injection or detection, and instead relies on spin accumulation. It might therefore appear that this paradigm will yield a larger separation between logic levels and result in a smaller bit error rate. Unfortunately, it turns out to be not the case. The input logic levels are encoded in mutually anti-parallel magnetizations of the contacts 1 and 5, which are indeed well-separated. However, the output logic levels are determined by the amplitudes of the transient current that flows through the capacitor in Fig. 16 corresponding to output states 0 and 1. The ratio of these two amplitudes is a mere factor of ~2 according to the calculations in ref. [112], which makes the separation between logic levels very small at the output port. This makes SAL worse than even the SPINFET or MTJ logic, and, in fact, makes it inapplicable in mainstream logic applications since the bit error rate will be too high.

## 5.2. Energy dissipation in SAL logic gates

The major problem with SAL is that it has standby power dissipation because the batteries $V_{bat}$ constantly drive currents between contacts 1 and 2, or 4 and 5, even when the gate is not processing information, i.e. not producing an output in response to input bits. In contrast, there is no standby power dissipation in a MISFET, or SPINFET, or SSL, or MQCA. Therefore, the spin accumulation logic is not at all energy-efficient and does not resonate with the rest of spintronics that focuses primarily on reducing energy dissipation. The Supplementary Information section accompanying ref. [112] has calculated the energy dissipation in a chip of $10^6$ gates to be ~ 35 Watts with a 1 GHz clock, so that the power dissipation per gate is 35 μWatts. Compare this with the power dissipated in a MISFET on the Pentium IV chip. Assuming that, on the average, a MISFET gate comprises three transistors, and that the Pentium IV chip has a MISFET density of $10^8$ transistors in each $cm^2$ area, this chip will have $3.3 \times 10^7$ gates. With a 2.8 GHz clock, and perhaps with an activity level of 5%, i.e. one in every twenty gate is switching at any given instant of time, this chip dissipates less than 50 Watts, so that the dissipation per gate is ~ 30 μWatts. This is still less than what an SAL gate dissipates at one-third the clock frequency. Thus, the spin accumulation logic gate appears to be actually less energy-efficient than a traditional charge based logic gate. It will also have a lower density (because it has 5 terminals instead of just 3 for a normal transistor) and it will be slower than a traditional transistor-based gate since magnetic reversal is unlikely to permit a clock speed exceeding 1 GHz in the near term. Its most debilitating drawback is the very small logic separation at the output node. All this currently precludes application in digital electronics.

## 6. SPIN WAVE BUS (SWB) TECHNOLOGY

The spin wave bus (SWB) technology [113-116] is a recently proposed paradigm where a spin wave is utilized to carry information between logic gates. Logic bits are encoded in the *phase* of the spin waves. A phase of $0^0$ encodes one logic bit (say, bit 1) and a phase of $180^0$ encodes the other logic bit (say, bit 0).

At any finite temperature, the spins in a ferromagnetic material are not perfectly ordered parallel to each other, but may fluctuate in time and space as shown in Fig. 17. This constitutes a *spin wave*, which can be guided along a ferromagnetic film, thereby transmitting information from one location to another, if information can be embedded in the wave. In a spin wave bus (SWB) circuit, logic information (binary bits 0 and 1) is encoded in the phase of the spin wave, and a suitable waveguide ferries this information between different devices that can extract and decode the phase.

A SWB logic device is schematically shown in Fig. 18. It is adapted from ref. [113] and has a ferromagnetic film that hosts a spin wave. The waves are launched from asymmetric coplanar strips (ACPS) transmission lines on top of the structure. The ACPS lines are periodically spaced with a period of *d*. They also act as detectors and detect the amplitude of spin waves passing directly underneath them through the voltage induced on the line by the wave.

The sign of the voltage on the exciting ACPS line determines the initial phase of the spin wave launched from that line. The device is designed such that a positive voltage of +*V* results in an initial phase $\phi = 0^0$ and a negative voltage of −*V* results in an initial phase $\phi = 180^0$.

Ref. [113] has described the implementation of various logic gates using spin wave bus technology. An inverter (NOT gate) is configured by tapping the output voltage from an ACPS line located at distance *d* from the exciting line after a time *t* such that

$$t = \frac{\pi d}{v_{ph}} \quad (30)$$

where $v_{ph}$ is the phase velocity of the spin wave. In this case, the phase of the spin wave at the detector ACPS line is $180^0$ out of phase with that in the exciting ACPS line. Therefore, if the exciting voltage was positive and represented logic 1, the induced voltage on the detector line would be negative and represent logic bit 0. This realizes an inverter (NOT gate). In fact, a reference voltage $V_{ref}$ is introduced such that if the induced voltage $V_{ind} > V_{ref}$, then the induced voltage is read as logic 1; otherwise, as logic 0. For implementing the NOT gate, the reference voltage can be zero.

An AND gate is configured by launching spin waves from two ACPS lines that are equidistant from the detector. Time resolved measurement of the voltage on the detector is made after a time 2*t* (where *t* is given in Equation (30)) and $V_{ref}$ is set equal to *V*/2. If both inputs were 1, the phases of the waves interfere constructively at the detector which sees the crest of the interfering waves so that $V_{ind} > V_{ref}$, and the output is read as 1. If either of the inputs were logic bit 0, then the waves interfere destructively and $V_{ind} < V_{ref}$, so that the output is read as 0. If both inputs were logic 0, then the waves interfere constructively at the detector, but the detector sees the trough of the interfering waves so that $V_{ind} < V_{ref}$, and the output is again read as 0. It is easy to see that the input-output relations satisfy the truth table of the AND gate in Table II, so that indeed this is an AND gate.

An OR gate can be configured in exactly the same way, provided we assign $V_{ref} = -V/2$.

## 6.1. Is the SWB Technology Energy-Efficient?

Energy dissipation in spin wave bus has four components: (i) energy dissipated in the wave launcher, (ii) energy dissipated in the ferromagnetic medium during propagation, (iii) energy dissipated in the detector, and (iv) energy dissipated in the clock. The first three energies are technology and materials dependent and the fourth depends primarily on the clock amplitude. As before, we will assume that the clock dissipation per cycle is 100 $kT$, which is 2.6 eV at room temperature. Ref. [114] provides an estimate of the energy dissipated per bit for a 10 GHz clock during propagation and this energy is $10^{-18}$ Joules or 5.26 eV (~200 kT at room temperature). Therefore, the energy dissipated (excluding those dissipated in the launcher and detector, which are like writers and readers) is ~ 10 times higher than in MQCA, but still significantly less than what transistors dissipate today.

## 6.2. Experimental Status of SWB Technology

To our knowledge, there has been but a single experimental demonstration of a SWB device where an output voltage, sensitive to the relative phases of spin waves, was produced [116]. However, this experiment does not provide enough details to estimate power efficiencies and speed achieved.

## 6.3. Shortcomings of the SWB Technology

The SWB paradigm depends on the precise control and detection of the *phase* of a spin wave. All phase-based devices and circuits suffer from certain general maladies which afflict the SWB devices as well. We list some of them here.

1. Phase is a delicate entity and is easily disrupted by scattering of the wave by imperfections. If the phase is disrupted differently between different sections of a spin bus, it presents a problem with reliability and increases the bit error rate. The phase of any wave is also extremely sensitive to dimensions of the waveguide (e.g. the distance *d* which is the period of the ACPS lines) and geometry. As a result, phase devices have very little fault tolerance [6, 7].

2. *Dispersion* is a serious problem. The phase velocity $v_{ph}$ of the spin wave is frequency-dependent [117]. It is extremely difficult, if not impossible, to launch strictly monochromatic spin waves from ACPS lines or any other transducer. In reality, a spectrum of frequencies is launched. Nonlinearities in the waveguide (ferromagnetic film) can generate additional frequencies through harmonic generation. Different frequency components travel with different phase velocities $v_{ph}$ because the medium is dispersive. Therefore, the time *t* in Equation (30) is *not* constant, which presents a serious problem in making the time resolved measurements.

   The bit error probability caused by dispersion alone will be approximately $\frac{\Delta t}{t}$ where $\Delta t$ is the spread in the time *t* caused by a variation in the phase velocity due to dispersion. From Equation (30), we get $\frac{\Delta t}{t} = -\frac{\Delta v_{ph}}{v_{ph}}$. In a ferromagnet, the energy dispersion relation of a spin wave is approximately parabolic [117], so that $-\frac{\Delta v_{ph}}{v_{ph}} = -\frac{\Delta q}{q} = \frac{\Delta f}{2f}$ where *f* is the center frequency of the spin wave and *Δf* is the bandwidth. Therefore, for a center frequency of 1- 10 GHz and a bandwidth of 100 kHz, the bit error probability will be $10^{-5}$ - $10^{-4}$ due to dispersion alone.

3. Phase is a continuous variable, not a discrete binary variable. In SSL, the spin polarization is made bistable with a global magnetic field, and in MQCA, the magnetization orientation is made bistable by exploiting shape anisotropy. Therefore, the spin polarization and the magnetization orientation become *natural binary variables* that can naturally encode the logic bits 0 and 1. This does not happen with phase. Since phase is a continuous variable, it must be transduced into voltage at the ACPS line and that voltage must be rendered "bistable" by the use of a non-linear element (such as a transistor) that has an S-type non-linear transfer characteristic [118]. This is the common approach used in electronics where current and voltage (which are also continuous variables) are used to encode binary logic bits. The element with an S-type transfer characteristic serves to effectively discretize the continuous input variable into a discrete (binary) output variable. More importantly, such a device will restore logic signal at circuit nodes and correct for errors that occur due to a current or voltage fluctuations in a noisy environment. No equivalent for this device exists for spin waves. Without that, the SWB technology will be very error-prone if used in digital electronics.

The performance estimates for SWB technology are summarized in Table V.

## 7. CONCLUSION

In this review, we have attempted to provide a synopsis of our current understanding of those spin-based logic devices, gates and architectures that have attracted the most attention in the engineering and applied physics community. We have found that no spin-based approach is perfect; all have serious shortcomings.

Spin-transistors, which are an evolutionary approach to replacing charge-based electronics with spin-augmented electronics, have two major drawbacks: (1) insufficient energy efficiency, and (2) poor conductance ON/OFF ratio (or small logic level separation) that causes a high bit error rate in a noisy environment. Since energy efficiency ultimately determines device density on a chip (less dissipative devices can be packed more densely without overwhelming heat sinking technologies), these transistors are unlikely to yield more processing power per unit area than traditional MISFETs used today. Additionally, the poor conductance on-off ratio, accruing from inefficient spin injection and detection, makes them error-prone and unreliable. Among all the proposed variations of spin-transistors discussed here, only the transit-time-spin-field-effect-transistor and a few other related devices have been experimentally demonstrated, while all others await demonstration.

The MTJ logic device also has a small logic level separation owing to inefficient spin injection and detection at ferromagnet/spacer interfaces. Therefore, they too are error-prone and unreliable, but may be somewhat more energy-efficient than SPINFETs. The MTJ logic gate and related gates based on giant magneto-resistance devices have been experimentally demonstrated.

SSL, MQCA and domain wall logic are more revolutionary approaches to spin-based electronics which address complete architectures as opposed to mere discrete devices (like

SPINFETs) or logic gates (like MTJ). In SSL, the logic device is essentially a single electron and in MQCA it is a single magnetic nanograin. Therefore, understandably, they dissipate very little energy when they switch between logic states. Consequently, these paradigms are extremely energy-efficient and can reduce energy dissipation per bit flip by a factor of 1000 -10,000 compared to MISFETs. Their distinguishing feature is that the two logic levels are very well separated (upspin and downspin in SSL, or magnetizations parallel and anti-parallel to the easy axis of a magnetic grain in MQCA), but that does not necessarily translate to very low bit error rates. Only in SSL, the bit error rate can be made low (bit error probability of $10^{-9}$), but it requires cryogenic temperatures. In MQCA, the misalignment problem precludes a small bit error rate. In domain wall logic, the bit error rate can be significantly smaller than in MQCA, even at room temperature, but still may not be small enough. Thus, none of these paradigms are very fault-tolerant and reliable [21]. Both MQCA and domain wall logic gates have been demonstrated while SSL remains to be demonstrated.

The Spin Accumulation Logic (SAL) paradigm has standby power dissipation and in the end is considerably more power hungry than even traditional charge based electronics. Furthermore, the separation between logic levels is poor which will lead to unacceptable bit error rates. To our knowledge, there has been no experimental demonstration of SAL. At this time, this paradigm does not seem to be particularly adaptable to digital logic, but future improvements may some day make it more competitive.

---

[21] Domain wall motion however has formed the basis of a now famous memory architecture known as "race-track memory" which is both non-volatile and relatively fast [see, for example, S. S. P. Parkin, M. Hayashi and L. Thomas, Science, 320, 190 (2008) and S. S. P. Parkin, Scientific American, 300, 76 (2009)]. Since we have constrained ourselves to logic architectures only, and avoided any discussion of memory, we do not discuss it here.

The Spin Wave Bus (SWB) is a phase based approach that appears to be unsuitable for digital electronics since phase is not naturally bistable and, in fact, is extremely delicate. However, it may have applications in analog signal processing. It too is an energy-efficient paradigm like SSL, MQCA and domain wall logic. Preliminary demonstration of a SWB detector has been reported.

In the end, only time will tell if any of these ideas ultimately makes any inroads into commercial electronics and begins to displace the celebrated MISFET. The latter has 60 years of research and investment behind it, and displacing it from mainstream electronics is always a very tall order.

**Acknowledgement:** S. B. is indebted to Prof. Kang L. Wang and Dr. Alexander Khitun for illuminating discussion regarding the SWB technology. His work was supported by the National Science Foundation under grant CCF 0726373.

**References**

1. The International Technology Roadmap for Semiconductors, available at http://www.itrs.net.

2. Suman Datta, private communication (2006).

3. D. B. Tuckerman and R. F. W. Pease. IEEE Trans. Elec. Dev., $\underline{28}$, 1230 (1981).

4. G. E. Moore, Electronics Magazine (McGraw Hill, New York, 1965).

5. S. Datta and B. Das, Appl. Phys. Lett., $\underline{56}$, 665 (1990).

6. R. Landauer, Philos. Trans. Royal Soc. London, Ser. A, $\underline{353}$, 367 (1995).

7. S. Subramaniam, S. Bandyopadhyay and W. Porod, J. Appl. Phys., $\underline{68}$, 4861 (1990).

8. M. I. D'yakonov and V. I. Perel', Sov. Phys. Solid State, $\underline{13}$, 3023 (1972).

9. R. J. Elliott, Phys. Rev., $\underline{96}$, 266 (1954).

10. A. G. Mal'shukov abnd K. A. Chao, Phys. Rev. B, $\underline{61}$, R2413 (2000).

11. A. W. Holleitner, V. Shih, R. C. Myers, A. C. Gossard and D. D. Awschalom, New J. Phys., $\underline{9}$, 342 (2007).

12. S. Pramanik, S. Bandyopadhyay and M. Cahay, IEEE Trans. Nanotech., $\underline{4}$, 2 (2005).

13. M. Cahay and S. Bandyopadhyay, Phys. Rev. B., $\underline{69}$, 045303 (2004).

14. M. Cahay and S. Bandyopadhyay, Phys. Rev. B., $\underline{68}$, 115316 (2003).

15. Yu. A. Bychkov and E. I. Rashba, J. Phys. C., $\underline{17}$, 6039 (1984).

16. S. Bandyopadhyay and M. Cahay, *Introduction to Spintronics* (CRC Press, Boca Raton, 2008).

17. S. Datta, *Electronic Transport in Mesoscopic Systems* (Cambridge University Press, Cambridge, 1995).

18. A. Trivedi, S. Bandyopadhyay and M. Cahay, IET Circuits, Devices and Syst., 1, 395 (2007).

19. G. Salis, R. Wang, X. Jiang, R. M. Shelby, S. S. P. Parkin, S. R. Bank and J. S. Harris, Appl. Phys. Lett., 87, 262503 (2005).

20. S. Bandyopadhyay and M. Cahay, Appl. Phys. Lett., 85, 1814 (2004).

21. J. Schliemann, J. C. Egues and D. Loss, Phys. Rev. Lett., 90, 146801 (2003).

22. X. Cartoixá, D. Z. Y. Tang and Y-C Chang, Appl. Phys. Lett., 83, 1462 (2003).

23. K. C. Hall and M. E. Flatté, Appl. Phys. Lett., 88, 162503 (2006).

24. S. Bandyopadhyay and M. Cahay, Appl. Phys. Lett., 85, 1433 (2004).

25. A. S. Sedra and K. C. Smith, *Microelectronic Circuits*, 5$^{th}$ edition (Oxford University Press, New York, 2004).

26. A. M. Bratkovsky, Rep. Prog. Phys., 71, 026502 (2008).

27. J. Nitta, T. Akazaki, H. Takayanagi, and T. Enoki, Phys. Rev. Lett., 78, 1335 (1997).

28. J. H. Kwon, H. C. Koo, J. Cmang, S. H. Han and J. Eom, J. Korean Phys. Soc., Part 1, 53, 2491 (2008).

29. S. Bandyopadhyay and M. Cahay, Physica E, 25, 399 (2005).

30. I. Appelbaum and D. J. Monsma, Appl. Phys. Lett., 90, 262501 (2007).

31. D. J. Monsma, J. C. Lodder, Th. J. A. Popma and B. Dieny, Phys. Rev. Lett., 74, 5260 (1995).

32. D. J. Monsma, L. Vlutters and J. C. Lodder, Science, 281, 407 (1998).

33. B. Huang, D. J. Monsma and I. Appelbaum, Appl. Phys. Lett., 91, 072501 (2007).

34. I. Appelbaum, B, Huang and D. J. Monsma, Nature (London), 447, 295 (2007).

35. J. Fabian, I. Žutić and S. Das Sarma, Appl. Phys. Lett., 84, 85 (2004).

36. M. E. Flatte, Z. G. Yu, E. Johnston-Halperin and D. D. Awschalom, Appl. Phys. Lett., 82, 4740 (2003).

37. M. E. Flatté and G. Vignale, Appl. Phys. Lett., 78, 1273 (2001).

38. S. Bandyopadhyay and M. Cahay, Appl. Phys. Lett., 86, 133502 (2005).

39. M. Johnson, Science, 260, 320 (1993).

40. M. Johnson, IEEE Spectrum, 31, 47 (1994).

41. K. Mizushima, T. Kinno, T. Yamauchi and K. Tanaka, IEEE Trans. Magn., 33, 3500 (1997).

42. P. LeMinh, H. Gokcan, J. C. Lodder and R. Jansen, J. Appl. Phys., 98, 076111 (2005).

43. R. Jansen, H. Gokcan, O. M. J. van't Erve, F. M. Postma and J. C. Lodder, J. Appl. Phys., 95, 6927 (2004).

44. N. P. Vasil'eva and S. I. Kastakin, Russ. Microelectron., 26, 406 (1997).

45. M. M. Hassoun, W. C. Black, Jr., E. K. F. Lee and R. L. Geiger, IEEE Trans. Magn., 33, 3307 (1997).

46. J. Shen, IEEE Trans. Magn., 33, 4492 (1997).

47. A. T. Hanbicki, R. Magno, S-F Cheng, Y. D. Park, A. S. Bracker and B. T. Jonker, Appl. Phys. Lett., 79, 1190 (2001).

48. R. Richter, L. Bär and J. Wecker, Appl. Phys. Lett., 80, 1291 (2002).

49. A. Ney, C. Pampuch, R. Koch and K. H. Ploog, Nature (London), 425, 485 (2003).

50. S. Ikeda, J. Hayakawa, Y. Ashizawa, Y. M. Lee, K. Miura, H. Hasegawa, M. Tsunoda, F. Matsukura and H. Ohno, Appl. Phys. Lett., 93, 082508 (2008).

51. Sin-itoro Tomonoga, *The Story of Spin*, Translated by Takeshi Oka (The University of Chicago Press, Chicago, 1997).

52. B. Kanchibotla, S. Pramanik, S. Bandyopadhyay and M. Cahay, Phys. Rev. B., <u>78</u>, 193306 (2008).

53. V. V. Zhirnov, R. K. Cavin, J. A. Hutchby and G. I. Bourianoff, Proc. IEEE, <u>91</u>, 1934 (2003); K. Galatsis, A. Khitun, R. Ostroumov, K. L. Wang, W. R. Dichtel, E. Plummer, J. F. Stoddart, J. I. Zink, J. Y. Lee, Y. H. Xie and K. W. Kim, IEEE Trans. Nanotech., <u>8</u>, 66 (2009).

54. R. K. Cavin, V. V. Zhirnov, J. A. Hutchby and G. I. Bourianoff, Fluctuations and Noise Letters, <u>5</u>, C29 (2005).

55. D. E. Nikonov, G. I. Bourianoff and P. Gargini, J. Supercond. Novel Magnetism, <u>19</u>, 497 (2006).

56. J. J. Welser, G. I. Bourianoff, V. V. Zhirnov and R. K. Cavin, J. Nanopart. Res., <u>10</u>, 1 (2008).

57. C. S. Lent, M. Liu and Y. Lu, Nanotechnology, <u>17</u>, 4240 (2006). See also the comment on this paper by V. V. Zhirnov and R. K. Cavin, Nanotechnology, <u>18</u>, 298001 (2007).

58. S. Bandyopadhyay, B. Das and A. E. Miller, Nanotechnology, <u>5</u>, 113 (1994).

59. S. N. Molotkov and S. S. Nazin, JETP Lett., <u>62</u>, 256 (1995).

60. D. Rugar, R. Budakian, H. J. Mamin and B. H. Chui, Nature (London), <u>430</u>, 329 (2004).

61. M. Xioa, I. Martin, E. Yablonovitch and H. W. Jiang, Nature (London), <u>430</u>, 435 (2004).

62. J. M. Elzerman, et al., Nature (London), <u>430</u>, 431 (2004).

63. H. Agarwal, S. Pramanik and S. Bandyopadhyay, New J. Phys., <u>10</u>, 015001 (2008).

64. S. Bandyopadhyay and V. P. Roychowdhury, Jpn. J. Appl. Phys. Pt. 1, <u>35</u>, 3350 (1996).

65. D. K. Schroder, *Advanced MOS Devices*, Modular Series on Solid State Devices, Eds. G. W. Neudeck and R. F. Pierret (Addison-Wesley, Reading, MA, 1987).

66. S. Bandyopadhyay, Superlat. Microstruct., <u>37</u>, 77 (2005).

67. R. Landauer, IBM J. Res. Develop., <u>5</u>, 183 (1961).

68. R. Landauer and R. W. Keyes, IBM J. Res. Develop., <u>14</u>, 152 (1970).

69. L. B. Kish, Phys. Lett. A, <u>305</u>, 144 (2002).

70. D. V. Melnikov and J-P Leburton, Phys. Rev. B, <u>73</u>, 155301 (2006)

71. X. W. Zhang, W. J. Fan, S. S. Li and J. B. Xia, Appl. Phys. Lett., <u>90</u>, 193111 (2007).

72. S. Bandyopadhyay and V. P. Roychowdhury, Superlat. Microstruct., <u>22</u>, 411 (1997).

73. R. de Sousa and S. Das Sarma, Phys. Rev. B., <u>67</u>, 033301 (2003). X. Hu, R. de Sousa and S. Das Sarma, *Foundations of Quantum Mechanics in the Light of New Technology*, Eds. Y. A. Ono and K. Fujikawa (World Scientific, Singapore, 2003).

74. S. Amasha, K. MacLean, I. P. Radu, D. M. Zumbühl, M. A. Kastner, M. P. Hanson and A. C. Gossard, Phys. Rev. Lett., <u>100</u>, 046803 (2008).

75. S. Pramanik, C-G, Stefanita, S. Patibandla, S. Bandyopadhyay, K. Garre, N. Harth and M. Cahay, Nature Nanoetch., <u>2</u>, 216 (2007).

76. W. L. Wang, O. V. Yazyev, S. Meng and E. Kaxiras, Phys. Rev. Lett., 102, 157201 (2009).

77. B. Meurer, D. Heitmann and K. Ploog, Phys. Rev. Lett., 68, 1371 (1992).

78. M. Ciorga, A. S. Sachrajda, P. Hawrylak, C. Gould, P. Zawadzki, S. Jullian, Y. Feng and Z. Wasilewski, Phys. Rev. B., 61, R16315 (2000).

79. M. Piero-Ladriere, M. Ciorga, J. Lapointe, P. Zawadzki, M. Korukusisnki, P. Hawrylak and A. S. Sachrajda, Phys. Rev. Lett., <u>91</u>, 026803 (2003).

80. C. Livermore, C. H. Crouch, R. M. Westerveldt, K.L. Campman and A. C. Gossard, Science, <u>274</u>, 1332 (1996); A. W. Holleitner, R. H. Blick, A. K. Huttel, K. Eberl and J. P. Kotthaus, Science, <u>297</u>, 70 (2001).

81. T. H. Oosterkamp, T. Fujisawa, W. G. van der Wiel, K. Ishibashi, R. V. Hijman, S. Tarucha and L. P. Kouwenhoven, Nature(London), <u>395</u>, 873 (1998).

82. N. J. Craig, J. M. Taylor, E. A. Lester, C. M. Marcus, M. P. Hanson and A. C. Gossard, Science, <u>304</u>, 565 (2004).

83. R. Hanson, B. Witkamp, L. M. K. Vandersypen, L. H. W. vanBeveren, J. M. Elzerman and L. P. Kouwenhoven, Phys. Rev. Lett., <u>91</u>, 196802 (2003).

84. J. R. Petta, A. C. Johnson, J. M. Taylor, E. A. laird, A. Yacoby, M. D. Lukin, C. M. Marcus, M. P. Hanson and A. C. Gossard, Science, <u>309</u>, 2180 (2005).

85. K. C. Nowack, F. H. L. Koppens, Yu. V. Nazarov and L. M. K. Vandersypen, Science, <u>318</u>, 5855 (2007).

86. J. Berezovsky, M. H. Mikkelsen, N. G. Stoltz, L. A. Coldren and D. D. Awschalom, Science, <u>320</u>, 5874 (2008).

87. B. Doris, et al., Technical Digest of the IEEE Electron Device Meeting, San Fracisco, USA (2002).

88. G. S. Huang, X. L. Hu, Y. Xie, F. Kong, Z. Y. Zhang, G. G. Siu and P. K. Chu, Appl. Phys. Lett., <u>87</u>, 151910 (2005).

89. R. P. Cowburn, D. K. Koltsov, A. O. Adeyeye, M. E. Welland and D. M. Tricker, Phys. Rev. Lett., <u>83</u>, 1042 (1999).

90. R. P. Cowburn and M. E. Welland, Science, <u>287</u>, 1466 (2000).

91. R. Street and J. C. Woolley, Proc. Phys. Soc., Sec. A, 62, 562 (1949); W. F. Brown, Phys. Rev., 130, 1677 (1963).

92. P. Gaunt, J. Appl. Phys., 48, 3470 (1977).

93. A. Perumal. H. S. Ko and S. C. Shin, Appl. Phys. Lett., 83, 3326 (2003).

94. S. Salahuddin and S. Datta, Appl. Phys. Lett., 90, 093503 (2007).

95. S. Wolfram, *Cellular Automata and Complexity: Collected Papers* (Westview Press, © Stephen Wolfram, 1994).

96. D. A. Allwood, G. Xiong, C. C. Faulkner, D. Atkinson, D. Petit and R. P. Cowburn, Science, 309, 1688 (2005).

97. G. Csaba, A. Imre, G. H. Bernstein, W. Porod and V. Metlushko, IEEE Trans. Nanotech., 1, 209 (2002).

98. C. H. Bennet, Int. J. Theor. Phys., 21, 905 (1982).

99. B. Behin-Aein, S. Salahuddin and S. Datta, IEEE Trans. Nanotech., 8, 505 (2009).

100. W. J. Carr, J. Appl. Phys., 45, 394 (1974); L. Berger, J. Phys. Chem. Solids, 35, 947 (1974); P. P. Freitas and L. Berger, J. Appl. Phys., 57, 1266 (1985); L. Berger, Phys. Rev. B, 54, 9353 (1996).

101. J. C. Slonczewski, J. Magn. Mater., 159, L1 (1996).

102. E. B. Myers, D. C. Ralph, J. A. Katine, R. N. Louie and R. A. Buhrman, Science, 285, 867 (1999).

103. R. P. Cowburn, Nature Mat., 6, 255 (2007).

104. K. Yamada, S. Kasai, Y. Nakatani, K. Kobayashi, H. Kohno, A. Thiaville and T. Ono, Nature Mat. 6, 269 (2007).

105. A. Imre, G. Csaba, L. Ji, A. Orlov, G. H. Bernstein and W. Porod, Science, 311, 205 (2006).

106. G. A. Prinz, Science, 282, 1660 (1998).

107. C. Chappert, A. Fert and F. N. van Dau, Nature Mat., 6, 813 (2007).

108. S. Mao, J. Nanosci. Nanotech., 7, 1 (2007).

109. D. A. Allwood, G. Xiong, C. C. Faulkner, D. Atkinson, D. Petit and R. P. Cowburn, Science, 309, 1688 (2005).

110. D. A. Allwood, G. Xiong and R. P. Cowburn, J. Appl. Phys., 100, 123908 (2006).

111. D. A. Allwood, G. Xiong and R. P. Cowburn, Appl. Phys. Lett., 89, 102504 (2006).

112. H. Dery, P. Dalal, L. Cywinski and L. J. Sham, Nature (London), 447, 573 (2007).

113. A. Khitun and K. L. Wang, Superlat. Microstruct., 38, 184 (2005)

114. A. Khitun, D. E. Nikonov, M. Bao. K. Galatsis and K. L. Wang, Nanotechnology, 18, 465202 (2007); A. Khitun, D. E. Nikonov, M. Bao. K. Galatsis and K. L. Wang, IEEE Trans. Elec. Dev., 54, 3418 (2007).

115. T. Schneider, A. A. Serga, B. Leven, B. Hillebrands, R. L. Stamps and M. P. Kostylev, Appl. Phys. Lett., 92, 022505 (2008).

116. A. Khitun, M. Bao, Y. Wu, J-Y Kim, A. Hong, A. P. Jacob, K. Galatsis and K. L. Wang, Mater. Res. Soc. Symp. Proc., 1067, B01-04 (2008).

117. N. W. Ashcroft and N. D. Mermin, *Solid State Physics* (Saunders College, Philadelphia, 1976).

118. D. A. Hodges and H. G. Jackson, *Analysis and Design of Digital Integrated Circuits*, 2nd edition, (McGraw Hill, New York, 1988), Chapter 1, p. 2.

**Figure Captions**

Fig. 1: Structure of the Datta-Das Spin Field Effect Transistor (SPINFET).

Fig. 2: Ideal transfer characteristic of the Datta-Das SPINEFT with a one-dimensional channel, no spin relaxation in the channel, no spin-mixing in the channel, and 100% spin injection and detection efficiencies.

Fig. 3: Plot of maximum conductance on-off ratio versus spin injection or detection efficiency for a SPINFET that relies on spin injection and detection at a ferromagnet/semiconductor interface. Reproduced from ref. [18] with permission from the Institution of Engineering and Technology. © IET.

Fig. 4: Schematic diagram of the Transit Time Spin Field Effect Transistor adapted from ref. [30].

Fig. 5: The conduction energy band profile of a heterostructured $n^+$-$p^+$-n spin bipolar junction transistor biased in the active region of operation. The base is ferromagnetic and the spin splitting energy in the base $2\Delta$. We assumed that the emitter band gap is widest and the base band gap is narrowest. The type-I alignment of the conduction band edge $E_c$ and the valence band edge $E_v$ at the emitter-base junction allows doping both emitter and base heavily, which makes the base access resistance small and increases the transistor's switching speed, without sacrificing the emitter injection efficiency.

Fig. 6: (a) A magnetic tunnel junction logic gate adapted from ref. [49]; (b) the INITIALIZATION STEP; and (c) the SET step.

Fig. 7: (a) Realization of an SSL NAND gate. A linear array of three quantum dots, each containing a single electron, is placed in a global magnetic field. There is exchange coupling between nearest neighbors. The spin of any electron can be either parallel or anti-parallel to the global magnetic field. The two peripheral dots are input ports and the central dot is the output port. We assume that spin parallel to the global magnetic field always represents logic bit 1 and spin anti-parallel represents logic bit 0. If we orient the input spins in desired directions for various combinations of the two inputs (0,0), (0,1), (1,0) and (1,1), the output bit is always the NAND function of the input bits as shown. This happens provided two conditions are fulfilled: (i) the exchange interaction strength $J$ is larger than one-fourth of the Zeeman splitting energy in any dot due to the global magnetic field, and (ii) the Zeeman splitting in the input dots due to the locally applied magnetic field that aligns the spins along the desired directions is much larger than $J$. (b) A "spin wire" that is split to provide fan out.

Fig. 8: Unidirectional transmission of a "spin-bit" along a spin wire from left to right by sequentially raising the potentials of gates pairwise using a 3-phase clock. Input data can be pipelined.

Fig. 9: An MQCA realization of an AND gate with nanomagnets. A magnetic field pulse whose direction is shown by the thick arrow is one of the two inputs to the gate. Its two possible directions, right and left, encode binary bits 1 and 0, respectively. An oscillating square wave

magnetic field with a large negative dc component acts as the other input. The positive cycle encodes logic bit 1 and the negative cycle logic bit 0. The magnetization direction of the grains encodes the output bit, with right oriented magnetization representing logic bit 1 and left oriented magnetization representing logic bit 0. This scheme is adapted from ref. [90].

Fig. 10: An alternate implementation of an MQCA AND gate, where the magnetic field pulse in Fig. 9 is replaced by the magnetization orientation of an elongated grain to serve as one of the binary inputs to the gate. This scheme is adapted from ref. [96].

Fig. 11: (Top panel) The two logic bits encoded in up and down magnetizations of anisotropic nanomagnets. (Bottom panel) An attempted implementation of a NAND gate using nanomagnets with shape anisotropy. The easy axis of magnetization is along the length which makes the magnetization bistable. The two possible orientations represent logic bits 0 and 1. There is nearest neighbor dipole-dipole interaction that renders the ground state anti-ferromagnetic. The NAND realization fails since when the two inputs are logic complements of each other, the output bit is indeterminate since influences from right and left are equally strong and no agent exists to break the tie.

Fig. 12: Unidirectional signal propagation along an MQCA chain using a "granular" Bennet clock. (a) Ground state of a 4-nanomagnet array showing anti-ferromagnetic ordering. Each nanomagnet is magnetized along the easy axis in the ground state; (b) the first nanomagnet is flipped (input to the chain altered) and the second nanomagnet is stuck in an indeterminate state since influences from the left and the right oppose and are equally strong; (c) the third

nanomagnet is magnetized along the hard axis by a *local* magnetic field pulse which acts as a local clock, (d) the second nanomagnet flips to the lowest energy state because the influences from left and right are unequal. Signal has propagated through the second nanomagnet; (e) the clock pulse applied to the third nanomagnet subsides and the fourth nanomagnet is clocked to place its magnetization along the hard axis, whereupon the third cell's magnetization flips to assume anti-ferromagnetic ordering. Signal has now propagated through the first three cells. By sequentially clocking the cells, the signal can be made to propagate unidirectionally from left to right in a domino-like fashion. Note that the input to the first cell can be changed after the signal has propagated through the first three cells, without affecting the propagation of the original input bit down the line. In other words, the architecture is *pipelined*.

Fig. 13: The global clocking scheme of ref. [97]. (a) A global horizontal magnetic field is first applied as a clock to reset the magnetization of every cell along the hard axis, (b) the input bit is written in the leftmost cell; (c) The second cell finds itself in an asymmetric environment with its left neighbor magnetized along the easy axis and right neighbor along the hard axis. The influence from the left is stronger, so that the second cell flips along the easy axis to assume anti-ferromagnetic ordering; (d) the third cell flips to preserve anti-ferromagnetism; (e) then the fourth cell flips to maintain anti-ferromagnetic ordering and so on. Thus, signal propagates unidirectionally from left to right in a domino-like fashion and the input bit is ultimately replicated in every odd-numbered cell. The next global clock pulse cannot be applied until all the cells have flipped to assume the anti-ferromagnetic ordering, i.e. until the input bit has propagated down the entire length of the chain. Otherwise, the already ordered cells will be disrupted by the clock. Therefore, the clock period is *M* times the flipping time of each cell,

where *M* is the number of cells in the line. There is no point in changing the input before the old input has propagated down the entire chain, since the new input will not propagate before the application of the next clock pulse. Therefore, we cannot alter the input before the output has been produced, i.e. the architecture with a global clock is *not* pipelined.

Fig. 14: The "misalignment problem". Different nanomagnets are oriented somewhat differently during fabrication. (a) The clock field is not exactly aligned with the hard axis of the magnet, but subtends an angle $\theta$ with it. Here, we show the case of a granular clock where the magnetic pulse can be oriented in slightly different directions in each nanomagnet to reduce $\theta$ since each nanomagnet has its own private clock. The clock field's direction is designated by an arrow. (b) The problem is much more severe with a global clock since all nanomagnets have a common clock. Here, some nanomagnets can be misaligned by a large angle $\theta$ and no corrective action is possible.

Fig 15: Domain wall logic. (a) A cusp acts as a NOT gate, and (b) a fork serves to provide fan out. Adapted from ref. [109].

Fig. 16: Depiction of a Spin Accumulation Logic (SAL) NAND gate. The magnetization directions in ferromagnetic layers 1 and 5 are the two inputs and the transient current measured by the meter M connected to the capacitor is the output bit. Adapted from ref. [112].

Fig. 17: Schematic depiction of a spin wave in a ferromagnet.

Fig. 18: A spin wave bus (SWB) gate. Reproduced from ref. [114] with permission from Institute of Physics.

**Table I: Comparison between SPINFET, TTSFET and MISFET for binary logic switch application**

|  | SPINFET[22] | TTSFET | MISFET |
|---|---|---|---|
| Conductance ON/OFF ratio | Low due to inefficient spin injection and detection. | Low due to inefficient ballistic spin filtering | High (higher by ~ 4 orders of magnitude) |
| Energy efficiency | Worse than MISFET | Worse than MISFET | - |
| Bit error probability | High | High | Low (because of higher conductance ON/OFF ratio) |
| Speed | Same | Slightly higher | Same |
| Device density | Same | Slightly lower | Same |
| Cost | Same | Slightly higher | Same |

---

[22] The SPINFET includes the Datta-Das device and all its clones mentioned in Section 2.1.2. The clones do not improve on the original Datta-Das device in any significant way, and therefore their performances are not any better.

**Table II: Input-output relations (truth table) of an AND gate**

| Input 1 | Input 2 | Output |
|---------|---------|--------|
| 1 | 1 | 1 |
| 1 | 0 | 0 |
| 0 | 1 | 0 |
| 0 | 0 | 0 |

**Input-output relations (truth table) of a NAND gate**

| Input 1 | Input 2 | Output |
|---------|---------|--------|
| 1 | 1 | 0 |
| 1 | 0 | 1 |
| 0 | 1 | 1 |
| 0 | 0 | 1 |

**Table III: Representative performance figures for SSL**

| | |
|---|---|
| Bit error probability | $10^{-9}$ |
| Max. clock frequency | 1 GHz |
| Temperature of operation | 1 K |
| Energy dissipated per bit flip | $kT\ln(1/p) = 2J = g\mu_B B = 2$ meV at 1 K |
| Total power dissipated per unit area with a 1 GHz clock and bit density of $10^{11}$ cm$^{-2}$ at 1 K | 0.18 Watts/cm$^2$ (250 times smaller than in Pentium IV with a 1000 times larger bit density and 3 times lower clock speed) |

**Table IV: Representative performance figures for MQCA**

| | |
|---|---|
| Bit error probability | $2\times10^{-5}$ ideally, but much larger because of the misalignment problem |
| Max. clock speed | ~ 1 GHz (granular clock). < 10 kHz (global clock) assuming a chip area of 1 cm$^2$ and a bit density of $10^{10}$ cm$^{-2}$. Larger chip area or higher bit density would make the clock slower |
| Temperature of operation | 300 K |
| Energy dissipated in switching a bit | 0.8 eV at 300 K, 4 meV at 1 K |
| Total power dissipated per unit area with a 1 GHz granular clock and bit density of $10^{10}$ cm$^{-2}$ at 300 K | ~10 – 20 Watts/cm$^2$ (2-5 times smaller than in Pentium IV with a 100 times larger bit density and 3 times lower clock speed) |

**Table V: Representative performance figures for SWB**

| | |
|---|---|
| Bit error probability | $10^{-5}$ - $10^{-4}$ due to dispersion alone |
| Max. clock speed | 1 – 10 GHz |
| Temperature of operation | 300 K |
| Energy dissipated in switching a bit | 5-6 eV at 300 K with 10 GHz clock |
| Total power dissipated per unit area with a 10 GHz clock and bit density of $10^{10}$ cm$^{-2}$ at 300 K | Assuming that energy dissipated launching, detecting and clocking is the same as in bit flip: $3.2 – 3.8$ kW/cm$^2$ (70 times higher than Pentium IV with a 4 times higher clock frequency and 100 times higher bit density) |

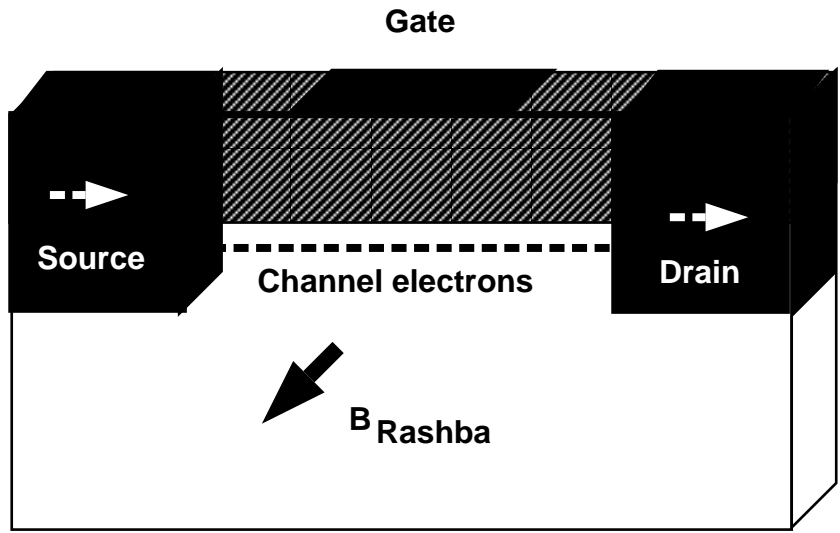

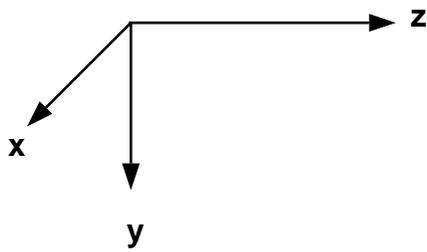

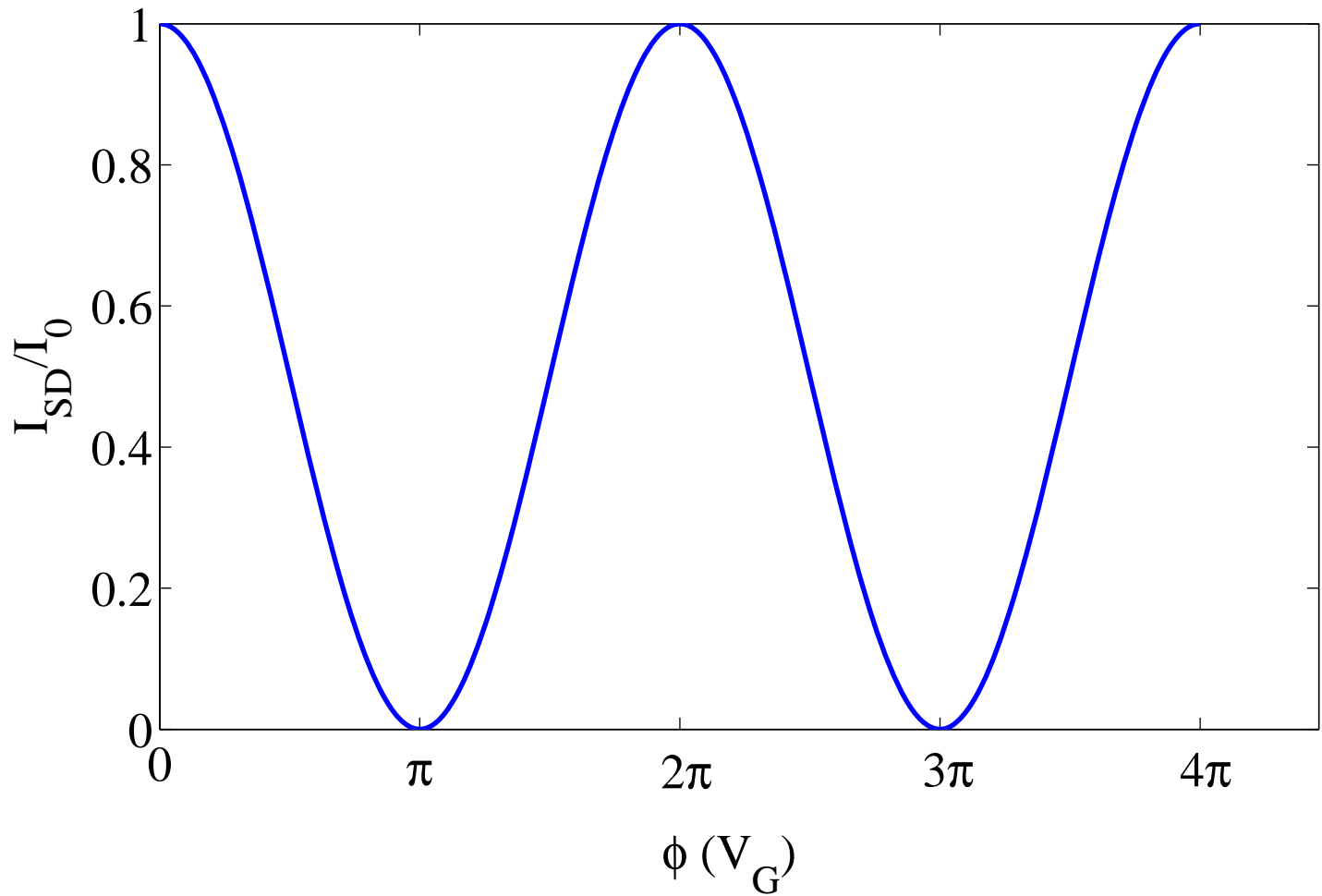

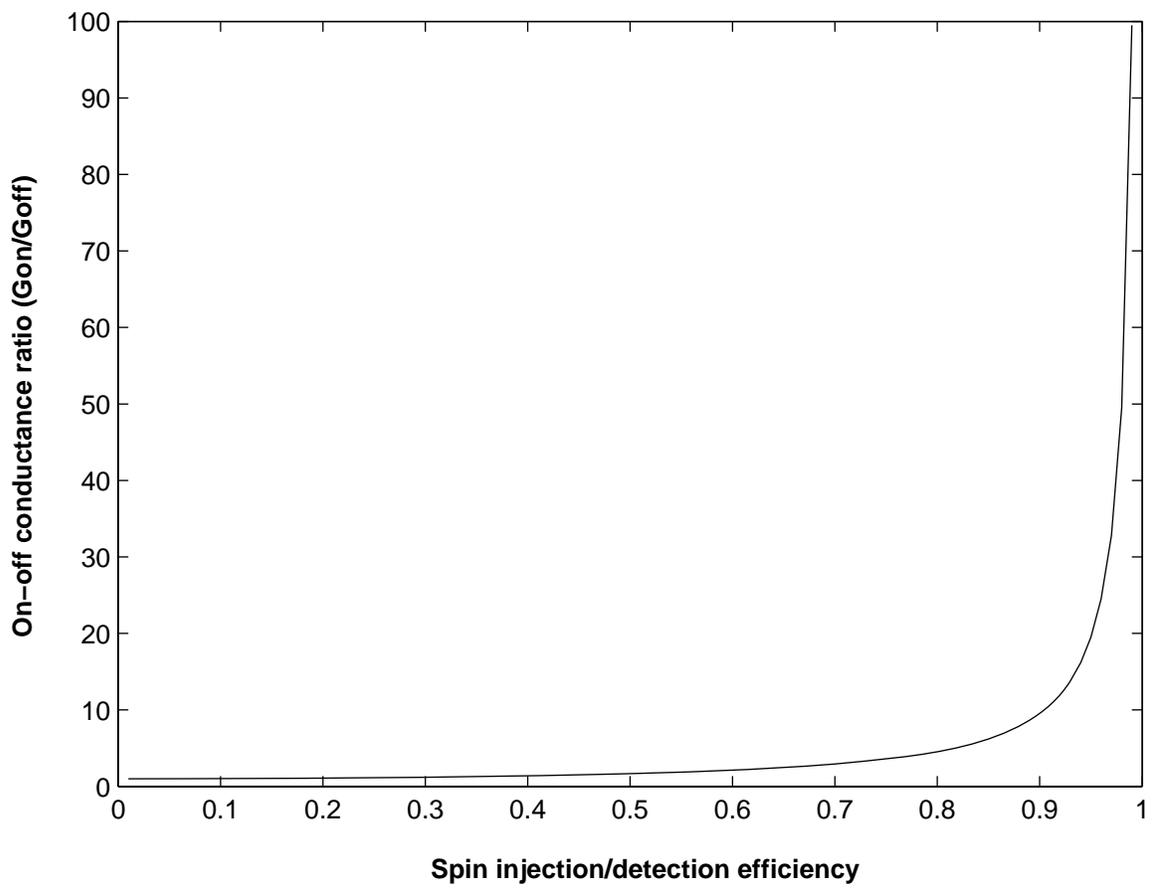

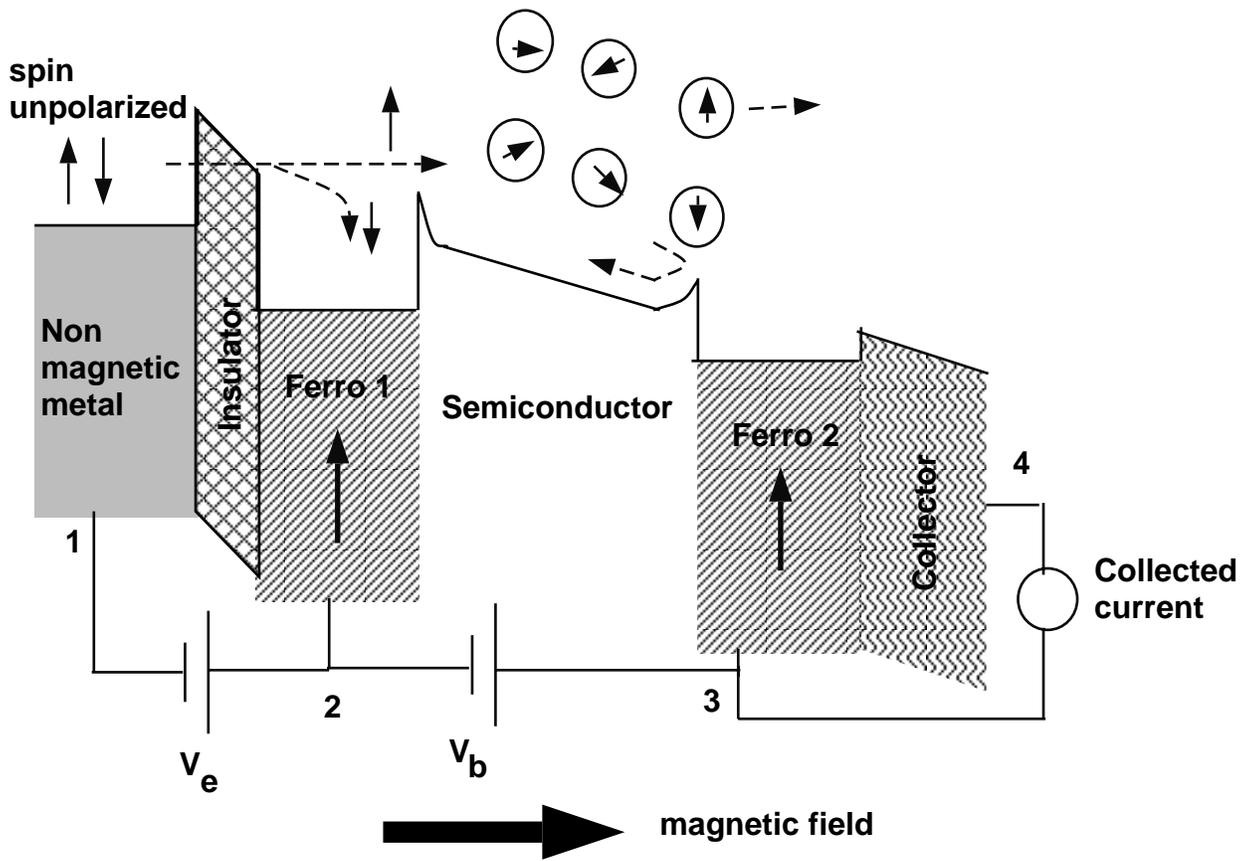

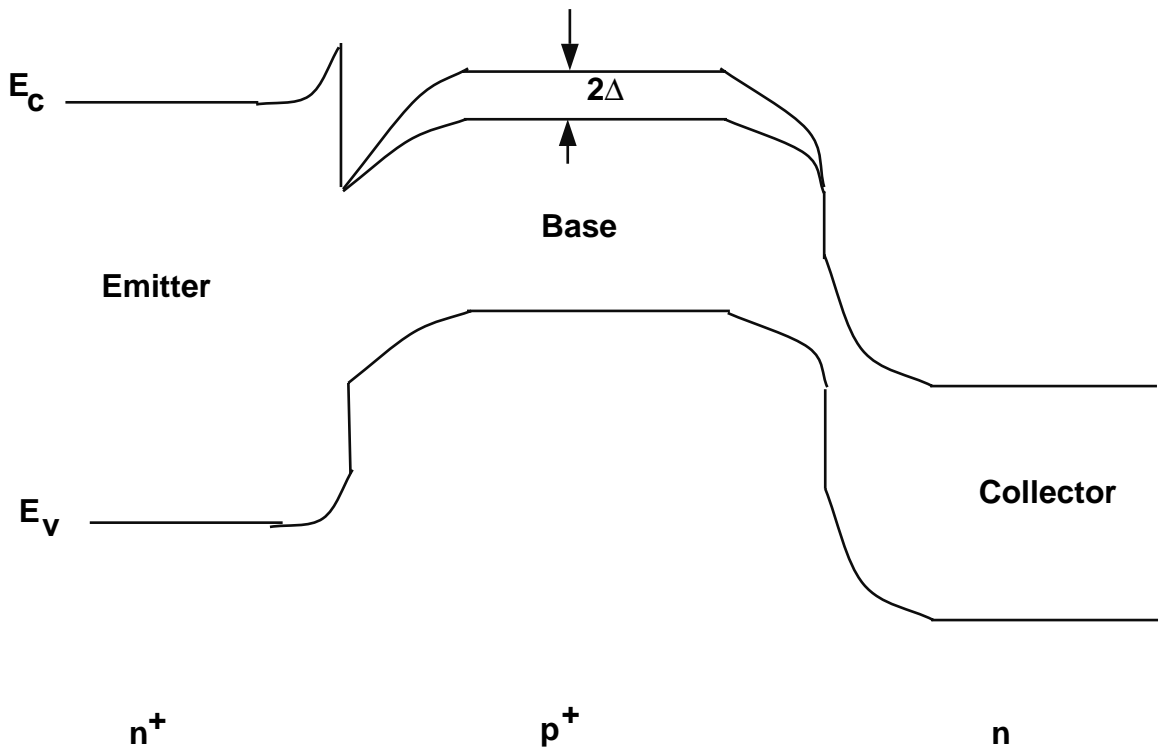

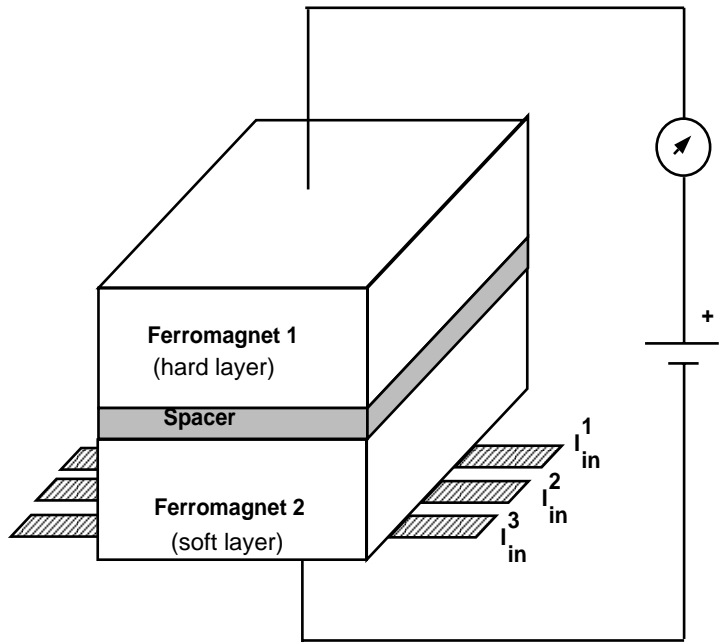

(a)

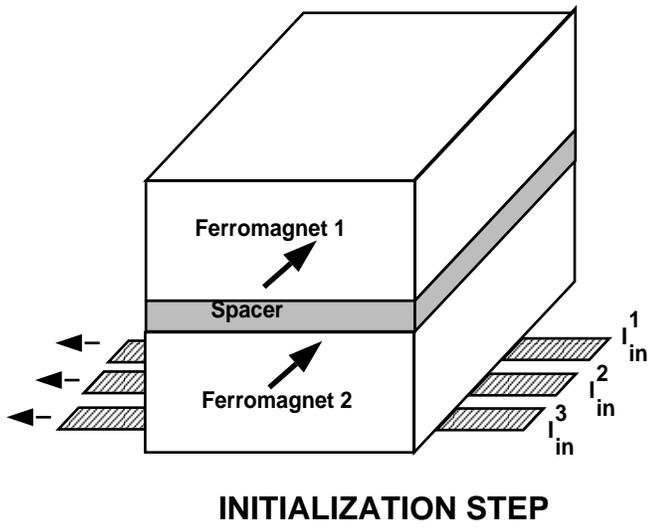

**INITIALIZATION STEP**

(b)

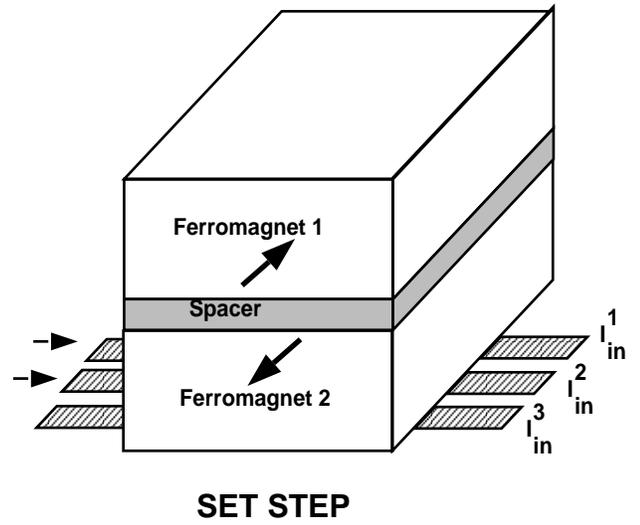

**SET STEP**

(c)

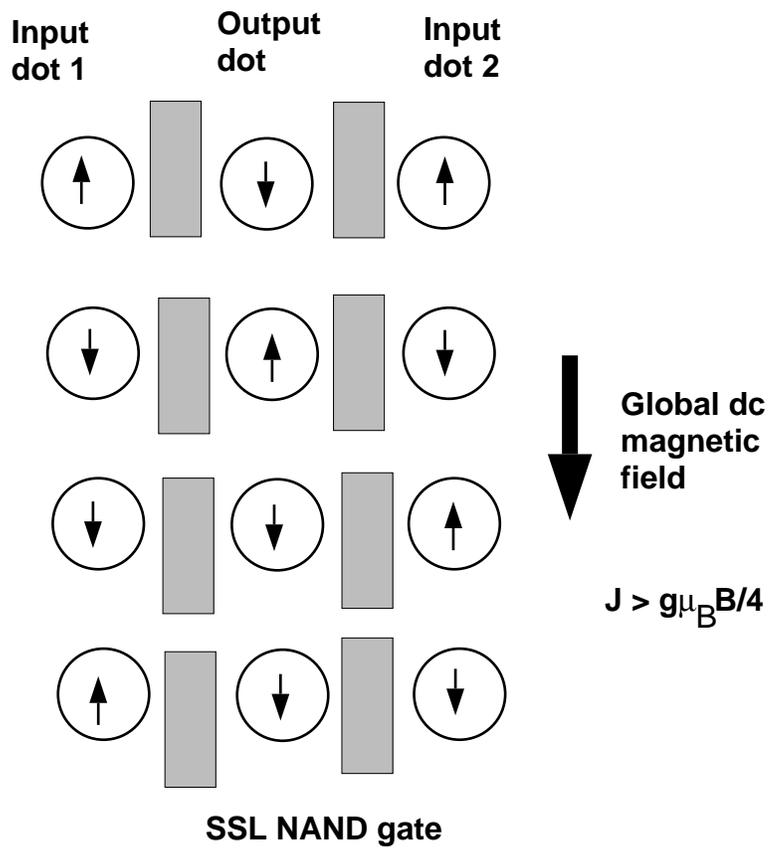

**(a)**

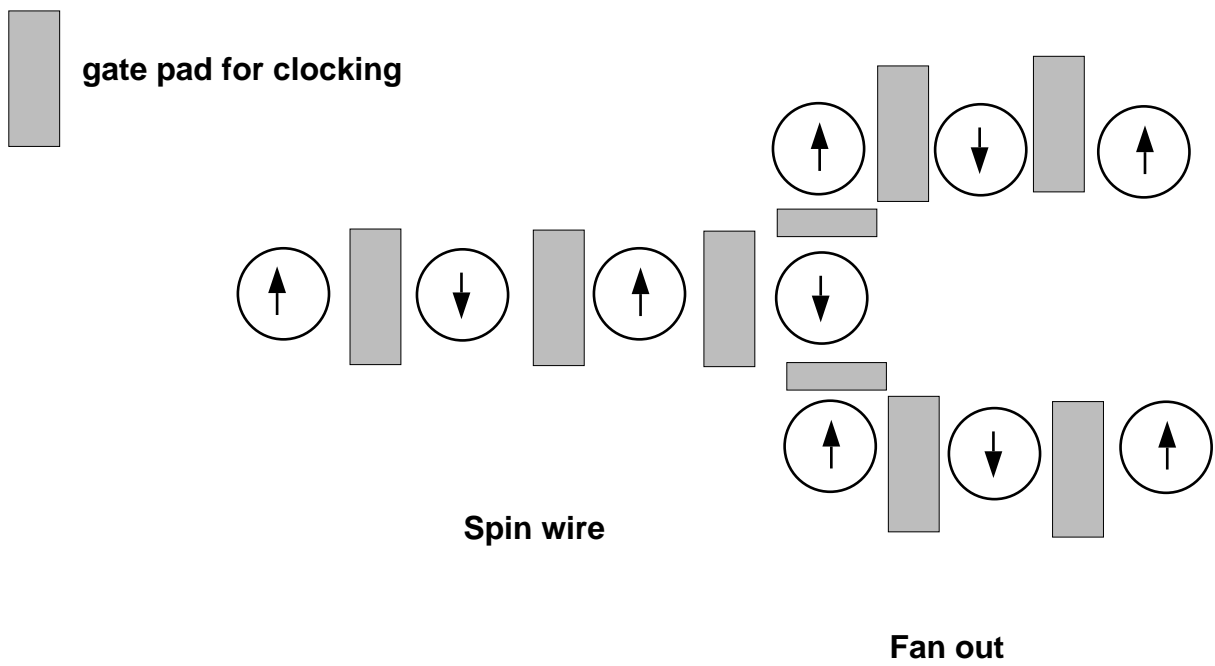

**(b)**

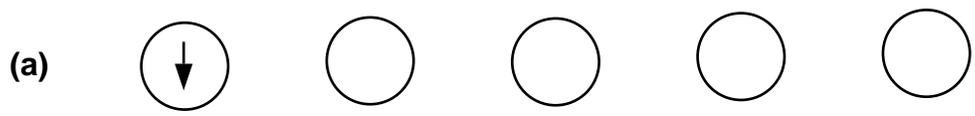
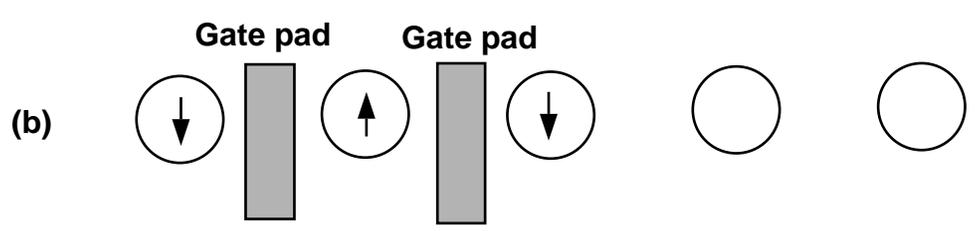
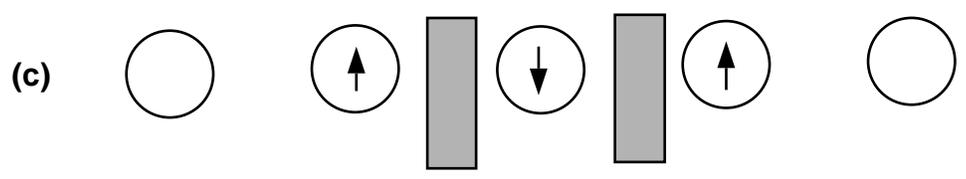
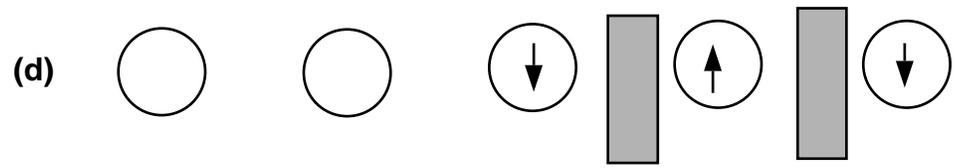

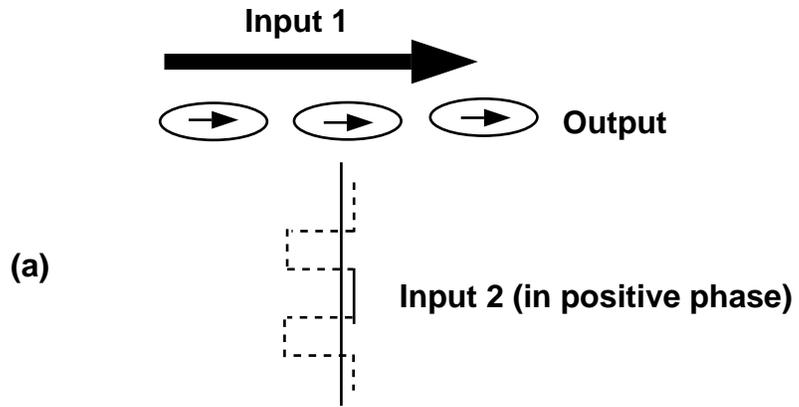

**(a)**

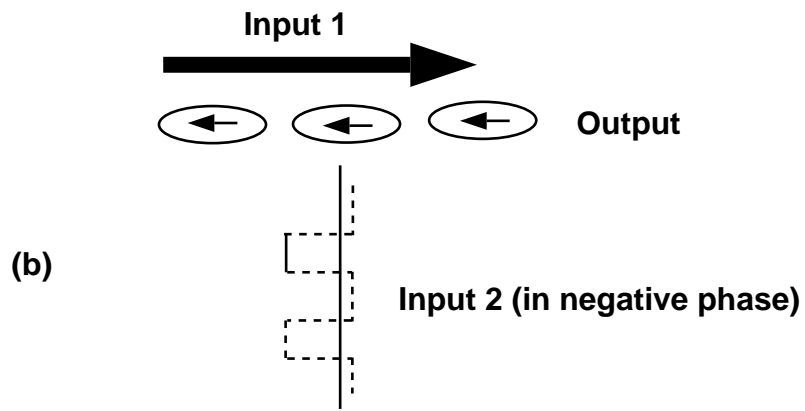

**(b)**

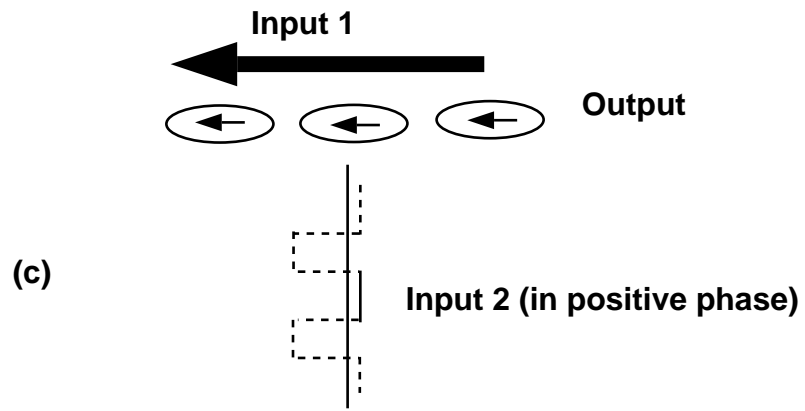

**(c)**

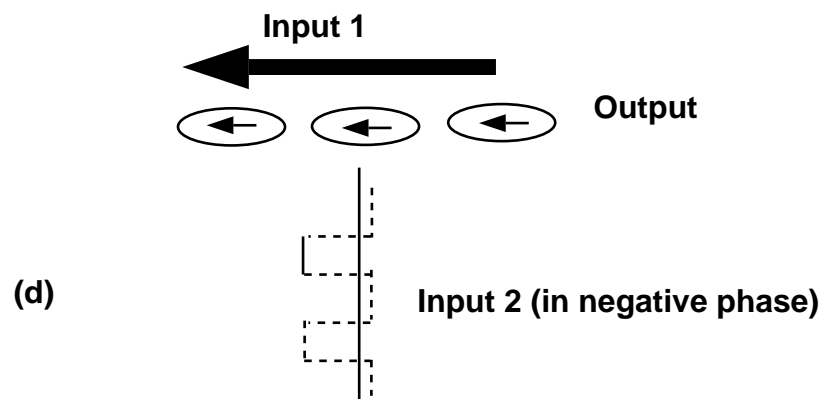

**(d)**

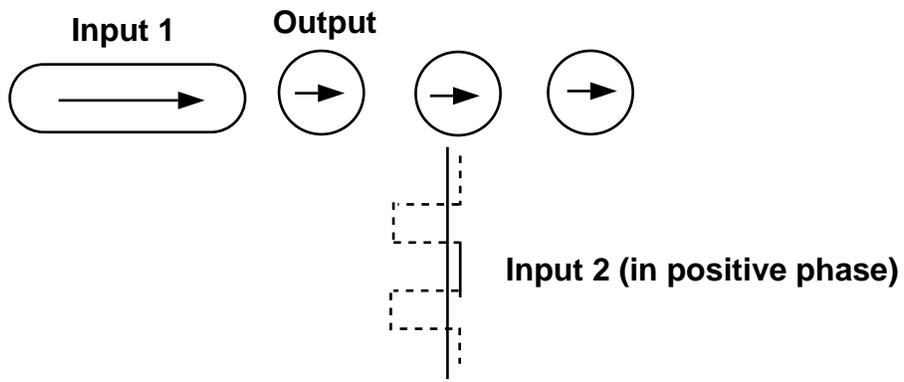
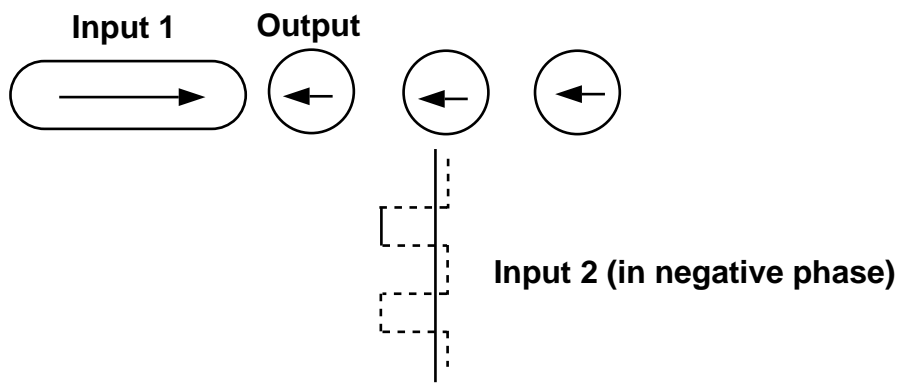
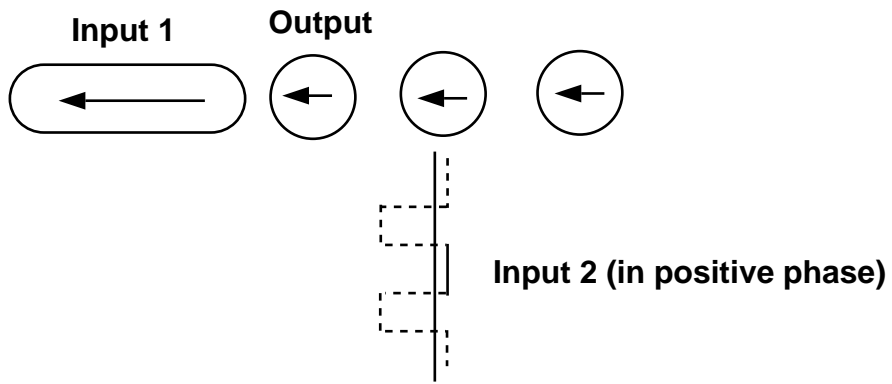
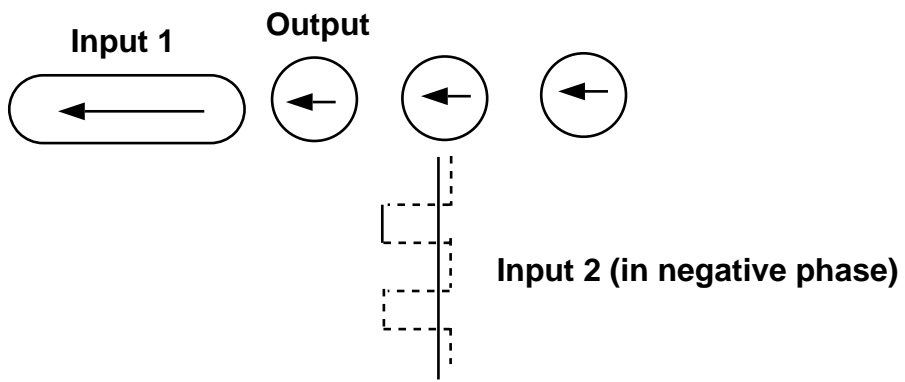

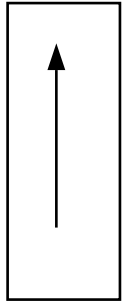

**logic 1**

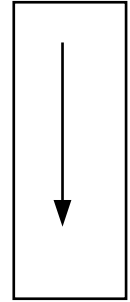

**logic 0**

**Input 1**   **Output**   **Input 2**

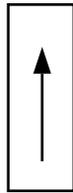 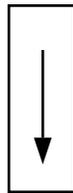 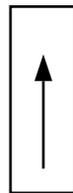

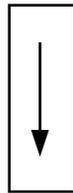 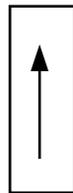 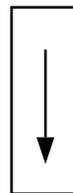

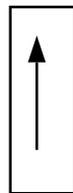 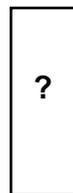 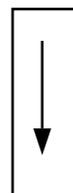

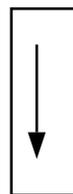 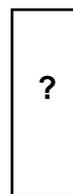 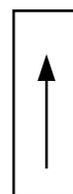

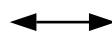 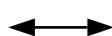

**dipole-dipole interaction**   **dipole-dipole interaction**

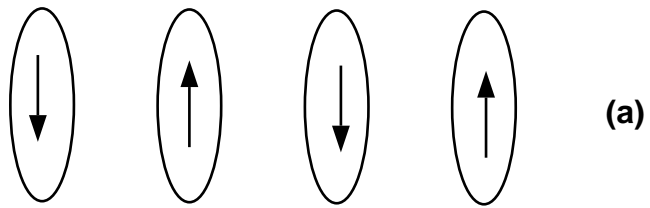

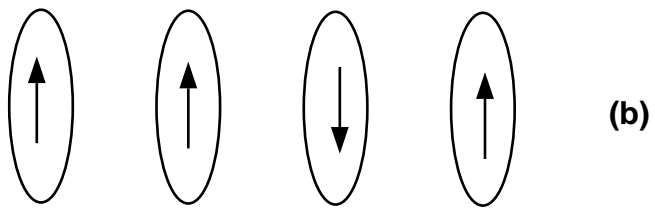

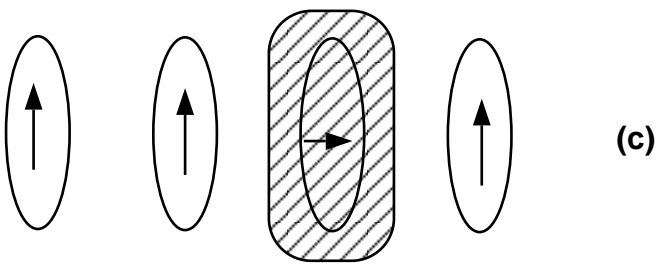

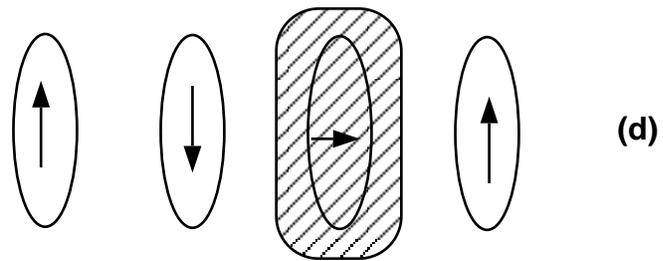

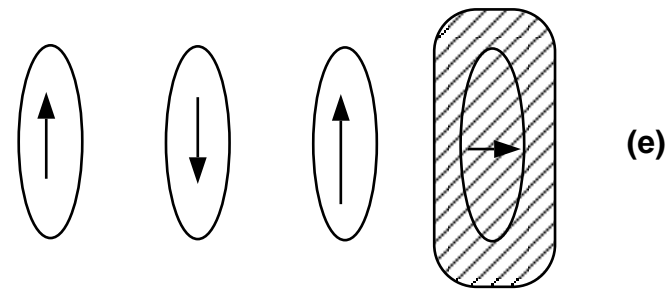

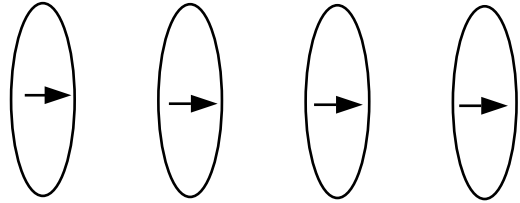

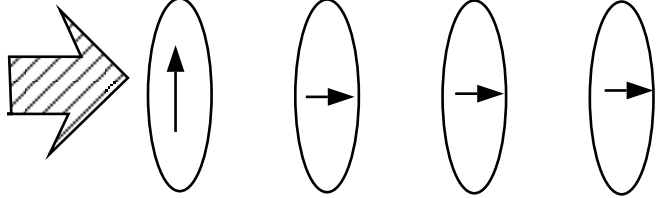

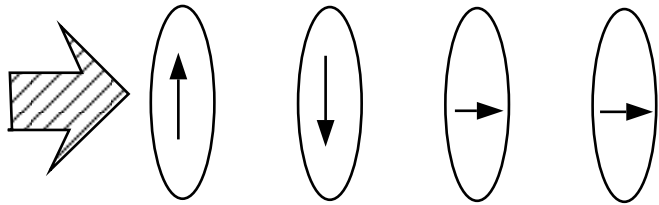

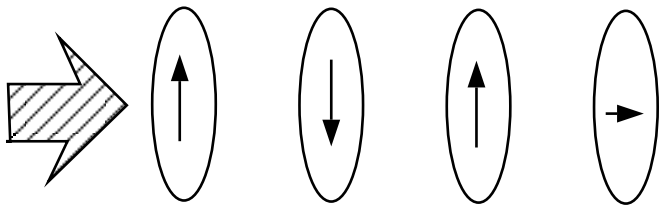

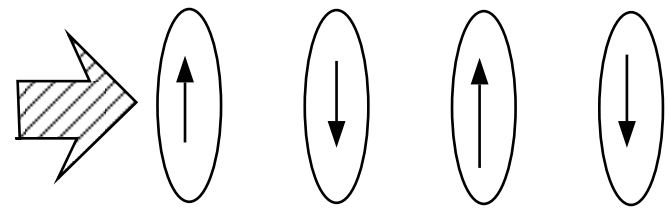

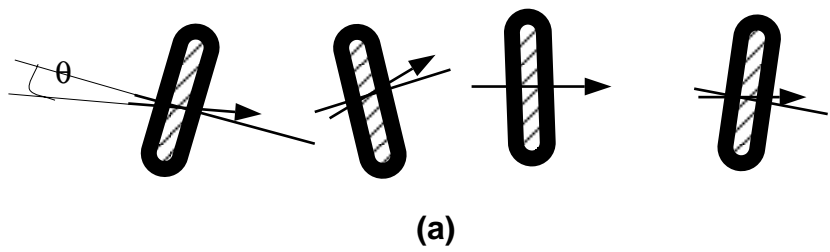

**(a)**

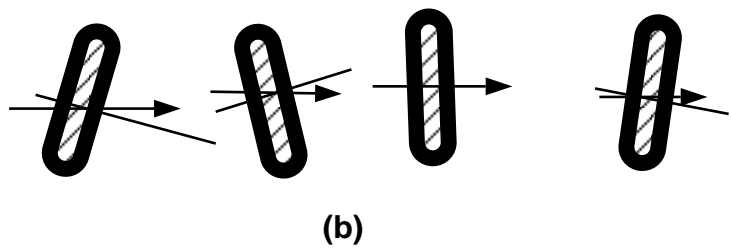

**(b)**

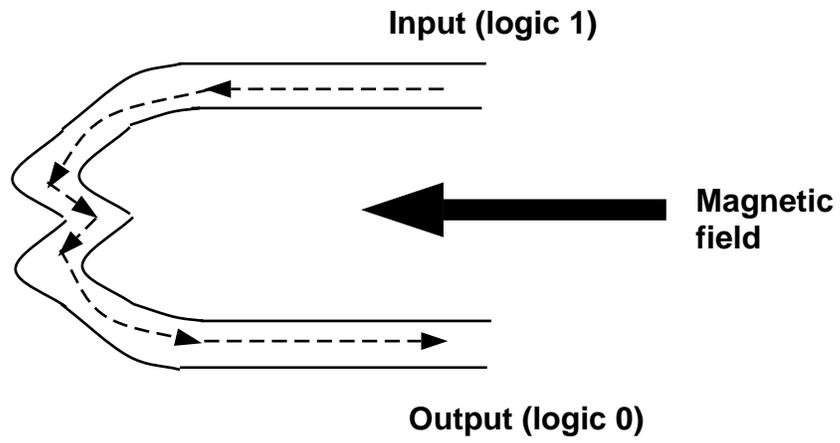

A nanowire with a cusp reverses the direction of domain wall motion and acts as a NOT gate

**(a)**

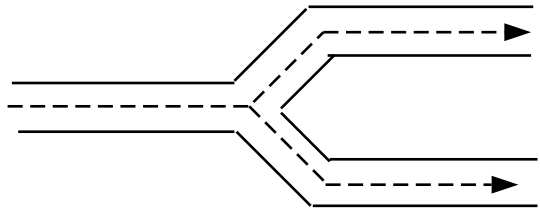

A fork provides fan out

**(b)**

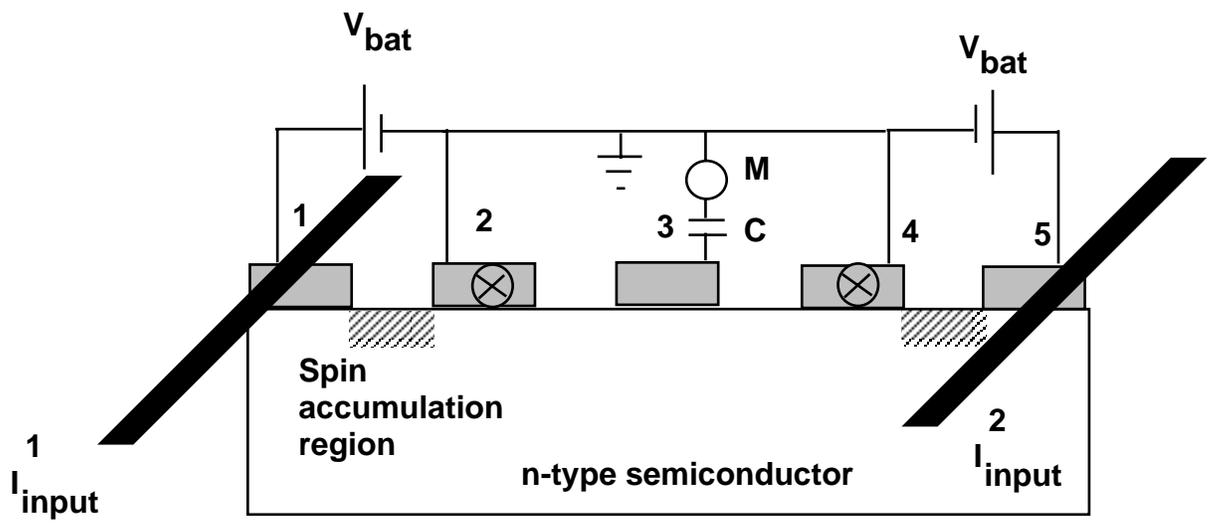

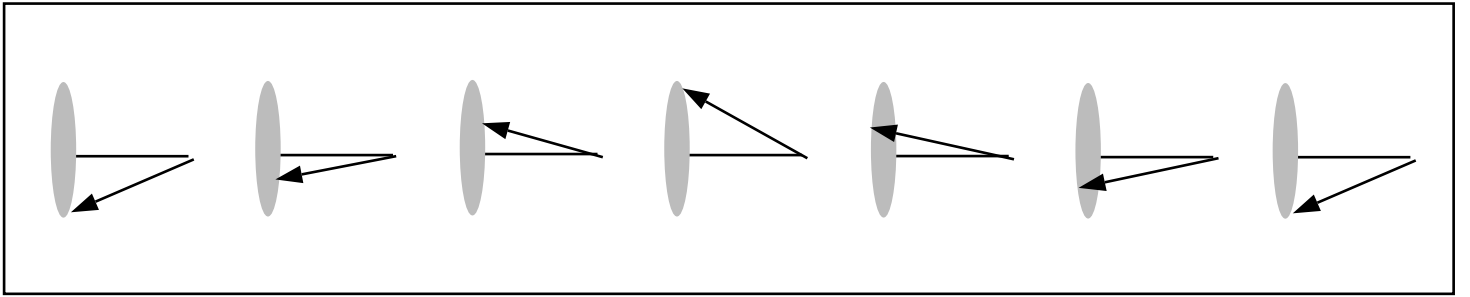

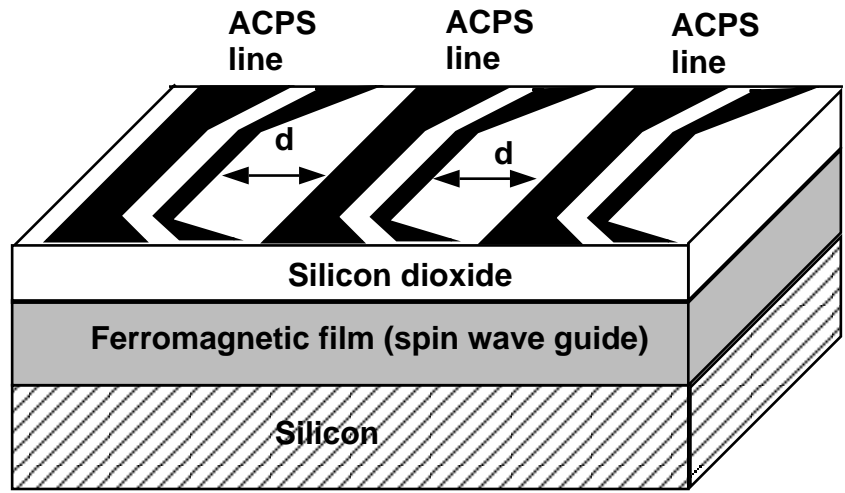